%% file: mscthesis_arxiv.tex
\documentclass[msc,amaths,a4paper,book,compact]{mcsthesis}

\usepackage{epsfig}
\usepackage[dvips]{color}
\usepackage{graphicx}
\usepackage{amssymb}

\newcommand{\eref}[1]{(\ref{eq:#1})}
\newcommand{\fref}[1]{fig.~\ref{fig:#1}}

\newcommand{\sref}[1]{\S\ref{sec:#1}}
\newcommand{\tref}[1]{table~\ref{tab:#1}}

\newcommand{\hlinevspace}{\\[-10pt]}

\renewcommand{\d}{\ensuremath{\mathrm{d}}}
\renewcommand{\O}[1]{\mathcal{O}\!\left[#1\right]}

\newcommand{\url}[1]{\texttt{#1}}

\newcommand{\sign}{\mathrm{sign}\:}
\def\rhoavg{\bar{\rho}}
\newcommand{\HI}{\textsc{Hi}}

\newcommand{\unit}[1]{\ensuremath{{\rm #1}}}
\newcommand{\fracunit}[2]{\ensuremath{\textstyle\frac{{\rm #1}}{{\rm
#2}}}}
\newcommand{\Msun}{\ensuremath{\mathcal{M}_\odot}}

\def\lint{\hbox{\Large $\displaystyle\int$}} 

\def\rr{\tilde{r}}

\def\rhoavg{\bar{\rho}}
\def\cphi{\varphi}
\def\Maple{{\sf Maple}}
\def\Mathematica{{\sf Mathematica}}

\title{Galactic halos and gravastars: static spherically symmetric spacetimes in modern general relativity and astrophysics}
\author{Tristan Faber}
\supervisor{Prof Matt Visser}
\date{January 19, 2006}

\begin{document}

\begin{abstract}
The crucial role played by pressure in general relativity is explored in the mathematically simple context of a static spherically symmetricc
geometry. By keeping all pressure terms, the standard formalisms of rotation curve and gravitational lensing observations are extended to a first
post-Newtonian order. It turns out that both post-Newtonian formalisms encode the gravitational field differently. Therefore, combined observations of
rotation curves and gravitational lensing of the same galaxy can in principle be used to infer \emph{both} the density \emph{and} pressure profile of
the galactic fluid, whereas the currently employed quasi-Newtonian formalisms only allow us to deduce the density profile.

If a suitable decomposition of the galaxy model is used to separate the dark matter from the galactic fluid, the newly introduced post-Newtonian
formalism might allow us to make inferences about the equation of state of dark matter. While the Cold Dark Matter paradigm is currently favoured in
the astrophysics and cosmology communities, the formalism presented herein offers an unprecedented way of being able to directly observe the
equation of state, and therefore either confirm the CDM paradigm or gain new insight into the nature of galactic dark matter.

In a logically distinct analysis, I investigate the effects of negative pressure in compact objects, motivated by the recently introduced 
\emph{gravastar} model. I find that gravastar like objects --- which have an equation of state that exhibits negative pressure at the core of the object ---
 can in principle mimic the external gravitational field of a black hole. Unlike a black hole, however, gravastars neither exhibit a pathological
curvature singularity at the origin nor do they posess an event horizon. Instead they are mathematically well defined everywhere. Finally, another
exotic option is considered as a mathematical alternative to black holes: The anti-gravastar, which is characterized by a core that has a negative
mass-energy density.
\end{abstract}

\frontmatter

\maketitle

\include{tc_acknowledgements}

\tableofcontents

\listoffigures
\addcontentsline{toc}{chapter}{List of figures}

\listoftables
\addcontentsline{toc}{chapter}{List of tables}

\mainmatter

\include{tc_introduction}

\include{tc_basics+definitions}

\include{tc_galactic_halo}
\include{tc_compact_objects}
\include{tc_conclusions}

\include{tc_bibliography}

\appendix

\include{ta_refractive_index_arxiv}
\include{ta_gravastar_anisotropy_arxiv}
\include{ta_galaxy_eos_arxiv}


\end{document}

%% file: tc_acknowledgements.tex
\chapter*{Acknowledgements}
\addcontentsline{toc}{chapter}{Acknowledgements}

I would like to thank everyone who supported me while writing this thesis. For many interesting and helpful discussions I want to thank C\'eline Catto\"en and Silke Weinfurtner. Special thanks to my supervisor, Prof.~Matt Visser for his tremendous support, guidance, suggestions and assistance.

\vspace{3mm}
I also thank Rowan McCaffery, Ginny Nikorima and Prema Ram for their cheerful support in the administration office.

\vspace{3mm}
Finally I want to thank my parents for encouraging my studies abroad and my wonderful wife Sheyna for her endless emotional support.

\vspace{3mm}
This work was supported by the J.~L.~Stewart Scholarship and a Victoria University of Wellington Postgraduate Scholarship for Master's Study.

%% file: tc_introduction.tex
\chapter{Introduction}

Almost a century after Einstein's publications developing his general theory of relativity, it is astounding that the mathematics of the simplest case of static spherical symmetry still is not completely exhausted. In this thesis, I am going to discuss various aspects of general relativity that are of importance to today's research and that should be considered for the interpretation of current observations.

Before I elaborate on these issues, I introduce the necessary mathematical framework of general relativity in \sref{definitions}. This includes mention of the basic ideas of general relativity, i.e. the equivalence principle, the equations of motion in general relativity, Einstein's field equations, and stellar structure in general relativity. I will also introduce curvature coordinates, 
which are mostly used throughout this work, and discuss the individual parts of the stress-energy tensor, as they are important for the following arguments.

The first topic that I want to discuss in a general relativistic context is the gravitational field of galaxies. Today, the assumption that every galaxy has a dark matter halo is commonplace, and usually justified by the observed rotation curves. However, the nature of this dark matter is completely unknown, and the only property that has been pinned down by observations so far is the density of dark matter in galaxies. This density has been extracted from two kinds of observations: (i) galactic dynamics in which the gravitational field is inferred from the relative speed of involved particles, stars, gas, etc.~and (ii) gravitational lensing where the deflection of light is used to deduce the gravitational field of the host galaxy. Since the gravitational field of a galaxy is weak, both methods are usually analysed in a (quasi-)Newtonian approximation.

I will show how to interpret both kinds of observations in a general relativistic weak field approximation that extends the usual quasi-Newtonian picture. As a consequence, combined measurements of galaxy dynamics and gravitational lensing can in principle yield additional information beyond the density of the galactic fluid, which consists of the visible parts of a galaxy and the dark matter. Combining both methods also allows one to deduce the pressure of the galactic fluid. If the accuracy of these measurements is sufficient, it may be possible to infer both the density and the pressure of galactic dark matter, and therefore gain information about the dark matter's equation of state. This could turn out to be unprecedented observational evidence for Cold Dark Matter (CDM), or even an indication of a new form of astrophysical field, such as e.g.~Scalar Field Dark Matter (SFDM). Independent of what the actual result will be, in \sref{galactic_halo} I present a new formalism that is in principle compatible with current data reduction techniques and yields both the density \textit{and} pressure of a galactic fluid from combined measurements of galactic dynamics (\sref{rotcurve}) and gravitational lensing (\sref{grav_lensing}).

The second aspect of spherically symmetric general relativity which I want to discuss in this thesis is a mathematical alternative to black holes. Just like dark matter, black holes are also widely accepted in the astrophysics community. Black holes are the most dense and compact clumps of matter that can possibly exist. However, from a general relativistic point of view, black holes exhibit a pathological property: The mathematical solution of a black hole suggests that there is a curvature singularity at the center of the black hole, which basically translates to a failure of general relativity to describe this singular point appropriately. This singularity is only acceptable because it is hidden behind the event horizon, which is like a point of no return from which even light cannot escape. Therefore, one has no possible means (even in principle) of measuring anything behind that horizon. 

In \sref{compact_objects}, I will discuss these problems in more detail along with two mathematical alternatives to black holes that exhibit the same external gravitational field but do not show any of the previously mentioned pathological behaviour. The first alternative is the \textit{gravastar} (\sref{gravastar}), a concept that has been previously explored by Mazur and Mottola \cite{Mazur:2004fk}, or similarly by Laughlin \textit{et al.}~\cite{Chapline:2000,Chapline:2002}. I will extend their layered model to a continuous model and by that show that a basic property of a gravastar is the existence of anisotropic pressures, similar to a shear stress in an anisotropic crystal. The second alternative is the \textit{anti-gravastar} model which was developed by Matt Visser and myself. For the anti-gravastar (\sref{anti-gravastar}), anisotropic pressure is not necessary, but one needs to accept a negative density at the core instead. Both alternatives are of mathematical nature only, and as of January 2006 there is no physical or observational justification to prefer them over the usual black hole picture. The discussion in this thesis merely illustrates that there are alternatives to black holes that do not exhibit pathological behaviour of the aforementioned kind at all.

After the detailed dicussions in chapters \sref{galactic_halo} and \sref{compact_objects}, I will conclude in chapter \sref{conclusions} and point out the most important features of the new findings in this thesis. The following appendices contain several articles arising from the material in this thesis that have been published or submitted for publication.

%% file: tc_basics+definitions.tex
\chapter[Mathematical framework]{Mathematical framework of general relativity}
\label{sec:definitions}

In this chapter I give all definitions for the basic concepts and symbols I will use in later chapters.

\paragraph*{Geometric units} are used by default unless otherwise stated. Thus, the speed of light $c\equiv 1$, and Newton's constant of gravitation $G_N \equiv 1$. Hence, all of mass, length, time, energy and momentum will have units of length $[L]$ and density and pressure will have units $[L^{-2}]$. Multiplication by appropriate combinations of $c$ and $G_N$ yields SI-units.

\paragraph*{The Einstein summation} convention of omitting the sum symbol whenever a pair of contravariant (index up) and covariant indices (down) appears in one term is used throughout the whole thesis:
\begin{equation}
A_\mu \, B^\mu \equiv \sum_{\mu=0}^3 \, A_\mu \, B^\mu \, ; \qquad X_a \, Y^a \equiv \sum_{a=1}^3 \, X_a \, Y^a \, .
\end{equation} 
Greek indices indicate four spacetime dimensions ($\mu,\nu,\ldots=0,1,2,3$) and latin indices indicate three spatial dimensions ($a,b,\ldots=1,2,3$).

In places where the dimensionality of the index is unambiguous, I sometimes use the ``bullet notation'' to reduce the number of lettered indices and therefore increase the legibility. Bullets (``$\bullet$'' and ``$\circ$'') indicate indices that are used for contraction:
\begin{equation}
A_\bullet \, B^\bullet \equiv A_\mu \, B^\mu \, ; \qquad A_{\bullet\circ} \, B^{\mu\bullet} \, C^{\nu\circ} \equiv A_{\alpha\beta} \, B^{\mu\alpha} \, C^{\nu\beta} \, .
\end{equation}

\section{Equivalence principle}
The theory of general relativity is based on the principle of universal free fall, also called the (weak) equivalence principle. From the observation that the inertial mass and the gravitational mass are identical to remarkable precision, one can conclude that all objects ``fall'' in the same manner, i.e. the actual trajectory of a body under the influence of a gravitational force is independent of its mass. This is not true for other forces like e.g. the electromagnetic force.

Combined with the special theory of relativity, Einstein formulated the strong equivalence principle, also called the ``Einstein equivalence principle'', which states that gravity is represented by the Christoffel connexion on a Lorentzian manifold with an associated metric tensor $g_{\mu\nu}$, and that:
\begin{itemize}
\item the universality of free fall is given by the geodesics \eref{geodesic_eqns_general_parameter} of the Christoffel connexion \eref{christoffel_1stkind}, and
\item the rules of special relativity are recovered in a local rest frame, where the metric tensor takes Minkowskian form \eref{Minkowski_metric}.
\end{itemize}
This approach guarantees that the equivalence of masses is imposed at a very basic level. Since the equations of motion are given by the Christoffel connexion and hence, the metric tensor, Einstein still needed to find field equations \eref{Einstein_eqns} that relate the metric to the mass-energy which acts as the source of the gravitational field.

I will now summarize these fundamental aspects of general relativity to set the framework for the following chapters.

\section{Geodesics and affine parameters}

Geodesics are defined as the ``straightest possible'' curves in a manifold. Straight means that tangent vectors to a geodesic remain parallel to each other along the curve. Parallelism in a manifold is given by the notion of an affine connexion $\Gamma^\mu_{~\alpha\beta}$. Please refer to the standard literature about differential geometry and non-Euclidean spaces for further details, e.g. \cite{Hartle03,MTW,Visser:DGNotes,weinberg72}.

The covariant derivative $\nabla_\beta$ with respect to the coordinate $X^\beta$ is the generalization of taking derivatives in curved spaces. The departure from flat space is represented by the affine connexion $\Gamma^\mu_{~\alpha\beta}$. Applied to a vector $A^\alpha$, the covariant derivative is given by
\begin{equation}
\nabla_\beta \, A^{\alpha} \equiv \partial_\beta \, A^{\alpha} + \Gamma^\alpha_{~\bullet\beta} \, A^{\bullet} \, ,
\end{equation} 
where $\partial_\beta$ denotes the partial derivative with respect to the coordinate $X^\beta$. Without derivation, I note that a curve with an arbitrary parametrization $\chi$ is a geodesic iff its tangent vectors,
\begin{equation}
t^\mu \equiv \frac{\d X^\mu}{\d\chi} \, ,
\end{equation} 
satisfy the differential equation
\begin{equation}
\label{eq:geodesic_eqns_general_parameter}
\frac{\d^2 X^\mu}{\d\chi^2} + \Gamma^\mu_{~\alpha\beta}\, \frac{\d X^\alpha}{\d\chi}\, \frac{\d X^\beta}{\d\chi} = f(\chi) \, \frac{\d X^\mu}{\d\chi} \, ,
\end{equation} 
or equivalently,
\begin{equation}
\label{eq:geodesic_with_tangents}
t^\beta \, \nabla_\beta\, t^\mu = f(\chi)\, t^\mu \, ,
\end{equation} 
where $f(\chi)$ is an arbitrary function of the parameter $\chi$. If $\chi$ is chosen in such a manner that $f(\chi)$ vanishes, $\chi$ is called an affine parameter. In that case it is possible to write \eref{geodesic_with_tangents} in index free notation,
\begin{equation}
\label{eq:geodesic_affine_parameter_indexfree}
t \cdot \nabla t = 0 \, ,
\end{equation} 
to illustrate that the tangent vectors $t$ along a geodesic curve have constant length. This is exactly what an affine parameter does, it parametrizes the curve in such a way that the ``parameter speed'' along that curve is constant. For later reference, I note that in the usual component notation, the geodesic equations for an affine parameter $\chi$, \eref{geodesic_affine_parameter_indexfree}, take the form 
\begin{equation}
\label{eq:geodesic_affine_parameter}
\frac{\d^2 X^\mu}{\d\chi^2} + \Gamma^\mu_{~\alpha\beta}\, \frac{\d X^\alpha}{\d\chi}\, \frac{\d X^\beta}{\d\chi} = 0 \, .
\end{equation} 

\subsection{Metric spaces}
\label{sec:geodesics_in_metric_spaces}

The metric connexion without torsion is the usual connexion that is used in general relativity. In this case, the affine connexion is given by the Christoffel symbols of the first kind,
\begin{equation}
\label{eq:christoffel_1stkind}
\Gamma_{\mu\alpha\beta} \equiv \left\lbrace \mu ; \alpha\beta \right\rbrace \equiv \frac{1}{2} \left( \partial_\alpha \, g_{\mu\beta} + \partial_\beta \, g_{\mu\alpha} - \partial_\mu g_{\alpha\beta} \right) \, ,
\end{equation} 
or the Christoffel symbols of the second kind,
\begin{equation}
\Gamma^\mu_{~\alpha\beta} =
\left\lbrace \!\! \begin{array}{c} \mu\\ \alpha\beta \end{array} \!\! \right\rbrace
= \frac{1}{2}\,g^{\mu\bullet} \left( \partial_\alpha \, g_{\bullet\beta} + \partial_\beta \, g_{\bullet\alpha} - \partial_\bullet g_{\alpha\beta} \right) \, ,
\end{equation}
where the contravariant metric tensor $g^{\mu\nu}$ was used to raise the index $\mu$. To lower an index one uses the covariant metric $g_{\mu\nu}$:
\begin{equation}
g_{\mu\bullet} \, A^{\bullet\nu} = A_\mu^{~~\nu} = g^{\bullet\nu} \, A_{\mu\bullet} \, .
\end{equation}
In flat space and local orthonormal frames, the Minkowski metric is used to raise and lower indices:
\begin{equation}
\label{eq:Minkowski_metric}
\eta_{\mu\nu} = \eta^{\mu\nu} = \mathrm{diag} [-1,1,1,1] \, .
\end{equation} 
For more details on dual spaces and index gymnastics see the introductory literature to differential geometry and general relativity, e.g. \cite{Hartle03,MTW,Visser:DGNotes,weinberg72}.

In metric spaces the geodesic notion of ``straightest possible'' curve coincides with the shortest possible curve in Riemannian geometries (with positive definite metric tensor) and with an ``extremal distance'' in Lorentzian geometries (where the metric tensor has the signature $-++\,+$, i.e. that of the Minkowski metric $\eta_{\mu\nu}$).

Let $g_{\mu\nu}$ be the metric of the $3$- or $4$- (or even $n$-) dimensional Riemannian or Lorentzian space, and $\chi$ be a parameter of an arbitrary curve. Then I define arclength $s$ of that curve analogously to Euclidean arclength in flat space \cite{Visser:DGNotes},
\begin{equation}
\label{eq:def_arclength}
s = \int \frac{\d s}{\d\chi}\: \d\chi \equiv \lint \sqrt{ g_{\mu\nu} \, \frac{\d X^\mu}{\d\chi} \, \frac{\d X^\nu}{\d\chi} } \: \d\chi \, .
\end{equation} 
Applying the Euler-Lagrange equations of variational calculus yields a differential equation that characterizes the path which has the shortest arclength $s$ between the fixed endpoints of the curve along which the integral is evaluated:
\begin{equation}
\frac{\d}{\d\chi} \left[ \frac{\partial\,\sqrt{ g_{\mu\nu} \, \frac{\d X^\mu}{\d\chi} \, \frac{\d X^\nu}{\d\chi} }}{\partial (\d X^\alpha/\d\chi)}  \right] - \frac{\partial\,\sqrt{ g_{\mu\nu} \, \frac{\d X^\mu}{\d\chi} \, \frac{\d X^\nu}{\d\chi} }}{\partial X^\alpha} = 0 \, ,
\end{equation} 
which is
\begin{equation}
\label{eq:geodesic_inbetween}
\frac{\d}{\d\chi} \left[ \frac{1}{\sqrt{ g_{\mu\nu} \, \frac{\d X^\mu}{\d\chi} \, \frac{\d X^\nu}{\d\chi} }} \, g_{\alpha\beta}\, \frac{\d X^\beta}{\d\chi} \right] -
\frac{1}{2\sqrt{ g_{\mu\nu} \, \frac{\d X^\mu}{\d\chi} \, \frac{\d X^\nu}{\d\chi} }} \, \frac{\partial g_{\beta\gamma}}{\partial X^\alpha}\, \frac{\d X^\beta}{\d\chi}\, \frac{\d X^\gamma}{\d\chi} = 0 \, .
\end{equation}
If the curve is not a null curve in a Lorentzian space (for which $s=0$ everywhere), this can be reparametrized in terms of the arclength $s$. Hence, I multiply by
\begin{equation}
\frac{\d\chi}{\d s} = \frac{1}{\sqrt{ g_{\mu\nu} \, \frac{\d X^\mu}{\d\chi} \, \frac{\d X^\nu}{\d\chi} }}
\end{equation} 
which simplifies \eref{geodesic_inbetween} tremendously to
\begin{equation}
\frac{\d}{\d s} \left[ g_{\alpha\beta}\, \frac{\d X^\beta}{\d s} \right] -
\frac{1}{2} \, \frac{\partial g_{\beta\gamma}}{\partial X^\alpha}\, \frac{\d X^\beta}{\d s}\, \frac{\d X^\gamma}{\d s} = 0 \, .
\end{equation} 
Expanding the first term and rearranging the derivatives of the metric $g_{\alpha\beta}$ gives
\begin{equation}
g_{\alpha\beta}\, \frac{\d^2 X^\beta}{\d s^2} + \left\lbrace \frac{\partial g_{\alpha\beta}}{\partial X^\gamma}  - \frac{1}{2} \, \frac{\partial g_{\beta\gamma}}{\partial X^\alpha} \right\rbrace \, \frac{\d X^\beta}{\d s}\, \frac{\d X^\gamma}{\d s} = 0 \, ,
\end{equation} 
which is easily identified as the geodesic equation with the Christoffel symbols of the first kind \eref{christoffel_1stkind},
\begin{equation}
\label{eq:geodesic_eqns_1st_christoffel}
g_{\alpha\beta}\, \frac{\d^2 X^\beta}{\d s^2} + \Gamma_{\alpha\beta\gamma} \, \frac{\d X^\beta}{\d s}\, \frac{\d X^\gamma}{\d s} = 0 \, ,
\end{equation} 
or equivalently, using the Christoffel symbols of the second kind,
\begin{equation}
\frac{\d^2 X^\alpha}{\d s^2} + \Gamma^\alpha_{~\beta\gamma} \, \frac{\d X^\beta}{\d s}\, \frac{\d X^\gamma}{\d s} = 0 \, .
\end{equation} 
Hence, a geodesic describes the shortest possible path in a Riemannian space where arclength $s$ is defined as in \eref{def_arclength}. In a Lorentzian space, this calculation requires $\d s \neq 0$, hence it does not apply to null curves. For all timelike and spacelike curves, the geodesic is equivalent to the curve with a locally extremal arclength $s$. 

This calculation also shows that for metric spaces, arclength $s$ is an affine parameter, since \eref{geodesic_affine_parameter} is always satisfied for positive definite metrics or spacelike and timelike geodesics in Lorentzian spaces.

\subsection{Affine parameters for null curves}
The affine parameter for lightlike geodesics, also called null curves, cannot be easily specified in general, since the motion of massless particles at the speed of light is characterized by the vanishing of the invariant interval,
\begin{equation}
\label{eq:null_invariant_interval}
\d s^2 = g_{\mu\nu}\, \d X^\mu\, \d X^\nu = 0 \, .
\end{equation} 
Hence, the previous derivation of affine parameters for time- and spacelike curves is invalid for null curves, due to a division by zero. Instead the affine parameter has to be determined for each metric tensor $g_{\mu\nu}$. I will only consider static spacetimes, which are the only kind of spacetimes used in this thesis.

If a spacetime is stationary, the metric is independent of the time coordinate, $\partial_t\, g_{\mu\nu} =0$ and therefore, by the invariance of infinitesimal coordinate translations in the $t$-direction, $K^\mu=\delta^{\mu}_{~t}$ is a timelike Killing vector \cite[\S 25.2]{MTW}. From the Killing equation follows that the covariant derivative of the Killing vector is completely antisymmetric in its two indices and hence,
\begin{equation}
t^\bullet \nabla_\bullet (t^\circ K_\circ) = (t^\bullet \nabla_\bullet t^\circ) K_\circ + t^\bullet t^\circ (\nabla_\bullet K_\circ) = (t^\bullet \nabla_\bullet t^\circ) K_\circ \, ,
\end{equation} 
where $t^\mu= \d X^\mu/\d\chi$ is tangent vector to a null geodesic. But by \eref{geodesic_affine_parameter}, this must vanish if $\chi$ is an affine parameter. Consequently, the quantity
\begin{equation}
\label{eq:killing_conserved_stationary}
E \equiv t^\bullet K_\bullet = g_{\bullet\circ}\, t^\bullet K^\circ = g_{\bullet t}\, \frac{\d X^\bullet}{\d\chi} = \frac{\d t}{\d\chi} \left( g_{tt} + g_{ti}\,\frac{\d X^i}{\d t}  \right)
\end{equation}
is conserved along all geodesics with an affine parameter -- it does not matter whether the geodesic is null or not. However, since the affine parameter for null geodesics is still lacking an operational definition, I rearrange \eref{killing_conserved_stationary} into a definition for the affine parameter $\chi$ of null geodesics,
\begin{equation}
\d\chi \equiv -\, \d t \, \left( g_{tt} + g_{ti}\,\frac{\d X^i}{\d t}  \right) \, ,
\end{equation}
where the constant $-E$ was absorbed into the affine parameter $\chi$, as the overall normalization of an affine parameter is irrelevant. In the more restricted case of a static spacetime (with $g_{ti}=0$), the affine parameter of null curves takes an even simpler form:
\begin{equation}
\label{eq:null_affine_parameter}
\d\chi = -\, \d t \, g_{tt} \, .
\end{equation}
As a note on the side, this can also be expressed using the three-dimensional proper distance,
\begin{equation}
\d\chi = \sqrt{-g_{tt}} \, \sqrt{g_{ij} \, \d X^i \, \d X^j} \, ,
\end{equation}
which can be derived with the help of the invariant interval \eref{null_invariant_interval} for static spacetimes:
\begin{equation}
g_{tt}\,\d t^2 + g_{ij} \, \d X^i \, \d X^j = 0 \, .
\end{equation}

\section{Curvature coordinates}

In this work, I will concentrate on static, spherically symmetric fluid spheres. It is a standard result, see e.g. \cite[\S 23.2]{MTW} or also \cite[\S 8.1]{weinberg72}, that all geometries of this category are represented by a metric of the form:
\begin{equation}\label{eq:metric_standard_form}
\d s^{2}\equiv g_{\mu\nu} \, \d X^\mu \, \d X^\nu =
-A(r)\, \d t^{2}
+B(r)\, \d r^{2}
+r^{2}\, \d\Omega^2 \, ,
\end{equation} 
where $g_{\mu\nu}$ are the metric components and $X^\mu$ is the set of coordinates. The arbitrary functions $A(r)$ and $B(r)$ depend on $r$ only\footnote{Weinberg \cite[\S 8.1]{weinberg72} calls metrics of this type ``static and isotropic'', where \textit{static} refers to independence of the time coordinate (as usual) and \textit{isotropic} is the property that the metric is invariant under rotation -- which is identical to the term \textit{spherically symmetric} in this thesis. Weinberg's slightly unusual use of the term \textit{isotropic} is not to be confused with isotropic fluids that will be mentioned later on, nor with the distinct concept of an \textit{isotropic coordinate system} which is also commonly used in general relativity, and later in this work.}, $t$ is the time coordinate, $r$ is the space coordinate in the radial direction of the sphere and
\begin{equation}
\label{eq:metric_unit_sphere}
\d\Omega^2 \equiv \d\theta^{2} + \sin^2\theta\, \d\cphi^2
\end{equation} 
is the metric of the two-dimensional unit sphere with the two spherical polar coordinates $\theta$ and $\cphi$. Coordinates of this type are called ``curvature coordinates'' or sometimes ``Schwarzschild coordinates''. This choice of $r$ coordinate has the advantage of a physically clear meaning of $r=(\mathrm{proper~circumference})/2\pi$, where ``proper circumference'' refers to the integrated proper distance interval, $\d s$, on a closed circle with its center at the coordinate origin:
\begin{equation}
\hbox{(proper~circumference)} = \int_{\cphi=0}^{\cphi=2\pi} \d s = r \int_{0}^{2\pi} \d\cphi = 2\pi\,r \, .
\end{equation} 
Here $\d t=0$, $\d r=0$ and, due to spherical symmetry one can choose $\theta=\pi/2$ \cite[\S 23.3]{MTW}. While this seems to be trivial at first, note that the proper distance in radial direction is
\begin{equation}
\hbox{(proper~radius)} = \int_{r=r_1}^{r=r_2} \d s = \int_{r_1}^{r_2} \sqrt{B(r)}\, \d r \neq r_2 - r_1 \, .
\end{equation} 

Depending on the problem under investigation, it is generally advisable to replace the functions $A(r)$ and $B(r)$ by functions which are more descriptive in the context of that problem. I choose to use $A(r) \equiv \exp[2\Phi(r)]$ and $B(r) \equiv 1/(1-2m(r)/r)$ throughout this work unless otherwise stated:
\begin{equation}\label{eq:ss_metric}
\d s^{2}= 
-e^{2\Phi(r)}\, \d t^{2}
+\frac{\d r^{2}}{1-2m(r)/r}
+r^{2}\, \d\Omega^{2} \, .
\end{equation} 
To understand the physical meaning of the two metric functions, $\Phi(r)$ and $m(r)$, it is necessary to invoke the Einstein field equations.

\section{Einstein equations}

The Einstein field equations are derived in basically every standard textbook concerning general relativity, for example \cite{Hartle03,MTW,weinberg72}:
\begin{equation}\label{eq:Einstein_eqns}
R_{\mu\nu} - \frac{1}{2} g_{\mu\nu}\, R = 8\pi \, T_{\mu\nu}
\end{equation} 
where $R_{\mu\nu}$ is the Ricci curvature tensor, $R \equiv g^{\mu\nu}\,R_{\mu\nu}$ is the Ricci scalar and $T_{\mu\nu}$ is the stress-energy tensor. In SI-units, $8\pi$ is replaced by $8\pi\, G_N/c^4$. The combination
\begin{equation}
G_{\mu\nu} \equiv R_{\mu\nu} - \frac{1}{2} g_{\mu\nu}\, R
\end{equation} 
is called the Einstein tensor, which can be obtained through a tedious but straightforward calculation once the metric components $g_{\mu\nu}$ are given. Nowadays, this monotonous task is usually executed by a computer algebra program such as \Maple\footnote{~\copyright~ by Maplesoft, a division of Waterloo Maple Inc.} or \Mathematica\footnote{~\copyright~ by Wolfram Research, Inc.}. 

\paragraph*{Geometric part of the field equations.}

The left hand side of \eref{Einstein_eqns} represents the geometry of the spacetime and is given as a non-linear combination of the metric components $g_{\mu\nu}$ and their first and second derivatives. For completeness, I shall define the Riemann curvature tensor,
\begin{equation}
R^\mu_{~\nu\alpha\beta} \equiv \partial_\alpha\, \Gamma^\mu_{~\nu\beta} - \partial_\beta\, \Gamma^\mu_{~\nu\alpha} + \Gamma^\mu_{~\bullet\alpha}\, \Gamma^\bullet_{~\nu\beta} -  \Gamma^\mu_{~\bullet\beta}\, \Gamma^\bullet_{~\nu\alpha} \, ,
\end{equation} 
and the Ricci tensor which is given by the contraction over the first and third index of the Riemann tensor:
\begin{equation}
R_{\mu\nu} \equiv R^\bullet_{~\mu\bullet\nu} \, .
\end{equation}
As a purely geometric result of metric spaces, the Einstein tensor obeys the contracted Bianchi identity,
\begin{equation}
\label{eq:Bianchi_ID}
\nabla_\beta \, G^{\alpha\beta} = 0 \, .
\end{equation} 

\paragraph*{Components of the Einstein tensor for given metric.}

Since the Einstein tensor is completely determined by the metric, I shall give the components of the Einstein tensor, as derived from \eref{ss_metric}, for further reference. The significance of second rank tensors with one index up and one index down is explained later in \sref{iupidn}. Prime denotes the partial derivative with resepect to the $r$-coordinate: $' \equiv \d/\d r$. The follwing components of the Einstein tensors are the only non-zero components of the metric \eref{ss_metric} which were derived using a computation by {\sf Maple} and manipulations by hand:
\begin{eqnarray}
\label{eq:Gtt}
G^t_{~t} &=& - \frac{2\, m'}{r^2} \\
\label{eq:Grr}
G^r_{~r} &=& - \frac{2}{r^2}\, \left[ \frac{m}{r} - r\,\Phi'\left( 1-\frac{2\,m}{r} \right)  \right]  \\
\label{eq:Gthetaphi}
G^\theta_{~\theta} = G^\cphi_{~\cphi} &=& - \frac{(m'\,r-m)(1+r\,\Phi')}{r^3} + \left( 1-\frac{2\,m}{r} \right) \left[ \frac{\Phi'}{r} + \Phi'^2 +\Phi'' \right] 
\end{eqnarray}

\section{Stress-energy tensor}
\label{sec:stress-energy}

The righthand side of \eref{Einstein_eqns} is the source term in form of a stress-energy tensor that describes the matter or field which is creating the curvature in spacetime. The
stress-energy tensor --- also called more accurately the \textit{energy momentum tensor} --- is generally the sum of all different kinds of stress-energy present in the system under investigation, e.g. fluid stress-energy, electro-magnetic stress-energy, stress-energy arising from other fields, etc. :
\begin{equation}
T^{\mu\nu} = T^{\mu\nu}_{\mathrm{fluid}} + T^{\mu\nu}_{\mathrm{e-m}} + T^{\mu\nu}_{\mathrm{fields}} + \cdots \, .
\end{equation} 
The stress-energy tensor of a swarm of $n$ particles indexed by the particle number $i$ is given by \cite[2.8.5a]{weinberg72}
\begin{equation}
\label{eq:set_swarm}
T^{\alpha\beta}_{\mathrm{swarm}}(X^\mu) = \sum_{i=1}^n m_{0,i}\, \int V_i^\alpha\, V_i^\beta\, \delta^4(X^\mu-X_i^\mu(\tau))\, \d\tau
\end{equation}
where $\delta^4(\cdot)$ denotes the four-dimensional delta-function and $m_{0,i}$ is the rest-mass of particle $i$ with the four-velocity $V_i^\alpha \equiv \d X_i^\alpha / \d\tau$ at the position given by the coordinates $X_i^\mu$. Furthermore $\tau$ is the proper time that parametrises the trajectories of all particles. Instead of referencing the individual particles by their indices $i$, one can write the stress-energy tensor in terms of the local number density $n(X^\mu)$, the mass of the particle at $X^\mu$, $m(X^\mu)$, and the local four-velocity $V^\alpha(X^\mu)$ of the particle at $X^\mu$ \cite[22.17]{Hartle03}:
\begin{equation}
\label{eq:set_cont_fluid}
T^{\alpha\beta}_{\mathrm{swarm}}(X^\mu) = m\,n\, V^\alpha\, V^\beta \, .
\end{equation}
At this point some amount of averaging takes place: the mass of the particle at $X^\mu$ is actually the average mass of the particles in the immediate vicinity of $X^\mu$, using the number density of the associated species as a weight for averaging. If one is only dealing with one species of particles in the fluid, $m(X^\mu)$ is of course a constant.

From now on, I will drop the index ``swarm'' of the stress-energy tensor, since it is the only type that is considered in the rest of the thesis. However, I will use it in a slightly different form that corresponds to a continuous fluid. 

To see that, let us consider a small volume $\Delta V$ at rest in the observer's orthonormal frame which is part of a constant-$t$ three-surface with an associated timelike normal vector $\hat{v}_\beta=(1;0;0;0)$. The product \cite[22.13]{Hartle03}
\begin{equation}
\Delta P^\alpha = T^{\hat\alpha\hat\beta}\, \hat{v}_\beta\, \Delta V = T^{\hat\alpha\hat{t}}\, \Delta V\end{equation} 
is identified with the four-momentum $P^\mu \equiv m_0 V^\mu$ of the fluid in that small volume \cite[\S 5.3]{MTW} so that we can attribute physical meaning to the components of the stress-energy tensor in the observer's orthonormal frame (denoted by the hats on the indices):
\begin{eqnarray}
T^{\hat{t}\hat{t}} &=& \frac{\Delta P^t}{\Delta V} \; = \; \rho \quad \mbox{(mass-energy density)} \\
T^{\hat{i}\hat{t}} &=& \frac{\Delta P^i}{\Delta V} \; = \; \pi^i \quad \mbox{(momentum density in direction $i$)} \, .
\end{eqnarray} 
Similarly, we can look at a three-volume $\Delta t\Delta X^2\Delta X^3$ with a spacelike normal vector in the $X^1$ direction \cite[22.18]{Hartle03}:
\begin{equation}
\Delta P^\alpha = T^{\hat\alpha\hat{1}}\, \Delta t\Delta A^1
\end{equation}
where $\Delta A^1 = \Delta X^2\Delta X^3$ is the spacelike two-area normal to the $X^1$ direction. We can repeat this procedure for all three space directions and find:
\begin{eqnarray}
T^{\hat{t}\hat{j}} &=& \frac{\Delta P^t}{\Delta A^j\Delta t} \; = \; \frac{\Delta P^t}{\Delta V}\,\frac{\Delta X^j}{\Delta t} \; = \; \pi^j \quad \mbox{(energy flux in direction $j$)} \\
T^{\hat{i}\hat{j}} &=& \frac{\Delta P^i/\Delta t}{\Delta A^j} \; = \; \frac{\mbox{(force)}}{\mbox{(area)}} \; = \; f^{ij} \quad \mbox{(stress tensor)} \, .
\end{eqnarray} 
Due to the equivalence of mass and energy in general relativity, one realises that momentum density and energy flux are actually the same physical quantity $\pi^i$. This has to be so since the stress-energy tensor is symmetric in its two indices, as can easily been seen from \eref{set_swarm} and \eref{set_cont_fluid}. The components of the 3-dimensional stress tensor $f^{ij}$ are the $i$ components of a force per unit area -- also called a stress in classical mechanics -- exerted across a surface with normal in direction $j$ \cite[22.22]{Hartle03}.

The easiest example of a stress tensor is pressure in a fluid as seen by a comoving observer: the fluid is at rest relative to the observer and thus, the forces exerted is constant in all directions and the force per unit area is simply the pressure $p$ \cite[22.25]{Hartle03}:
\begin{equation}
\label{eq:set_perfect_stress}
T^{\hat{i}\hat{j}} = p\, \delta^{ij} \, ,
\end{equation} 
where $\delta^{ij} = \mathrm{diag} [1,1,1]$ is the Kronecker delta with two contravariant indices.

Again, some more averaging took place in going from a particle swarm to the notion of a continuous fluid. When we are talking about a force that is exerted across a surface, we need to realise that in a fluid, that force is created by momentum transfer between the interacting particles. So we either keep track of the exact motion of all particles, as is suggested by \eref{set_swarm} and even \eref{set_cont_fluid}, and the stress tensor drops out implicitly; or we average over the small fluctuating motion that creates the pressure (Brownian motion) and insert the stress tensor explicitly into the stress-energy tensor. Due to quantum fluctuations, there will always be some amount of random motion in the fluid. If one wants to speak about a fluid at rest -- which is really a fluid in dynamical equilibrium -- one has to average over the small fluctuations and introduce the pressure explicitly. If there is no interaction between the particles whatsoever, no pressure can be exerted within the ``fluid''.

The notion of pressure is important in this context because it contributes to generating the gravitational field as I will soon show. This contribution arising from the pressure will play a vital role in the formalism which I will introduce in \sref{galactic_halo}.

\subsection{Perfect fluid}

An isotropic fluid, like the one that was considered in \eref{set_perfect_stress}, is said to be perfect when heat conduction, viscosity or other transport and dissipative processes are negligible \cite[\S 22.2]{Hartle03}. Such a fluid is free of shear stress in the rest frame which implies that the stress tensor is diagonal and has 3 identical eigenvalues \cite[\S 5.5]{MTW}:
\begin{equation}
T^{\hat{i}\hat{j}} \propto \delta^{ij} \, .
\end{equation} 
Hence, a perfect fluid in an orthonormal rest frame takes the form
\begin{equation}
\label{eq:set_diag_perfluid}
T^{\hat\mu\hat\nu} = \mathrm{diag} [\rho,p,p,p] = \rho\, \delta_{~t}^\mu\delta_{~t}^\nu + p\, \delta_{~i}^\mu\delta_{~j}^\nu\delta^{ij} \, ,
\end{equation} 
which can also be written in terms of the flat space Minkowski metric of the orthonormal frame, $\eta^{\mu\nu}=\mathrm{diag} [-1,1,1,1]$, and the four-velocity of the observer at rest $u^\alpha=(1;0;0;0)$:
\begin{eqnarray}
\nonumber
T^{\hat\mu\hat\nu} &=& \rho\, u^\mu u^\nu + p\, (\eta^{\mu\nu}+u^\mu u^\nu) \\
&=& (\rho+p)\, u^\mu u^\nu + p\, \eta^{\mu\nu} \, ,
\end{eqnarray} 
The natural generalisation of this perfect fluid stress-energy tensor is to permit arbitrarily moving observers with corresponding four-velocity $u^\alpha$ and to replace the locally flat spacetime metric of the orthonormal frame, $\eta^{\mu\nu}$, with the metric of the curved spacetime, $g^{\mu\nu}$ (we can now drop the hats, since we are no longer in orthonormal coordinates) \cite[22.39]{Hartle03}:
\begin{equation}
\label{eq:set_gen_perfect_fluid}
T^{\mu\nu} = (\rho+p)\, u^\mu u^\nu + p\, g^{\mu\nu} \, .
\end{equation}
We can see that this generalisation is consistent with the Einstein equations \eref{Einstein_eqns} by noting that because of the general relativistic Euler and continuity equations, \eref{set_gen_perfect_fluid} satisfies the covariant conservation of stress-energy,
\begin{equation}
\label{eq:set_cov_cons}
\nabla_\nu \, T^{\mu\nu} = 0 \, ,
\end{equation} 
which is necessary to fulfill the contracted Bianchi identity \eref{Bianchi_ID}.

\subsection{Diagonal metric}
\label{sec:iupidn}

When working with a diagonal metric $g_{\mu\nu}$, one can benefit from using second rank tensors with one contra- and one covariant index. A short calculation shows that the components of a diagonal second rank tensor with one index up and one index down do not change when the coordinates are transformed into an orthonormal basis, i.e.
\begin{equation}
\label{eq:set_1up1down}
A^{\hat\alpha}_{~\hat\beta} = A^\alpha_{~\beta}\,\delta^{\hat\alpha}_{~\alpha}\, \delta^{\beta}_{~\hat\beta} = A^\alpha_{~\beta} \, .
\end{equation} 
An orthonormal basis $(e_{\hat\alpha})^\alpha$ in a space with signature $-++\,+$ is characterized by the conditions \cite[\S 7.8]{Hartle03}
\begin{equation}
g_{\alpha\beta}\, (e_{\hat\alpha})^\alpha\, (e_{\hat\beta})^\beta = \eta_{\hat\alpha\hat\beta}
\end{equation} 
where the indexing convention has to be understood in the following way: Indices without a hat label components in the curved space, the realm of $g_{\alpha\beta}$. Indices with a hat indicate components of the flat space (with metric $\eta_{\hat\alpha\hat\beta}$) that is tangent space to the curved space. The index $\hat\alpha$ in $(e_{\hat\alpha})^\alpha$ is a labeling index that enumerates the vectors in the basis while the index $\alpha$ is a coordinate index that refers to the basis-vector's $\alpha$-coordinate in the curved space.

If the metric $g_{\alpha\beta}$ is diagonal, finding an orthonormal basis is particularly easy since the basis vectors of $g_{\alpha\beta}$ are already orthogonal. The only step left to do is to normalize the basis vectors at every point in the spacetime:
\begin{equation}
(e_{\hat\alpha})^\alpha = \sqrt{\frac{\eta_{\hat\alpha\hat\alpha}}{g_{\alpha\alpha}}}\, \delta_{\hat\alpha}^{~\alpha}\, ; \qquad (e^{\hat\alpha})_\alpha = \sqrt{\frac{\eta^{\hat\alpha\hat\alpha}}{g^{\alpha\alpha}}}\, \delta^{\hat\alpha}_{~\alpha} \, .
\end{equation}
The transformation of a second rank tensor with one contra- and covariant index each is then given by
\begin{equation}
\label{eq:set_trafo_1u1d}
A^{\hat\alpha}_{~\hat\beta} = A^{\alpha}_{~\beta}\, (e^{\hat\alpha})_\alpha\, (e_{\hat\beta})^\beta = A^{\alpha}_{~\beta}\, \sqrt{\frac{\eta^{\hat\alpha\hat\alpha}\,\eta_{\hat\beta\hat\beta}}{g^{\alpha\alpha}\,g_{\beta\beta}}}\, \delta^{\hat\alpha}_{~\alpha}\, \delta_{\hat\beta}^{~\beta} \, .
\end{equation} 
If the tensor $A^{\alpha}_{~\beta}$ is diagonal, the square root in \eref{set_trafo_1u1d} equals unity for every non-vanishing component and \eref{set_1up1down} is obtained. If the signature of the tensor $A$ with both indices either up or down is assumed to be $+++\,+$, then it follows from
\begin{equation}
A^{\hat\alpha}_{~\hat\beta} = \eta_{\hat\bullet\hat\beta}\, A^{\hat\alpha\hat\bullet} = \eta^{\hat\alpha\hat\bullet}\, A_{\hat\bullet\hat\beta} \, ,
\end{equation} 
that the signature of the tensor $A$ with one index up and one index down is $-++\,+$. Hence, the stress-energy tensor \eref{set_diag_perfluid}, for example, takes the form
\begin{equation}
T^\alpha_{~\beta} = \left[ \begin{array}{cccc}
-\rho & 0 & 0 & 0 \\
0 & p & 0 & 0 \\
0 & 0 & p & 0 \\
0 & 0 & 0 & p \\
\end{array} \right] = \mathrm{diag} [-\rho,p,p,p] = T^{\hat\alpha}_{~\hat\beta} \, .
\end{equation} 
The minus sign in front of the density $\rho$ does not indicate a negative density, but is a mere reflection of the tensor's signature $-++\,+$. This slight awkwardness is the price for the complete absence of physically irrelevant coordinate artefacts in the components of the tensor.

\subsection{Canonical forms of the stress-energy tensor}

The stress-energy tensor of the perfect fluid was one particularily simple example of how matter is represented in general relativity. In general, the stress-energy tensor has more than two independent components. Hawking and Ellis \cite[\S 4.3]{Hawking73} distinguish four possible canonical types of stress-energy in an orthonormal basis. The following classification is more or less a direct quotation from \cite{Hawking73}.

\paragraph{Type I} is the general case where the stress-energy tensor has one timelike eigenvector that is unique (unless $\rho=-p_i$). It can be expressed as
\begin{equation}
\label{eq:stressenergy_typeI}
T^{\hat\alpha\hat\beta} = \left[ \begin{array}{cccc}
\rho & 0 & 0 & 0 \\
0 & p_1 & 0 & 0 \\
0 & 0 & p_2 & 0 \\
0 & 0 & 0 & p_3 \\
\end{array} \right] \, .
\end{equation} 
The eigenvalue $\rho$ represents the energy-density as measured by an observer at rest, i.e. the observer's four-velocity is $u_{\mathrm{obs}}^{\hat\alpha}=(1;0;0;0)$ in the orthonormal basis. The eigenvalues $p_i$ represent the principal pressures in the three spacelike directions. All observed fields with a non-vanishing rest mass are of type I, as are all zero rest mass fields with the exception of the special cases that are represented by stress-energy tensors of type II.

\paragraph{Type II} is the special case where the stress-energy tensor has a double null eigenvector:
\begin{equation}
T^{\hat\alpha\hat\beta} = \left[ \begin{array}{cccc}
\nu+\kappa & \nu & 0 & 0 \\
\nu & \nu-\kappa & 0 & 0 \\
0 & 0 & p_1 & 0 \\
0 & 0 & 0 & p_2 \\
\end{array} \right] \, , \qquad \nu \neq 0 \, .
\end{equation} 
The only observed occurrence of this form is for zero rest mass fields that represent radiation that is travelling only in the direction of the double null eigenvector.

\paragraph{Type III \& Type IV} are the cases where the stress-energy tensor has a triple null eigenvector and no timelike or null eigenvector. Hawking and Ellis \cite{Hawking73} are not aware of any physical occurence or relevance of types III and IV, and neither am I.

In this thesis I will only focus on stress-energies of type I which represent the majority of relevant matter distributions. For simplicity, I will further restict the contents of this thesis to spherically symmetric cases.

The nature of the involved pressure terms is irrelevant for the discussion presented. That is, it does not matter whether the pressure or stress of that stress-energy arises from the random motion of particles or from a field of some sort (see e.g. \sref{non-cdm_fields}).

\subsection{Spherically symmetric non-perfect fluids (anisotropic fluids)}

In spherically symmetric coordinates, the rotation invariance of the sphere ensures that the physical properties of the $\theta$- and $\cphi$-directions are identical. Mathematically this is established through the occurence of the unit sphere's metric \eref{metric_unit_sphere} in the metric of the spacetime \eref{ss_metric}. Hence, the $\theta\theta$- and $\cphi\cphi$-components of a spherically symmetric second rank tensor, that is derived from the metric and its derivatives, must be identical apart from coordinate artefacts. Thus, the Einstein tensor $G^{\alpha}_{~\beta}$ has identical $\theta\theta$- and $\cphi\cphi$-components, as it is already evident from our example \eref{Gtt} to \eref{Gthetaphi}.

Since the Einstein tensor is by \eref{Einstein_eqns} related to the stress-energy tensor, the latter one must also have identical $\theta\theta$- and $\cphi\cphi$-components to represent a matter distribution that sources a spherically symmetric gravitational field. Generally, any second rank tensor that decribes any spherically symmetric property must have identical $\theta\theta$- and $\cphi\cphi$-components. We can easily check this fact by noting that the components of such a tensor in an orthonormal frame must be invariant under rotation of space about the $r$-direction. Such a rotation is given by the rotation matrix
\begin{equation}
R^\alpha_{~\beta}(\gamma) = \left[ \begin{array}{cccc}
1 & 0 & 0 & 0 \\
0 & 1 & 0 & 0 \\
0 & 0 & \cos\gamma & -\sin\gamma \\
0 & 0 & \sin\gamma & \cos\gamma \\
\end{array} \right] \, .
\end{equation} 
Rotating the tensor $A^\alpha_{~\beta}=\mathrm{diag} [-a,b,c,d]$ by an angle $\gamma$ gives
\begin{equation}
(A_{rot})^\alpha_{~\beta}=(R^{-1})^\alpha_\bullet \, A^\bullet_{~\circ} \, R^\circ_\beta = \left[ \begin{array}{cccc}
-a & 0 & 0 & 0 \\
0 & b & 0 & 0 \\
0 & 0 & c\cos^2\gamma + d\sin^2\gamma & (c-d)\sin\gamma\cos\gamma \\
0 & 0 & (c-d)\sin\gamma\cos\gamma & c\sin^2\gamma + d\cos^2\gamma \\
\end{array} \right] \, ,
\end{equation} 
so that rotation invariance is only given when $c=d$.

Hence, a stress-energy tensor that describes a spherically symmetric matter distribution as measured by an observer at rest is given by
\begin{equation}
\label{eq:set_ss_fluid}
T^\alpha_{~\beta}=\left[ \begin{array}{cccc}
-\rho & 0 & 0 & 0 \\
0 & p_r & 0 & 0 \\
0 & 0 & p_t & 0 \\
0 & 0 & 0 & p_t \\
\end{array} \right] \, ,
\end{equation} 
where $p_r$ and $p_t$ are the pressure or stress in the radial and transverse direction.

Please note that ``an observer at rest'' is equivalent to saying that the observer's four-velocity has only components in the time direction, $u_{\mathrm{obs}}^{\alpha}=(1;0;0;0)$. Therefore, \eref{set_ss_fluid} is still a good approximation when the observer's three-velocity relative to the fluid's motion is small compared to the speed of light.

\paragraph*{The stress-energy tensors in this thesis} are of the form \eref{set_ss_fluid}, which represents the whole class of type I stress-energy of the Hawking and Ellis classification \cite{Hawking73} subjected to spherical symmetry. This includes ordinary baryonic matter as well as some dark matter candidates and scalar fields as they appear later in \sref{dark_matter_candidates}.

All static spherically symmetric fields and stress-energies of type I can be reinterpreted in the terminology of energy-density $\rho$ and principal pressures $p_r$ and $p_t$, and thus, are subject to statements made in this thesis.

\section{Stellar structure equations}
\label{sec:stellar_structure}

The structure of any static spherically symmetric system that obeys the rules of general relativity is given by the ``stellar'' structure equations, see for example \cite[\S 23.5]{MTW}. The name stellar structure equations is commonly used since these equations are generally used to describe compact objects like ordinary fusioning stars and neutron stars, etc. They do, however, also apply to larger systems like galaxies (in the spherically symmetric approximation).

The structure is determined by five functions of the radial coordinate $r$: the metric functions $\Phi(r)$, $m(r)$, the density $\rho(r)$ and the radial and transverse pressures $p_r(r)$ and $p_t(r)$. Therefore, to uniquely determine the structure, an appropriate model needs to supply five equations plus boundary conditions. The Einstein equations \eref{Einstein_eqns} provide three equations in static spherically symmetric coordinates: The $tt$-field equation yields
\begin{equation}
\label{eq:stelstruc_mass}
m(r)=4\pi\int_0^r \rho(\rr)\,\rr^2\,\d\rr + m(0)
\end{equation} 
and the $rr$-field equation gives
\begin{equation}
\label{eq:stelstruc_grav}
g(r) \equiv \Phi'(r)=\frac{m(r)+4\pi p_r(r)\, r^3}{r^2 \left[ 1-2m(r)/r \right] }
=\frac{4\pi\,r}{3}\, \frac{\rhoavg(r)+3p_r(r)}{1-2m(r)/r}
\end{equation} 
where the ``average density'' $\rhoavg(r)\equiv m(r)/(\frac{4}{3}\pi r^3)$ was introduced. Please note that $r$ is just a coordinate for the radial direction and not ``proper radius''! The name ``average density'' is thus somewhat misleading, but very illustrative.

From now on, in the interests of legibility, I discontinue indicating the explicit $r$-dependence of all relevant functions. The last non-zero Einstein equation could be used as a structure equation, but its complicated form is highly inconvenient. Indeed, one generally replaces the remaining field equation by the covariant conservation of stress-energy \eref{set_cov_cons}. To see that this is a valid way to proceed, let us count the equations at hand: The 10 components of the Einstein tensor by construction obey the four contracted Bianchi identities \eref{Bianchi_ID} and consequently imply the covariant conservation of stress-energy \eref{set_cov_cons}. Thus, one can choose to disregard up to four Einstein equations by demanding that the Bianchi identities -- or equivalently the covariant conservation of stress-energy -- are also satisfied. The Einstein equations supply 10 equations, of which the six equations corresponding to the off-diagonal elements of the stress energy tensor are vacuous (``$0=0$'') in the employed coordinate system with the diagonal metric \eref{ss_metric}. The two equations corresponding to the $\theta$- and $\cphi$-coordinates are identical due to spherical symmetry. The remaining three equations automatically satisfy the covariant conservation of stress-energy,
\begin{equation}
\label{eq:stel_struc_set_cons}
p_r' = -(\rho+p_r)\, \Phi' + \frac{2\left( p_t - p_r \right)}{r} \, ,
\end{equation}
and hence, we can choose to replace the $\theta\theta$-field equation with \eref{stel_struc_set_cons}.

The physical interpretation of $\rho$, $p_r$ and $p_t$ was already discussed in \sref{stress-energy}. The interpretation of $m(r)$ is given by \eref{stelstruc_mass}: $m(r)$ is the ``total mass-energy inside the coordinate radius $r$'' and this includes contributions from ``rest mass-energy'', ``internal energy'' and ``gravitational potential energy''. For details, please refer to \cite[Box 23.1]{MTW}.

To find the meaning of $g\equiv \Phi'$, I point out that the $g_{tt}$ component of an accelerating observer's proper reference frame is generally given by \cite[eq. 13.71]{MTW}
\begin{eqnarray}
g_{tt} &=& -(1+2\,a_{\hat{j}}\,x^{\hat{j}}) + \O{|x^{\hat{j}}|^2} \\
 &=& -(1+2\,a_j\,x^j) + \O{|x^j|^2} \, ,
\end{eqnarray} 
where $a_{\hat{j}}$ is the physical acceleration felt by the observer. For comparison, the $g_{tt}$ component of the metric \eref{ss_metric} in the vicinity of $r_0$ is to first order in $r$
\begin{equation}
g_{tt} = -e^{2\Phi(r_0+\delta r)} = -e^{2\Phi(r_0)} \left[ 1 + 2\,\Phi'(r_0)\, \delta r \right] + \O{r^2} \, ,
\end{equation} 
which leads, in a properly normalized reference frame with origin at $r=r_0$, to
\begin{equation}
g_{\hat{t}\hat{t}}(r_0+\delta r) =  -\left[ 1 + 2\,\Phi'(r_0)\, \delta r \right]  + \O{r^2} \, .
\end{equation}
One can now easily identify $g\equiv \Phi' = a_r$ as the local gravitational acceleration felt by an observer in a proper reference frame, which is --- for positive $g$ --- pointing towards the coordinate origin in the Newtonian picture and away from the origin in Einstein's interpretation. In analogy to Newton's gravity, we shall call $\Phi(r)$ the \textit{potential} or even the \textit{gravitational potential}.

So far I have given you three of the necessary five structure equations and physical interpretations of all involved functions. The two remaining equations are given by the properties of the matter or field in our system. They usually occur in the form of equations of state
\begin{eqnarray}
p_r &=& p_r(\rho, r, n, s, \ldots) \\
p_t &=& p_t(\rho, r, n, s, \ldots)
\end{eqnarray} 
but can they can generally also be replaced by other equations, see appendix \sref{gravastar_eos} for a more extensive discussion. I indicated that the equations of state generally depend on a variety of physical properties, like density $\rho$, number density $n$, entropy per particle $s$, or also chemical composition $X$, or even position $r$, etc. For each new function that adds complexity to the system, a further equation is necessary to determine the system uniquely. Within the scope of this thesis, I am only going to explore systems that are determined by the five aforementioned functions.

\section[The Tolman-Oppenheimer-Volkoff equation]{The Tolman-Oppenheimer-Volkoff (TOV) 
equation}

In most discussions, the derivative of the potential $\Phi'$ is eliminated from the structure equations by inserting \eref{stelstruc_grav} into \eref{stel_struc_set_cons}. Thus, one obtains the \textit{anisotropic TOV equation}:
\begin{equation}
\label{eq:TOV_aniso}
\frac{\d p_r}{\d r}  = -\frac{(\rho + p_r) \, (m+4\pi p_r\, r^3)}{r^2 \left[ 1-2m/r \right] } 
+\frac{2\left( p_t - p_r \right)}{r} \, .
\end{equation}
In the case of isotropic pressures $p=p_r=p_t$ this leads to the (standard) \textit{ TOV equation}, which can be written in any of the equivalent forms
\begin{equation}
\label{eq:TOV_iso}
\frac{\d p}{\d r} = 
-\frac{(\rho + p) \, (m+4\pi p\, r^3)}{r^2 \left[ 1-2m/r \right] } 
= -\frac{4\pi\,r}{3}\, \frac{(\rho+p)\,(\rhoavg+3p)}{1-2m/r} \, .
\end{equation}
To see the connection to Newton gravity, it is helpful to write the isotropic TOV in the following way:
\begin{equation}
\frac{\d p}{\d r} = 
-\frac{\rho \, m}{r^2} \left( 1 + \frac{p}{\rho} \right) \left( 1+\frac{4\pi p\, r^3}{m} \right) \left( 1-\frac{2m}{r} \right)^{-1} \, .
\end{equation}
This is the Newtonian equation of hydrostatic equilibrium multiplied by three relativistic corrections that are negligible in the Newtonian limit where $|p|\ll\rho$ and $m\ll r$. The interesting fact about this formula is that in Einstein's gravity it drops out automatically through the interaction between stress-energy and geometry while in Newton mechanics hydrostatic equilibrium is a seperate concept of fluid mechanics. Of course this only works because the appropriate fluid mechanics are already implemented in the way the stress-energy tensor was defined -- in Einstein's theory of gravity, it is not possible to separate fluid mechanics and gravity!


%% file: tc_galactic_halo.tex
\chapter{Galactic Dark Matter halos}
\label{sec:galactic_halo}

\input{ts_galaxies}
\input{ts_rotation_curves}
\input{ts_galaxy_lensing}
\input{ts_galaxy_rotcurves+lensing}



%% file: ts_galaxies.tex
\section{Gravitational field of a galaxy}
\label{sec:galaxy}

\subsection{Types of galaxies}
A galaxy is a gravitationally bound system of $10^{10}$ to $10^{12}$ stars \cite[\S 1.1]{binney87}. According to Hubble's revised classification of 1936 \cite{Hubble36}, one distinguishes four basic types of galaxies, elliptical (E0--E7), spiral (Sa--Sd), spiral bar (SBa--SBd) and irregular (Irr), and the transition class of lenticular galaxies (S0).

\begin{figure}[hbt]
\centerline{\epsfig{file=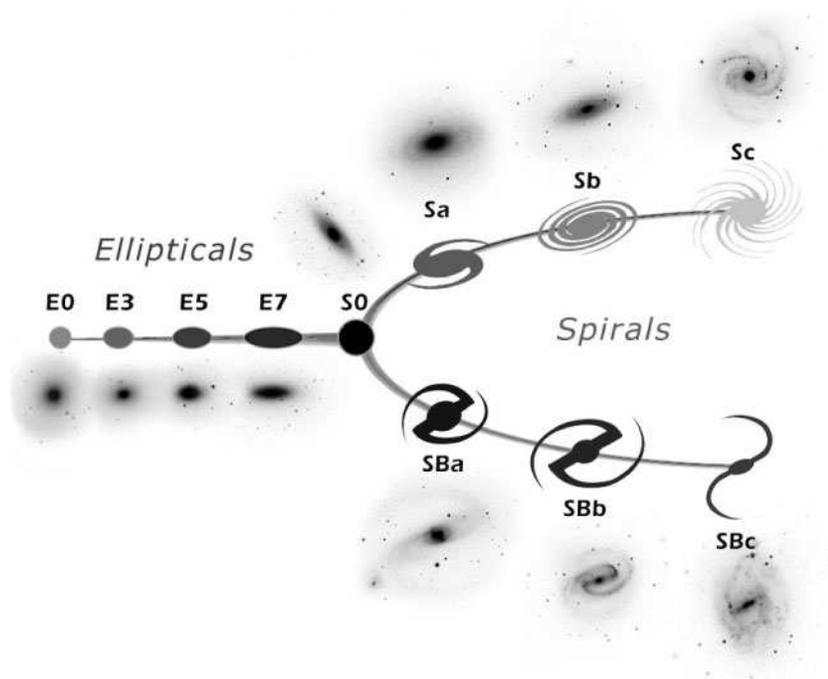,width=11cm}}
\caption[Hubble's classification of galaxies.]{Hubble's classification of galaxies\footnotemark.}
\label{fig:hubble_class}
\end{figure}
\footnotetext{Figure taken from \url{http://hubblesite.org}, reproduced with permission of NASA and STScI. 
}

Elliptical galaxies are smooth and featureless stellar systems that typically consist of old stars with low heavy metal abundances (Population II stars). Elliptical galaxies contain little or no interstellar gas or dust. In the classification, they are denoted by E$n$ where $n$ is ten times the ellipticity of the galaxy, thus E0 being spherical and E7 being highly elongated galaxies. \cite{binney87,eaa}

Lenticular galaxies are the class in transition between the elliptical and spiral galaxies. They are smooth featureless disks without gas, dust or bright young stars. They do obey the exponential surface brightness law of spiral galaxies and are labelled S0 in Hubble's classification. \cite{binney87}

Spiral galaxies, denoted by Sa--Sd, contain a rotating disk of relatively young metal-rich stars (Population I), gas and dust. Within the disk are dense filaments of bright young stars, gas and dust that demark active star formation areas. These filaments take the shape of two or more spiral arms and are responsible for the name of this class. Apart from the disk, spiral galaxies also contain a central bulge and a spheroidal halo of Population II stars that form globular clusters. Along the sequence Sa--Sd, the relative luminosity of the bulge decreases, the fraction of gas increases, and the arms become more loosely wound. The Milky Way is classified as Sbc, an intermediate of Sb and Sc. \cite{binney87,eaa}

Spiral bar galaxies are distinguished through the existence of a central bar from which the two main spiral arms begin. A bar is effectively an elongated central bulge. Apart from the added bar, spiral bar galaxies follow the sub-classification of normal spirals and are distinguished by the prefix SB. \cite{eaa}

Finally, the class of irregular galaxies (Irr) comprises all other galaxies that don't fit into the other classes. This includes galaxies that have been ``ripped apart'' by an encounter with another galaxy, as well as smaller satellite galaxies like the Magellanic Clouds. \cite{binney87}

This classification of galaxies, and also the distinction between Population I and II stars, is rather incomplete and coarse. There are more sophisticated classifications that take account of more distinct features, e.g. rings in spiral galaxies \cite{Vaucouleurs:1959}. The Hubble classification, however, is absolutely sufficient to outline the basic features of galaxies. In this chapter, I will be concentrating on (barless) spiral galaxies which are observationally interesting because of their clearly defined rotating disk.

\subsection{Anatomy of spiral galaxies}
\label{sec:galaxy_anatomy}

Spiral galaxies have a lot more obvious substructure than either elliptical or irregular galaxies. While ellipticals appear to be smooth and featureless, spiral galaxies exhibit several visible parts: The aforementioned disk which contains the spiral arms, a central bulge, in the cases of SB galaxies a central bar, and finally a spheroidal halo that extends beyond the disk. Recent observations have shown that generally there is also a supermassive black hole at the center of most if not all spiral galaxies. I will use figures of our Milky Way galaxy to illustrate the dimensions of a typical spiral galaxy.

\paragraph*{Disk and spiral arms.} \label{sec:spiral_disk_anatomy}
The differentially rotating disk contains the active star formation regions which form the spiral arms. Thus, it contains mostly young stars (Population I), as well as gas and dust \cite[\S 1.1]{binney87}. The spiral arms are not a manifestation of a differentially rotating disk, as one might na\"ively think, but are thought to be the result of two coexisting mechanisms: quasi-stationary density waves (according to the Lin-Shu hypothesis \cite{Lin:1966}), and self-propagating star formation regions where supernovae explosions create blast waves that trigger compression in nearby gas and dust clouds and therefore initiate star formation \cite{eaa}. Usually, the spiral arms trail the rotation of the disk, but in some cases, e.g. \textsc{NGC 4622}, the spiral arms are leading, i.e. pointing into the direction of the disk's rotation \cite{eaa}.

The measured rotation velocities are constant in the outer regions and even extend beyond the visible edge of the disk, indicating that the mass-to-light ratio $\Upsilon$ is not constant throughout the galaxy \cite[\S 10.1]{binney87}. A detailed discussion of the observed rotation curves follows in \sref{rotcurve}.

\addtocounter{footnote}{-1}
\begin{figure}[hbt]
\centerline{\epsfig{file=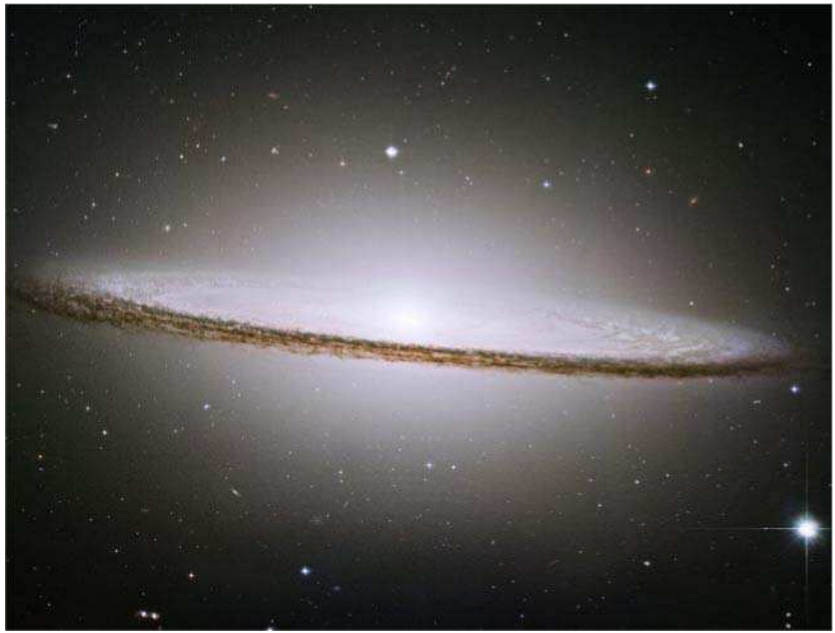,width=11cm}}
\caption[The Sombrero galaxy (M104, Type Sa).]{The Sombrero galaxy{\protect\footnotemark} (M104, Type Sa) has a large and extended central bulge of stars, and dark prominent dust lanes that appear in a disk that we see nearly edge-on. Billions of old stars cause the diffuse glow of the extended central bulge. Close inspection of the bulge in the above photograph shows many points of light that are actually globular clusters. M104's spectacular dust rings harbor many younger and brighter stars. The very center of the Sombrero glows across the electromagnetic spectrum, and is thought to house a large black hole. }
\label{fig:sombrero_galaxy}
\end{figure}
\footnotetext{Figure reproduced with permisson from the Hubble Heritage Team (AURA/STScI/NASA). Figure and (abridged) text taken from \url{http://antwrp.gsfc.nasa.gov/apod/ap031008.html} }

The characteristic thickness\footnote{The characteristic thickness of the galactic disk is defined as the ratio of the disk's surface density to its volume density at the galactic plane \cite[\S 1.1]{binney87}.} of the disk is different for older and younger stars. While the characteristic thickness for younger stars is about $200\unit{pc}$ in the Milky Way, for older stars, e.g. our sun, it is about $700\unit{pc}$. In general, the thickness of the disk is small compared to its radial extent, which is characterized by the Holmberg radius\footnote{The Holmberg radius is defined as the radius of the isophote with a surface brightness of $26.5\,\unit{mag\, arcsec^{-2}}$ in the B band, which roughly corresponds to 1\%-2\% of the sky's brightness \cite[\S 1.1]{binney87}.} of about $17\unit{kpc}$ for the Milky Way \cite[\S 10.1]{binney87}. Hence, the disk is usually modelled as a thin (two-dimensional) disk. Observationally it was found that the surface brightness $I$ of spiral galaxies follows an exponential law,
\begin{equation}
\label{eq:spiral_exp_disk}
I(r) = I_0\, e^{-r/r_d} \, ,
\end{equation} 
where the disk scale length $r_d=(3.5\pm 0.5)\,\unit{kpc}$ for the Milky Way \cite[\S 1.1]{binney87}. In this plane, the disk population follows near-circular orbits around the center of the galaxy \cite{eaa}.

\paragraph*{Bulge and bar.} According to Wyse \textit{et al.} \cite{Wyse:1997}, the bulge is commonly defined by allocating all ``non-disk'' light into the bulge. Generally, it has a nearly spherical shape and the flattening is consistent with the slow rotation of the bulge. It consists mainly of old Population II stars and is metal-poor compared to the typical disk population. The bulge of the Milky Way has an effective radius\footnote{The effective radius of the bulge is defined as the radius of the isphote containing half of the bulge's total luminosity \cite[\S 1.1]{binney87}.} of $r_e \approx 2.75\unit{kpc}$ \cite{Wyse:1997} and therefore extends above and below the galactic disk.

The term ``bar'' is widely used in the literature, generally without being defined precisely. In most cases the bar is regarded as an elongated bulge from whose end the two main spiral arms originate \cite{Wyse:1997}. Others consider the bar to be part of the galactic disk \cite{eaa}. Independent of the precise definition of the bar, it certainly breaks the azimuthal symmetry of galaxies near the center, that is otherwise given if one assumes the density perturbations of the spiral arms to be reasonably small.

\paragraph*{Central black hole.} It has long been suspected that the matter concentration at the center of a galaxy is high enough to harbour an extremely massive black hole. The center of the Milky Way is marked by the radio source {\sc Sgr A*}. Indeed, recent observations of several stars that orbit {\sc Sgr A*} have shown that a supermassive black hole is likely to be located at the center of the Milky Way. The technically correct deduction from the observations is that a mass of $(3.7\pm 1.5) \times 10^6 \Msun$ is located within a galactocentric sphere of radius $124\,\unit{AU}$ \cite{Schoedel:2002}. 

More recently, the diameter of the radio source {\sc Sgr A*} has been observed directly and is estimated at only $\sim 1\,\unit{AU}$ \cite{Shen:2005}. This provides very strong evidence that there is an extremely compact object at the center of the Milky Way. Such an object is commonly interpreted as a supermassive black hole, although there are ongoing discussions regarding whether other objects can bind that much mass in such a small volume, see e.g. \sref{gravastar} and \sref{anti-gravastar}.

\paragraph*{Halo.} The halo and bulge were long thought to be a single spheroid that follows a surface brightness profile $\propto r^{-1/4}$. In 1980, Bahcall and Soneira \cite{Bahcall:1980} tried to model the gravitational potential of the Milky Way with contributions from the disk and the bulge only. The density profiles were assumed to be given by the observed luminosities and a constant mass-to-light ratio $\Upsilon$ for each the disk and bulge. This model did not reproduce the observed ``flat'' rotation curves that exhibit a constant velocity for the largest observable radii. Only after adding an unseen spherical halo component, with a density profile that falls off as
\begin{equation}
\rho \propto r^{-2}
\end{equation} 
for large distances, did their model show the flat rotation curves. Using high-velocity Population II stars in the halo as an indicator for the gravitational field, the radius and mass of the spherical halo were bound to be at least $R_{\mathrm{halo}} \gtrsim 41\unit{kpc}$ and $M_{\mathrm{halo}} \gtrsim 4 \times 10^{11} \Msun$ \cite[\S 2.7]{binney87}. This shows that the galactic halo extends well beyond the disk and is at least six times as heavy as the visible parts of the Milky Way. This is one of the many indications of \textit{dark matter}, a form of matter whose existence is inferred only from its gravitational effects \cite[\S 10.1]{binney87}.

Within the last 15 years, it became apparent that the brightness of the bulge falls off steeper than that of the visible halo, indicating as well that bulge and halo should be considerent different structures \cite{Wyse:1997}. The halo consists of even older Population II stars than the bulge, which form globular clusters and a very thin population of scattered individual stars \cite{eaa}. 

For the last decade and longer, several different methods to measure the mass and extent of the Milky Way's dark matter halo have been employed. Most of them are discussed in a comprehensive review by Zaritsky (1999) \cite{Zaritsky:1999}: (i)~measurements of the disk rotation curve \cite{Fich:1991}, (ii)~mass estimates based on the escape speed of the halo \cite{Kochanek:1996}, (iii)~statistical analysis of the motion of satellite galaxies and globular clusters \cite{Zaritsky:1989}, (iv)~timing arguments for the dynamical system of the Milky Way and the Andromeda galaxy (\textsc{M31}) \cite{Einasto:1982,Peebles:1995,Shaya:1995}, (v)~analysis of orbital parameters of the Large Magellanic Cloud \cite{Lin:1995} and (vi)~motion of satellite galaxies of other galaxies similar to the Milky Way \cite{Zaritsky:1994}. A comparison of these measurements is displayed in \fref{halo_mass} and proves to be consistent with a halo model that exhibits a mass distribution which rises linearly with the radius. Although the figure shows data for up to $\sim 300\unit{kpc}$, the Milky Way's halo cannot extend much beyond $200\unit{kpc}$ due to the presence of our neighbour, the Andromeda galaxy. This should lead us to the conclusion that spiral galaxy halos are roughly spherical on a local scale and generally overlap on scales that include neighbouring galaxies \cite{Zaritsky:1999}.

\addtocounter{footnote}{-1}
\begin{figure}[hbt]
\centerline{\epsfig{file=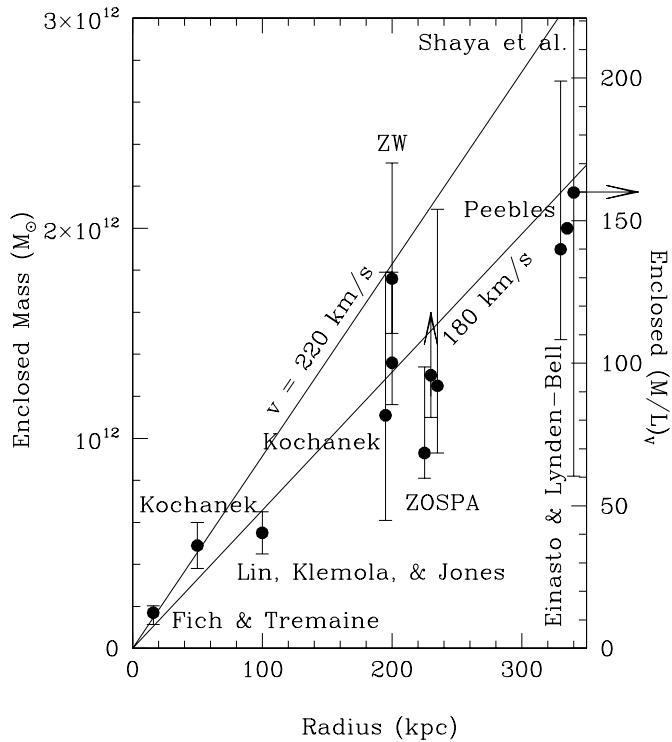,height=11cm}}
\caption[A comparison of various measurements of the enclosed mass of the Milky Way halo.]{A comparison of various measurements of the enclosed mass of the Milky Way halo\footnotemark. The solid lines denote the enclosed mass for dark matter halos which cause a constant circular velocities of $180$ and $220\,\unit{km/s}$. The data points are generally given with their 90\% confidence intervals. The methods in chronological order (see text): Einasto \& Lynden-Bell (iv, 1982), Zaritsky \textit{et al.} (iii, 1989, ZOSPA, 3 data points), Fich \& Tremaine (i, 1991), Zaritsky \& White (vi, 1994, 2 data points), Lin \textit{et al.} (v, 1995), Peebles and Shaya \textit{et al.} (both iv, 1995), Kochanek (ii, 1996, 2 data points).}
\label{fig:halo_mass}
\end{figure}
\footnotetext{Figure reproduced with permisson from Zaritsky (1999) \cite{Zaritsky:1999}.}

\label{sec:dm_halo_shape}
While the shape of the halo is generally assumed to be spherical, the actual observational situation is currently not conclusive \cite{Helmi:2004} and suggests that spherical as well as slightly oblate or prolate halos are possible. Recently, more data on the debris of the Sagittarius dwarf satellite galaxy suggests that the dark matter halo might actually be slightly prolate \cite{Helmi:2004a} while earlier microlensing observations suggest that the halo is more likely to be oblate \cite{Samurovic:1999}. Despite numerous articles that suggest techniques to measure the shape of the dark matter halo and others that predict halo shapes from numerical simulations, actual observations are very sparse. Hence, the observational situation can be described as somewhat vague, although most observations and predictions are consistent with a (at least close to) spherical halo.

\paragraph*{} Most of the observational evidence that was presented in this section applies only to the Milky Way, but the overall picture that has been drawn of a spiral galaxy is consistent with observations of other spiral galaxies apart from the Milky Way: The disk with its spiral arms and the bulge are clearly visible in other galaxies. Observations of nearby galaxies indicate that a central black hole is likely to be a standard feature of spiral galaxies \cite{Kormendy:2001}. A large sample of rotation curves ($\sim 1100$) showed that the flat rotation curve behaviour is a universal feature of spiral galaxies \cite{Persic:1996} which strongly suggests that a dark matter halo exists in every spiral galaxy. The same conclusion can be reached from large samples of weak gravitational lensing by galaxies \cite{Brainerd:1996}.

After outlining the structure of spiral galaxies, I shall now go on to discuss the contributions of the individual components to the gravitational field.

\subsection{Contributions to the gravitational field}
All of the aforementioned components of a galaxy have mass and thus contribute to the gravitational field of a galaxy. Therefore, constructing a comprehensive model that respects the effects of all parts is rather difficult. The dispute over which density profile to use for the individual components adds to the difficulty, and obscures an objective comparison of different galaxy models. Fortunately, the contribution of all parts is limited to certain distance scales, so that not all components have to be included in a reasonably realistic model. A model of the very center of the Milky Way, for example, should include the central black hole and the dense central region of the bulge \cite{Schoedel:2002}, while the effects of spiral arms or the dark matter halo almost certainly play no vital role.

The gravitational influence of the central black hole is noticeable only within a few parsec of the galactic center \cite{Schoedel:2002}, and can be completely neglected for studying the dynamics in the disk or the outer regions of the bulge. The bulge and bar dominate the innermost few kiloparsec of the galaxy and depending on their relative size, have to be accounted for in models of the galactic disk's dynamics. The spiral arms are density perturbations of the disk and thus only influence the gravitational field locally. When studying the motions of stars or gas in the disk, one should at least discuss the perturbations of the spiral arms, if they are not explicitly included in the model. Gas or dust clumps in the disk can have a similar influence on local disk dynamics \cite{Fraternali:2002}. 

The main contribution to the gravitational field in most of the disk originates, however, in the dark matter halo. While the disk mass definitely has to be accounted for, the major contribution in the outer region of the galactic disk is due to the halo. Dwarf galaxies are almost completely dominated by the dark matter halo and even the disk dynamics of the largest spirals cannot be modelled accurately without the halo \cite{Persic:1996}.

When modelling the gravitational field beyond the visible disk, out to several hundreds of kiloparsec, the dark matter halo is practically the only source of the gravitational field, see \fref{halo_mass}.

Summarizing, I note that the dark matter halo is generally the main contributor to the gravitational field of spiral galaxies. For very large galaxies, the disk mass plays an important, non-negligible role and the central region of a galaxy cannot be modelled acurately by the halo alone.

\subsection{Dark matter halo models in general relativity}
Choosing a model for the gravitational field of a galaxy depends of course on the planned application, i.e. the aspect of the research to be undertaken. In this thesis, I want to concentrate on the general relativistic aspects of the dark matter halo.

Strong gravitational fields, that require general relativity to describe the resulting dynamics accurately, occur only close to massive, compact objects, like stars and black holes. On a galactic scale, however, one cannot keep track of all $\sim 10^{11}$ individual stars and generally introduces a locally averaged density function. Because the absolute value of that density is very low\footnote{For example, the density in the solar neighbourhood  is $\rho_0 = (0.18\pm 0.03)\,\Msun\,\unit{pc^{-3}}$ (also called the Oort limit). \cite[\S 4.2.1]{binney87}} in galaxies, the induced gravitational fields are also weak. So why use general relativity to describe the weak gravitational field in galaxies?

Newtonian physics is obtained in the limit of general relativity where:
\begin{enumerate}
\renewcommand{\labelenumi}{(\roman{enumi})}
\item the gravitational field is weak,
\item the particle speeds involved are slow compared to the speed of light, and
\item the pressures and fluxes and stresses are small compared to the mass-energy density \cite[\S 17.4]{MTW}.
\end{enumerate}

While there is no doubt that all objects that make up a galaxy satisfy the conditions (i) and (ii), at present no one knows much about the nature of dark matter and hence, the possibility of dark matter being a high pressure fluid, or some sort of unknown field with high field tensions, cannot be excluded. It is this uncertainty about the dark matter's equation of state (see \sref{stellar_structure}) that makes it necessary to include general relativistic effects. Indeed, I will show in the following section that many different dark matter candidates exist which exhibit pressures that are comparable to or even dominate the energy density.

When performing measurements of the gravitational field of a galaxy, condition (ii) might play an important role. For rotation curves, the tracer particles surely exhibit subluminal speeds, but for gravitational lensing measurements, the particles directly influenced by the gravitational field are the photons of a background object which per definition travel at the speed of light. I will show how this influences the interpretation of gravitational lensing measurements in \sref{grav_lensing}.

If one does not require condition (iii), general relativity does not quite reduce to Newton's gravity. It is a standard result that using only condition (i) the gravitational potential $\Phi$ is generated by the $tt$-component of the Ricci tensor \cite[\S 17.4]{MTW}: 
\begin{equation}
\nabla^2 \Phi = R_{tt} \, .
\end{equation}
Invoking the Einstein field equations \eref{Einstein_eqns} with a stress-energy tensor of the form \eref{stressenergy_typeI} gives $R_{tt}$:
\begin{equation}
R_{tt} = 4\pi\,(\rho + p_1 + p_2 + p_3)\, ,
\end{equation}
where the $p_i$ are the principal pressures obtained by diagonalizing the space-space part of the stress energy tensor. Thus, in the case of a perfect fluid ($p=p_1=p_2=p_3=p_r=p_t$) we find the field equation
\begin{equation}
\label{eq:pressure_field_eq}
\nabla^2 \Phi = 4\pi\,(\rho + p_r + 2p_t) = 4\pi\,(\rho + 3p)\, ,
\end{equation}
which only reduces to Newtonian gravity if $p \ll \rho$, i.e. condition (iii). It is now quite obvious from \eref{pressure_field_eq} that the gravitational field is highly sensitive to the pressure if density and pressure are of the same order of magnitude.

A different reason to examine galactic dark matter in a general relativistic framework is the boundary problem. Since there is no observable boundary of the dark matter halo (see \fref{halo_mass}), it is possible that the halo is actually a phenomenon of cosmology. In that case, since the halo should then smoothly merge with a FLRW cosmology, a proper description of the halo is only possible in general relativity.

As a first approximation to the dark matter halo, I choose a spherically symmetric spacetime of the form \eref{ss_metric}. As I already discussed in \sref{dm_halo_shape}, most of the available measurements and simulations of halos are consistent with a spherical distribution of dark matter. Even if the halo was slightly aspherical, the high symmetry of \eref{ss_metric} simplifies the algebra tremendously and illustrates the results without obscuring them in too complicated a formalism. The basic principles that are discussed in this thesis, however, are basic in their nature and not specific to spherical symmetry.

The currently available data is still too vague to definitely identify a possible ellipticity of the halo. If it should turn out eventually that spherical symmetry is not suitable for an approximate description of the gravitational field of a galaxy, the results presented in this thesis can easily be adapted to a system with less symmetry.

Before we go on to see how the dark matter halos connect relativistically to the actual observations in \sref{rotcurve} and \sref{grav_lensing}, I will present some speculative candidates for dark mass-energy.

\subsection{The nature of dark matter: speculative models}
\label{sec:dark_matter_candidates}

As Binney and Tremaine \cite{binney87} point out, there is no \textit{a priori} reason to assume that mass and luminosity should be well correlated. Nonetheless it was long believed that in the universe, there is only matter where there is light. The first indication that there is more matter than just the visible kind, was found by Zwicky in 1933 \cite{Zwicky:1933}. He was observing the dynamics of galaxies in the Coma cluster and came to the conclusion that there is some 400 times more mass in that galaxy cluster than indicated by the luminosity of the galaxies and their corresponding mass-to-light ratio alone.

Since then, many more different indicators for dark matter have been found, see \cite{binney87} for a comprehensive overview. The question ``What \textit{is} dark matter?'', however, remained unanswered. Over five decades after Zwicky's discovery, Binney and Tremaine \cite{binney87} considered the nature of dark matter the ``probably single most important unresolved question in extragalactic astronomy''. Today, we still don't know the answer. 

Many different models exist for dark matter, all of which share the only known property: dark matter generates gravitational fields and does not emit photons in any measurable quantity. I am going to present the currently most discussed dark matter candidates and focus on their equation of state and typical pressure to density ratio,
\begin{equation}
\label{eq:doubleu}
w \equiv \frac{p_r+2p_t}{3\rho} = \frac{\bar{p}}{\rho} =\frac{\sum_i p_i}{3\rho} \, .
\end{equation} 
Please note that in this definition, $w$ is not necessarily a constant, but depending on the form of matter at hand, can vary with density and pressure. From equation \eref{pressure_field_eq} we see that general relativistic effects can only be neglected in observations if $|w| \ll 1$. In \sref{rotcurve} and \sref{grav_lensing} I will argue that a significant $w$ can in principle be observed when one keeps track of the pressure terms in a general relativistic analysis of the data.

\paragraph*{Current estimates of cosmological density parameters.}
\label{sec:comological_parameters}
Although the composition of dark matter is not known, reasonably tight constraints on the amount of different matter types exist today. A major boost in accuracy was achieved by fitting the cosmological standard model to the data of the Wilkinson Microwave Anisotropy Probe (WMAP) in 2003~\cite{Bennett:2003}. The density parameter for a matter species $x$ is given in terms of the critical density\footnote{The critical density, which is also known under the name of the Hubble density, is defined as $\rho_\mathrm{crit}\equiv\frac{3H_0^2}{8\pi\,G_N}$. If the average density of the universe equals $\rho_\mathrm{crit}$, i.e.~$\Omega_\mathrm{tot}=1$, \eref{friedman} predicts that the universe is spatially flat. \cite[\S 9.A]{binney87}}, $\Omega_x \equiv \left\langle\rho_x\right\rangle /\rho_\mathrm{crit}$, where $\left\langle\rho_x\right\rangle$ indicates the average over the whole universe. The first Friedmann equation can then be written as
\begin{equation}
\label{eq:friedman}
\Omega_\mathrm{tot} \equiv \sum_i \Omega_i = 1 + \frac{k}{a_0^2\,H_0^2} \, ,
\end{equation} 
where the sum runs over all species, $a_0$ is the curvature radius\footnote{$a_0$ and $H_0$ refer to the current epoch.} of the Friedmann-Lema\^itre-Robertson-Walker metric, $H_0$ is the Hubble parameter\footnotemark[\value{footnote}] and $k\in\{-1,0,1\}$ determines whether the FLRW space is a hyperbolic 3-plane, Euclidean or a hypersphere. The Hubble parameter is often given in terms of $h$, which is defined by $H_0 = 100\,h\, \unit{km\,s^{-1}\,Mpc^{-1}}$.

\begin{table}[htb]
\begin{center}
\begin{tabular}{lll}
\hline\hline\hlinevspace
Parameter & Symbol              & Value  \\ \hline \hlinevspace
Total average density of the universe      & $\Omega_\mathrm{tot}$ & $1.02 \pm 0.02$ \\
Density of cosmological constant $\Lambda$ & $\Omega_\Lambda$      & $0.77 \pm 0.06$ \\
Total matter density                       & $\Omega_\mathrm{m}$   & $0.25 \pm 0.04$ \\
Baryon density                             & $\Omega_\mathrm{b}$   & $0.043 \pm 0.006$ \\
Non-baryonic matter density                & $\Omega_\mathrm{nbm}$ & $0.21 \pm 0.04$ \\
Cosmic density of luminous matter (\S 20.3)& $\Omega_\mathrm{lum}$ & $0.0033 \pm 0.0002$ \\
Neutrino density (\S 21.1.2)               & $\Omega_\nu$          & $\sim 0.001$ \\
Radiation density (\S 21.1.4)              & $\Omega_\mathrm{rad}$ & $0.000046 \pm 0.000004$ \\ \hline\hline
\end{tabular} 
\end{center}
\caption[The cosmological density parameters as given by the Particle Data Group (2004) \cite{Eidelman:2004}.]{The cosmological density parameters as given by the Particle Data Group (2004) \cite{Eidelman:2004}. Fitting the first acoustic peak of the cosmic microwave background (CMB) angular power spectrum yields the values of $\Omega_\mathrm{tot}$, $\Omega_\mathrm{m}$ and $\Omega_\mathrm{b}$ (\S 21.1.4 and \S 21.3.3 of \cite{Eidelman:2004}). The dark energy density $\Omega_\Lambda$ is derived from $\Omega_\mathrm{tot}-\Omega_\mathrm{m}$, and $\Omega_\mathrm{nbm}$ is given by $\Omega_\mathrm{m}-\Omega_\mathrm{b}$ (\S 22.1.1). All values and $\sigma$-confidence intervals were obtained using a value of $h=0.73 \pm 0.03$.}
\label{tab:cosmo_densities}
\end{table}

The Particle Data Group \cite{Eidelman:2004} bi-annually provides the current best values of the density parameters, which are listed in \tref{cosmo_densities}. The data suggest that the most mass-energy in the universe is present in the form of dark energy, which is most typically modelled as cosmological constant with $w=-1$ \cite[\S 21.2.3]{Eidelman:2004}. The majority of matter in the universe is dark since the luminous fraction of matter is only $\Omega_\mathrm{lum}/\Omega_\mathrm{m} \approx 0.013$. Although the baryonic dark matter is dominating the visible matter, $\Omega_\mathrm{b}/\Omega_\mathrm{lum} \approx 13$, most of the dark matter must be present in non-baryonic form: $\Omega_\mathrm{nbm}/\Omega_\mathrm{m} \approx 0.83$.

So, not only do we not know what dark matter actually is, it also looks like there are at least two different kinds of dark matter -- if the cosmological standard model is correct. And as of 2004, it fits the data remarkably well \cite[\S 21.4]{Eidelman:2004}. I will now present some suggestions what dark matter actually might be.

\paragraph*{Baryonic dark matter} could be present as unobserved low-luminosity stars that do not have enough mass to burn hydrogen (brown dwarfs, $M\lesssim 0.08\,\Msun$) or stellar remnants of high mass stars, that now would be present as white dwarfs, neutron stars or black holes. Since all these objects do emit a small amount of in principle detectable radiation, their density can not exceed $\Omega_\mathrm{stel.rem.}\approx 0.03$ \cite[\S 10.4.1a]{binney87}.

Other examples of massive objects that are very difficult to detect are primordial black holes that were created during the big bang, or Jupiter mass objects ($\sim 10^{-3}\,\Msun$) \cite{eaa}. All of the aforementioned objects are generally referred to as MAssive Compact Halo Objects (MACHOs), and detection of their presence and estimation of their total mass is attempted by microlensing surveys \cite{Paczynski:1998}. Together with MACHOs, cold gas clouds have also been suggested as baryonic dark matter candidates \cite{Paolis:1995}.

All of these baryonic dark matter candidates have non-relativistic speeds and are therefore part of the class of Cold Dark Matter (CDM). The parameter $w$ for CDM can be estimated by considering an ideal gas of particles with mass $m_p$:
\begin{equation}
w = \frac{p}{\rho}  \approx \frac{k_B\,T}{m_p\,c^2} \approx \frac{1}{3}\,\left\langle \frac{v^2}{c^2} \right\rangle \, ,
\end{equation} 
where $k_B$ is Boltzmann's constant, $T$ is the ``temperature'' of an ideal gas as given by
\begin{equation}
\frac{3}{2}\,k_B\,T = \frac{1}{2}\,m\,\left\langle v^2 \right\rangle \, ,
\end{equation}
and $c$ the speed of light. Therefore, the non-relativistic particle speed of CDM indicates negligible pressure, i.e. $w \approx 0$.


\paragraph*{Exotic particles.} Massive neutrinos as well as Hot and Warm Dark Matter (HDM, WDM) are generally ruled out as major dark matter candidates, since their speed is usually too high to allow for the dark matter clumping which is observed in galaxies \cite{binney87,Navarro:1997}. 
 Consequently, if they are to be significant components of the dark matter, exotic particles should classify as CDM. The two most prominent exotic CDM particles in the literature at the moment are axions and Weakly Interacting Massive Particles (WIMPs) \cite[\S 22.1.2]{Eidelman:2004}.

Axions were first postulated as a means to solve the charge-parity problem of Quantum ChromoDynamics (QCD), but they also occur naturally in superstring models \cite[\S 22.1.2]{Eidelman:2004}. Two experiments that search for axions are currently collecting data, one at the Lawrence Livermore National Laboratory, California and the CARRACK experiment in Kyoto. Both of them have so far only excluded axions with certain rest masses, and have not yet actually found positive evidence for axions \cite[\S 22.2.2]{Eidelman:2004}. The current astrophysical observations, and laboratory experiments, indicate an upper limit on the axion mass of $m_a\lesssim 10^{-2}\,\unit{eV}$ \cite[p.393]{Eidelman:2004}. 

WIMPs are characterized as particles with a rest mass of roughly $10\,\unit{GeV}$ to a few $\unit{TeV}$ and a weak annihilation cross section that makes them difficult to detect. The currently favoured WIMP candidate is a Lightest SuperParticle (LSP) in supersymmetric models, and amongst the LSPs, the ``neutralino'' seems to be the most promising one \cite[\S 22.1.2]{Eidelman:2004}. 
With both, axions and WIMPs, being candidates for Cold Dark Matter, the expected pressure content of dark matter in these cases is negligible compared to the mass-energy density and hence, $w \approx 0$.

\paragraph*{Non-CDM fields.} \label{sec:non-cdm_fields}
Apart from the currently favoured Cold Dark Matter models, a few approaches consider scalar fields and Bose-Einstein condensates that exhibit a significant field pressure or tension.
Amongst these proposals for DM candidates is a ``string fluid'' which suggests $w \approx -1/3$ \cite{Soleng:1993} (though other models of string fluids also appear in the literature), and many different approaches to Scalar Field Dark Matter (SFDM) which is usually modelled through at least one scalar field in a general relativistic framework with a self-interaction potential. All SFDM models exhibit significant pressures, e.g. Schunck's model has a position dependent $w \in [0,1]$ \cite{Schunck:1999}, Peebles considers a scalar field with a purely quartic self-interaction potential and finds $w = 1/3$ \cite{Peebles:2000}, the scalar field model of Arbey \textit{et al.} (which is variant of a Bose-Einstein condensate) has a quadratic equation of state and thus\footnote{$\lambda$ is the quartic coupling parameter of the scalar field, $\hbar$ is Planck's constant divided by $2\pi$, $m$ is the mass associated with the scalar field, i.e. the boson mass of the Bose-Einstein condensate, and $c$ is the speed of light. In SI units, $w\approx c^2\,\left(\frac{\lambda}{10^{-2}}\right)\,\left(\frac{1\,\unit{eV}}{m}\right)^4\,\left(\frac{\rho}{9.3\times 10^{-14}\,\unit{kg/m^3}}\right)$. For comparison, the density of the solar neighbourhood is $\rho_0 = (0.18\pm 0.03)\,\Msun\,\unit{pc^{-3}} \approx 1.2 \times 10^{-20} \fracunit{kg}{m^3}$.} $w=\lambda\hbar^3/(4m^4c)\rho$ \cite{Arbey:2003} and Matos \textit{et al.} claim that their SFDM model even exhibits $w \approx 10^6$ \cite{Matos:2001}!

\paragraph*{Modified physics.}
A completely different approach to explain the dark matter problem is to demand that the theory of gravity must be modified for distance scales as big as those of the dark matter phenomena. This idea, generally referred to as MOdified Newton Dynamics (MOND), is formulated by Milgrom \cite{Milgrom:1983}. Until this theory is extended into a ``clean'' fully relativistic invariant theory, it can not be considered a complete solution to the dark matter problem.

\paragraph*{} Since there are competing dark matter candidates with different values for $w$, any way of actually observing $w$ would mean that it is possible to further restrict the possible dark matter candidates. In the following two sections, I will outline a way of in principle measuring $w$ in spiral galaxies.

%% file: ts_rotation_curves.tex
\section{Rotation curves}
\label{sec:rotcurve}

The term ``rotation curve'' generally refers to a set of circular rotation velocities of tracer particles in the disk of spiral galaxies that were obtained at different radii. Given the circular rotation velocity, one can infer the mass inside a sphere with the corresponding radius. A rotation curve then provides information about the radial matter distribution. If one additionally assumes a model for the 3D shape of the matter distribution in the spiral galaxy, i.e. respecting the different non-spherical contributors, the rotation curve determines the entire matter distribution.

The Doppler shift of emission lines makes the observation of the tracer particles' motion possible. Mostly the $21\,\unit{cm}$ emission line of neutral hydrogen (\HI) is observed at radio wavelengths. \HI{} gas observations generally extend to larger radii than other tracers. \cite[\S 10.1.6]{binney87}

\subsection{General shape of rotation curve profiles} 
\label{sec:rotcurve_shape}

In the early days of rotation curve measurements in the 1950s, the rotation curves were modeled after the exponential disk model of galaxies, see \eref{spiral_exp_disk}. This model implied that a rotation curve would exhibit three characteristic regions: (i) the central region where the velocity rises linearly with distance, (ii) a region where the velocity reaches its maximum, the so called \textit{turnover radius}, and (iii) the outer region, which was expected to show a \textit{Keplerian falloff} $\propto r^{-1/2}$ since the galaxy supposedly resembled a point mass at large distances. \cite{binney87}

With this expectation in mind, it is not a big surprise that the first observers assumed the Keplerian falloff to be there, even though the observation extended only up to a flat region of constant velocity, which was interpreted as the turnover radius. It was only in the 1970s that the improved instrumental sensitivity allowed observations out to larger radii and it was found that the flat region of the rotation curve extended further than agreement with the exponential disk model would allow. Instead, the flat region extended as far as the observations of the rotation curve reached, without any indication of declining rotational velocity. Consequently, Freeman suggested in 1970 that ``there must be [...]~additional matter which is undetected, either optically or at $21\,\unit{cm}$ [...], and its distribution must be quite different from the exponential distribution'' \cite{Freeman:1970}.

Today, several thousand rotation curves have been observed and are publicly available in electronic databases \cite{Sofue:2001}. Data analysis of about 1100 rotation curves shows common properties for all sizes of spiral galaxies: The most massive galaxies show a slightly declining, though non-Keplerian, profile in the outermost region; intermediate galaxies have a nearly flat region in the outer parts; and dwarf galaxies show a steadily increasing velocity profile throughout the observable region \cite{Persic:1996}. This analysis brought forth a completely empirical universal rotation curve for spiral galaxies that is only parameterized by the total luminosity $L$:
\begin{eqnarray}
v_c(r) &=& a(\lambda)\,\sqrt{b(\lambda)\,\frac{1.97 x^{1.22}}{(x^2+0.61)^{1.43}} + c(\lambda)\,\frac{x^2}{x^2+2.25\,\lambda^{0.4}}}  \\
\nonumber \mathrm{with} && \\
a(\lambda) &\equiv& 200\,\fracunit{km}{s} \times \lambda^{0.41}\left\lbrace0.80 + 0.49\log_{10}\lambda + \frac{0.75 \exp(-0.4\lambda)}{0.47+2.25\lambda^{0.4}}\right\rbrace^{-1/2} \\
b(\lambda) &\equiv& 0.72 + 0.44\log_{10}\lambda \\
c(\lambda) &\equiv& 1.6\, e^{-0.4\lambda} \\
x &\equiv& r/R_{opt} \simeq \frac{r}{\sqrt{\lambda}\,13\,\unit{kpc}} \\
\lambda &\equiv& L/L_* \\
L_* &\equiv& 1.0 \times 10^{10}\, h^{-2}\, L_\odot \, ,
\end{eqnarray} 
where $L_\odot$ is the Sun's luminosity in the visual band \cite{binney87}, and $h$ is the dimensionless Hubble parameter.
The $b(\lambda)$-term of the velocity arises from the stellar disk which is modeled as an approximation to the exponential disk model \eref{spiral_exp_disk}. The dark halo contribution of this universal rotation curve is given by the $c(\lambda)$-term and has an asymptotically constant rotation velocity of \cite{Persic:1996}
\begin{equation}
v_\infty\simeq 1.3\,a(\lambda)\,e^{-0.2\lambda} \, .
\end{equation}
This corresponds to a halo density that falls off as $\rho \propto r^{-2}$ and consequently implies an infinite total halo mass. Such a model is usually referred to as a ``singular isothermal sphere'' (SIS). The name arises from the solution of the Newtonian equation of hydrostatic equilibrium,
\begin{equation}
\frac{\d p}{\d r} = - \frac{M(r)\,\rho(r)}{r^2} \, ,
\end{equation} 
for an isothermal ideal gas with
\begin{equation}
\frac{\d p}{\d r} = \frac{k_B\,T}{m_p}\,\frac{\d \rho}{\d r} \, .
\end{equation} 
Combining both equations, one finds using $M'(r)=4\pi\,r^2\rho(r)$:
\begin{equation}
\frac{\d}{\d r} \left( r^2\, \frac{\d \ln{\rho}}{\d r} \right) = - \frac{4\pi\,m_p}{k_B\,T}\, r^2\,\rho(r) \, ,
\end{equation} 
with the solution
\begin{equation}
\rho(r) = \frac{k_B\,T}{2\pi\,m_p}\, \frac{1}{r^2} \, .
\end{equation} 
This density profile diverges for $r=0$, hence the name \textit{singular isothermal sphere}. Also the total mass of the SIS diverges linearly with radius.

Alternatively, numerical simulation of the evolution of CDM halos in a cosmological context suggest a different density profile, one that falls off as $\rho \propto r^{-3}$ for \textit{very} large distances. This still corresponds to a logarithmically divergent total mass. Such halo models are usually referred to as NFW-halos, named after Navarro, Frenk and White who conducted the relevant simulations \cite{Navarro:1996,Navarro:1997}. The associated rotation curve is asymptotically declining, although the decline only occurs in a region that lies beyond the region in which rotation curves of individual galaxies can be observed ($\gtrsim 100\,\unit{kpc}$). However, statistical analysis of large samples of satellite galaxies in orbit around their primary galaxies seem to support the notion of NFW-halos and the declining rotation curve at large distances \cite{Prada:2003}. Statistical methods that use weak gravitational lensing data (\sref{weak_lensing}) also seem to favour NFW halos \cite{Guzik:2002,Kleinheinrich:2003}.

Summarizing, the direct evidence of individually observed rotation curves exhibits a general behaviour of an asymptotically flat rotation curve up to the last observable data point. This corresponds to a density that falls off as $\rho \propto r^{-2}$. However, there are indirect indications from statistical methods that the rotation curve ultimately declines, which implies a density profile that goes asymptotically as $\rho \propto r^{-3}$ or even steeper. If the total mass of the galactic halo is to be \textit{finite}, then the density must asymptotically fall off \textit{faster} than $\rho \propto r^{-3}$.

\subsection{Measuring the rotation curve through direct observations}
\label{sec:measuring_rotcurves}

To calculate the relationship between the gravitational potential $\Phi$ and the observed redshift $Z$ of the tracer particles, I make two assumptions that will simplify the relevant algebra tremendously:
\begin{enumerate}
\item The tracer particles are in circular orbits around the center of the galaxy and are confined to the plane of the disk. \label{it:ass_plane}
\item The gravitational field is spherically symmetric. \label{it:ass_spsym}
\end{enumerate}
The first assumption agrees with what is generally known about spiral galaxies (see \sref{spiral_disk_anatomy}) and thus, the rigid restriction to motion within the plane ($\theta=\pi/2$) and circular orbits ($\d r=0$) is acceptable as a first approximation.
Assumption \ref{it:ass_spsym} is motivated by the approximately spherical shape of the dark halo. There are, however, also aspherical contributors to the gravitational field, which is in the region of interest mainly due to the stellar disk. But since we are for now only interested in the motion of the tracer particles within the galactic plane, the symmetry in the plane is circular, no matter whether the gravitational field is spherically or only azimuthally symmetric. Hence, I will utilize a metric of the form \eref{metric_standard_form} or equivalently \eref{ss_metric}.

The following derivation of the relation between the observable redshift $z_\pm$ and the metric function $\Phi(r)$ is based on a modification of the presentation of Lake \cite{Lake:2004} and Weinberg \cite{weinberg72}.

\subsubsection{Constants of motion}
The geodesic equations \eref{geodesic_affine_parameter} of the metric \eref{metric_standard_form} with an affine parameter $\chi$ are
\begin{eqnarray}
\label{eq:gde_t_0}
0 &=& \frac{\d^2 t}{\d\chi^2} +
\frac{A'(r)}{A(r)} \frac{\d t}{\d\chi} \frac{\d r}{\d\chi} \\
\label{eq:gde_r_0}
0 &=& \frac{\d^2 r}{\d\chi^2} +
\frac{A'(r)}{2B(r)} \left( \frac{\d t}{\d\chi} \right)^2 +
\frac{B'(r)}{2B(r)} \left( \frac{\d r}{\d\chi} \right)^2 -
\frac{r}{B(r)} \left( \frac{\d\theta}{\d\chi} \right)^2 -
\frac{r \sin^2\theta}{B(r)} \left( \frac{\d\cphi}{\d\chi} \right)^2 \\
\label{eq:gde_theta_0}
0 &=& \frac{\d^2\theta}{\d\chi^2} +
\frac{2}{r} \frac{\d\theta}{\d\chi} \frac{\d r}{\d\chi} -
\sin\theta \cos\theta \left( \frac{\d\cphi}{\d\chi} \right)^2 \\
\label{eq:gde_phi_0}
0 &=& \frac{\d^2\cphi}{\d\chi^2} +
\frac{2}{r} \frac{\d\cphi}{\d\chi} \frac{\d r}{\d\chi} +
2\cot\theta\, \frac{\d\cphi}{\d\chi} \frac{\d\theta}{\d\chi} \, ,
\end{eqnarray} 
where the shorthand symbol $' \equiv \d/\d r$ has been used. 
Due to assumption \ref{it:ass_plane}, \eref{gde_theta_0} is immediately satisfied, and we can disregard $\theta$ in the following discussion. The remaining geodesic equations simplify to
\begin{eqnarray}
\label{eq:gde_t}
0 &=& \frac{\d^2 t}{\d\chi^2} +
\frac{A'(r)}{A(r)} \frac{\d t}{\d\chi} \frac{\d r}{\d\chi} \\
\label{eq:gde_r}
0 &=& \frac{\d^2 r}{\d\chi^2} +
\frac{A'(r)}{2B(r)} \left( \frac{\d t}{\d\chi} \right)^2 +
\frac{B'(r)}{2B(r)} \left( \frac{\d r}{\d\chi} \right)^2 -
\frac{r}{B(r)} \left( \frac{\d\cphi}{\d\chi} \right)^2 \\
\label{eq:gde_phi}
0 &=& \frac{\d^2\cphi}{\d\chi^2} +
\frac{2}{r} \frac{\d\cphi}{\d\chi} \frac{\d r}{\d\chi} \, .
\end{eqnarray} 
Dividing \eref{gde_t} and \eref{gde_phi} by $\d t/\d\chi$ and $\d\cphi/\d\chi$ respectively, yields
\begin{eqnarray}
\frac{\d}{\d\chi} \left\lbrace \ln \left( A(r)\,\frac{\d t}{\d\chi} \right)  \right\rbrace &=& 0 \\
\frac{\d}{\d\chi} \left\lbrace \ln \left( r^2\,\frac{\d\cphi}{\d\chi} \right)  \right\rbrace &=& 0 \, ,
\end{eqnarray} 
which motivates the definition of two constants of motion:
\begin{eqnarray}
\label{eq:com_energy_null}
\tilde{\gamma} &\equiv& -g_{tt}(r)\,\frac{\d t}{\d\chi} = A(r)\,\frac{\d t}{\d\chi} = e^{2\Phi(r)}\,\frac{\d t}{\d\chi} \\
\label{eq:com_angmom_null}
\tilde{\ell} &\equiv& g_{\cphi\cphi}(r)\,\frac{\d\cphi}{\d\chi} = r^2\,\frac{\d\cphi}{\d\chi} \, ,
\end{eqnarray} 
where I stated $\tilde{\gamma}$ also in terms of the metric function $\Phi(r)$ of the geometry \eref{ss_metric}. These quantities are proportional to the conjugate momenta of the cyclical coordinates $t$ and $\cphi$, where the term ``cyclical'' refers to their absence in the metric components $g_{\alpha\beta}$.

With these definitions and multiplication by $2B(r)\,\d r/\d\chi$, the remaining geodesic equation \eref{gde_r} simplifies to
\begin{equation}
\frac{\d}{\d\chi} \left\lbrace B(r)\left( \frac{\d r}{\d\chi} \right)^2 - \frac{\tilde{\gamma}^2}{A(r)} + \frac{\tilde{\ell}^2}{r^2} \right\rbrace = 0 \, ,
\end{equation} 
which gives yet another constant of motion,
\begin{equation}
\label{eq:com_K}
{\cal K}^2 \equiv -B(r)\left( \frac{\d r}{\d\chi} \right)^2 + \frac{\tilde{\gamma}^2}{A(r)} - \frac{\tilde{\ell}^2}{r^2} \, .
\end{equation}
By inserting all four constants of motion, $\theta$, $\gamma$, $\ell$ and ${\cal K}$, into the metric \eref{metric_standard_form}, I find the proper time $\tau$ to be given by
\begin{equation}
\d\tau^2 \equiv -\d s^2 = {\cal K}^2\, \d\chi^2 \, .
\end{equation} 
This was expected, since proper time is in general an affine parameter for timelike particles. The constant ${\cal K}$ is not a dynamical constant of motion, but a mere manifestation of gauge-freedom to choose the affine parameter.

The special case ${\cal K}=0 \Rightarrow \d\tau=0 $ is characteristic of the motion of null-particles, i.e. photons or other particles that travel at the speed of light. For these particles, \eref{com_K} simplifies to
\begin{equation}
\label{eq:com_drdchi_null}
0= -B(r)\left( \frac{\d r}{\d\chi} \right)^2 + \frac{\tilde{\gamma}^2}{A(r)} - \frac{\tilde{\ell}^2}{r^2} \, , 
\end{equation}
where the affine parameter \eref{null_affine_parameter}, $\chi$, is generally chosen for convenience to satisfy
\begin{equation}
\frac{\d t}{\d\chi} \rightarrow 1 \quad \mathrm{for} \quad r \rightarrow \infty \, ,
\end{equation} 
if the particle actually reaches $r=\infty$ at any point of its trajectory.

All other values for ${\cal K}$ invoke a non-vanishing proper time interval $\d\tau \neq 0$ and thus correspond to the motion of timelike particles. Defining
\begin{equation}
\tilde{\gamma} \equiv {\cal K}\, \gamma \, ; \qquad \tilde{\ell} \equiv {\cal K}\, \ell \, ,
\end{equation} 
for all timelike particles, makes \eref{com_K} independent of ${\cal K}$ (as long as ${\cal K} \neq 0$):
\begin{equation}
\label{eq:com_drdtau}
1 = -B(r)\left( \frac{\d r}{\d\tau} \right)^2 + \frac{\gamma^2}{A(r)} - \frac{\ell^2}{r^2} \, .
\end{equation}
At the same time, the constants of motion $\gamma$ and $\ell$ gain a familiar interpretation: 
\begin{equation}
\label{eq:com_energy}
\gamma = e^{2\Phi(r)}\,\frac{\d t}{\d\tau} 
\end{equation} 
is easily identified as the \textit{energy per unit mass at infinity} and
\begin{equation}
\label{eq:com_angmom}
\ell = r^2\,\frac{\d\cphi}{\d\tau}
\end{equation} 
as the \textit{angular momentum per unit mass} \cite[\S 25.3]{MTW}. From now on, I choose ${\cal K}=1$ for timelike particles so that $\d\tau=\d\chi$. For null trajectories, however, the affine parameter $\chi$ must still be specified.

\subsubsection{Circular orbits of timelike particles}
For circular orbits, $r=r_e$ is another constant of motion, where $r_e$ labels the radius of the emitting particle's orbit. Thus, $\d r=0$ and using \eref{com_energy} and \eref{com_angmom}, the geodesic equation \eref{gde_r} simplifies to
\begin{equation}
\label{eq:gde_r_circ_orbit}
\frac{A'(r)}{2A(r)^2}\,\gamma^2 - \frac{\ell^2}{r^3} \; = \; 0 \; = \; \gamma^2\,e^{-2\Phi(r)}\,\Phi'(r) - \frac{\ell^2}{r^3} \, ,
\end{equation}
where again, I used the metric function $\Phi(r)$ instead of $A(r)$. Combining the geodesic equation \eref{gde_r_circ_orbit} with \eref{com_drdtau} for circular orbits gives the energy and angular momentum for the emitter at radius $r_e$:
\begin{eqnarray}
\label{eq:com_energy_emit}
\gamma_e &=& \frac{e^{\Phi(r_e)}}{\sqrt{1-r_e\,\Phi'(r_e)}} \\
\label{eq:com_angmom_emit}
\ell_e &=& \frac{r_e\sqrt{r_e\,\Phi'(r_e)}}{\sqrt{1-r_e\,\Phi'(r_e)}}
\end{eqnarray} 
Hence $\Phi'>0$, from the condition
\begin{equation}\label{eq:orbit_condition}
  0 \le r\,\Phi'(r) < 1
\end{equation}
for the existence of circular orbits. 

\subsubsection{Observing light emitted by particles on circular orbits}
\label{sec:obs_light}

The redshift of light is given by \cite[p. 47ff]{schroedinger56}
\begin{equation}\label{eq:general_redshift}
  1+z \equiv \frac{\lambda_o}{\lambda_e} = \frac{\left[ g_{\alpha\beta}\, V^\alpha\, k^\beta \right]_e}{\left[ g_{\alpha\beta}\, V^\alpha\, k^\beta \right]_o}
\end{equation}
where $\lambda$ is the wavelength, $e$ denotes the event of emission, $o$
the event of observation, $V^\alpha \equiv \d x^\alpha/\d \tau$ is tangent to the worldline of the emitting particle / observer and $k^\beta \equiv \d x^\beta/\d \chi$ tangent to the null
geodesic of the emitted / observed light. Furthermore, $\tau$ is proper time, $\chi$ is a suitable affine parameter for null curves, and $g_{\alpha\beta}$ is the contravariant metric. The affine parameter for the photon trajectory in a static spacetime \eref{null_affine_parameter} yields a constant energy at infinity $\tilde{\gamma}_N$ \eref{com_energy_null} for the null curve:
\begin{equation}
\d \chi = -g_{tt}\,\d t = e^{2\Phi(r)}\,\d t \, \quad \Longrightarrow \quad \tilde{\gamma}_N=1 \, .
\end{equation} 
Let us assume the observer to be at rest at infinity in an asymptotically flat space time with $\Phi(\infty)=0$, so that the only non-zero component of the observer's tangent vector is
\begin{equation}
  \left[\frac{\d x^t}{\d \tau}\right]_o = \left[ V^t \right]_o = e^{-\Phi(\infty)} = 1 \, .
\end{equation}
The emitting particle performs a circular orbit, thus $\d r=\d\theta=0$ and from \eref{com_angmom} and the invariant proper time interval, $\d\tau^2=-\d s^2$ \eref{ss_metric}, follow the non-zero components of the emitter's tangent vector
\begin{eqnarray}
\left[\frac{\d x^\cphi}{\d \tau}\right]_e &=& \left[ V^\cphi \right]_e = \frac{\ell_e}{r_e^2} \, ; \\
\left[\frac{\d x^t}{\d \tau}\right]_e &=& \left[ V^t \right]_e =
 \sqrt{1+\frac{\ell_e^2}{r_e^2}}\: e^{-\Phi(r_e)} \, .
\end{eqnarray}
The observed redshift then takes the form
\begin{equation}\label{eq:redshift1}
  1+z = \frac{ \left[ g_{tt}\, \sqrt{1+\frac{\ell_e^2}{r_e^2}}\, e^{-\Phi(r_e)} \frac{\d t}{\d \chi} +
  g_{\cphi\cphi}\, \frac{\ell_e}{r_e^2}\, \frac{\d \cphi}{\d \chi} \right]_e}{
  \left[ g_{tt}\, 1\, \frac{\d t}{\d \chi} \right]_o } \, .
\end{equation}
Inserting the affine parameter $\d \chi=-g_{tt}\,\d x^t$ yields
\begin{equation}\label{eq:redshift2}
  1+z = \left[ \sqrt{1+\frac{\ell_e^2}{r_e^2}}\, e^{-\Phi(r_e)} +
  \frac{g_{\cphi\cphi}}{g_{tt}} \frac{\ell_e}{r_e^2} \left. \frac{\d \cphi}{\d t}\right|_N \right]_e
\end{equation}
where $\left.\d \cphi/\d t\right|_N$ refers to the null path of the light, conveying information about the emitting particle to the observer. Recalling the constants of motion $\tilde{\gamma}_N$ and $\tilde{\ell}_N$ given by \eref{com_energy_null} and \eref{com_angmom_null}, which are defined along the null trajectory of the photon, I define a new constant of motion, the \textit{impact parameter}
\begin{equation}\label{eq:impact_param}
  b \equiv \frac{\tilde{\ell}_N}{\tilde{\gamma}_N} = \frac{r_N^2 \frac{\d\cphi}{\d\chi}}{e^{2\Phi(r_N)} \frac{\d t}{\d\chi}} = \frac{r_N^2}{e^{2\Phi(r_N)}} \left. \frac{\d\cphi}{\d t}\right|_N \, .
\end{equation}
The indices $N$ emphasize the fact that $b$ is constant for all points along the null worldline of any photon.

\begin{figure}[htb]
\begin{center}
\input{figures/322_impact_parameter.pstex_t}
\end{center}
\caption[Classical definition of impact parameter.]{\label{fig:impact_parameter}
A particle with mass $m$ on a straight trajectory with velocity $\vec{v}$ and position vector $\vec{r}$ has the impact parameter $b$. The absolute value of the particle's angular momentum is given by $L=|m\, \vec{r} \times \vec{v}| = m\, v\, r \sin\cphi$.}
\end{figure}
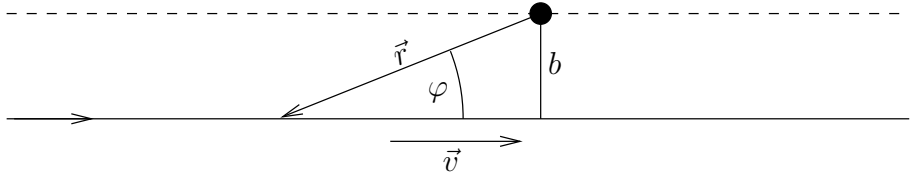
To see that $b$ is the impact parameter at infinity (i.e. the
observer), remember the classical definition of the impact
parameter (\fref{impact_parameter})
\begin{equation}\label{eq:class_impact_param}
  L = |m\, \vec{r} \times \vec{v}| = m\, v\, r
  \sin(\sphericalangle(\vec{r},\vec{v}))= m\, v\, b \quad\Rightarrow\quad b =
  \frac{L}{m\,v}= \frac{L}{p} = \frac{\tilde{\ell}}{\tilde{\gamma}} \, .
\end{equation}
Please note that for photons $\tilde{\gamma}=p\,c$ and in geometrical units
$c=1$. Since it is not possible to measure $r_e$ directly, it is
necessary to rely on observing the \emph{apparent radius}, which
is the impact parameter $b$ at infinity.

It follows now from \eref{redshift2}, using \eref{com_angmom_emit} for
the angular momentum of the emitting particle, $\ell_e$, and the
metric components $g_{\cphi\cphi}$ and $g_{tt}$ of \eref{ss_metric} at the radius $r_e$ that
\begin{eqnarray}
  \nonumber
  1+z &=& \sqrt{1+\frac{r_e^3\Phi'(r_e)}{r_e^2(1-r_e\Phi'(r_e))}}\, e^{-\Phi(r_e)} +
  \frac{r_e^2}{-e^{2\Phi(r_e)}}\,\frac{r_e\sqrt{r_e\Phi'(r_e)}}{r_e^2\sqrt{1-r_e\Phi'(r_e)}}  \left. \frac{\d \cphi}{\d t}\right|_N \\
  \nonumber
  &=& \frac{e^{-\Phi(r_e)}}{\sqrt{1-r_e\Phi'(r_e)}} -
  \frac{\sqrt{r_e\Phi'(r_e)}}{r_e\,\sqrt{1-r_e\Phi'(r_e)}}\, \frac{r_e^2}{e^{2\Phi(r_e)}} \left. \frac{\d \cphi}{\d t}\right|_N \, .
\end{eqnarray}
In terms of the impact parameter $b$, given by \eref{impact_param}
for the event of emission ($r_N=r_e$), this is finally
\begin{equation}\label{eq:total_redshift}
  1+z_\pm = \frac{1}{\sqrt{1-r_e\Phi'(r_e)}} \left( \frac{1}{e^{\Phi(r_e)}}
  - \frac{\pm |b| \sqrt{r_e\Phi'(r_e)}}{r_e} \right) \, .
\end{equation}

The terms involved in \eref{total_redshift} have a simple
interpretation: $z_\pm$ is the \emph{total redshift} of light that
was emitted by particles moving on circular orbits in a geometry
of the form \eref{ss_metric}, where $z_+$ is the redshift of
approaching and $z_-$ that of receding particles. The second term
vanishes when the light ray moves radially outwards ($\d x^\cphi /
\d\chi = 0$), perpendicular to the circular trajectory of the
emitter, which can easily been seen from \eref{redshift1}.
Therefore, it represents \emph{Doppler redshift}. The first term
is independent of the emitter's motion and only dependent on the
the gravitational potential $\Phi(r_e)$. It represents
\emph{gravitational redshift}. Thus, I define gravitational
redshift
\begin{equation}\label{eq:grav_redshift}
  1+z_g \equiv 1 + \frac{z_- + z_+}{2} = \frac{e^{-\Phi(r_e)}}{\sqrt{1-r_e\Phi'(r_e)}}
\end{equation}
and Doppler redshift (where $b>0$ for the rest of the discussion)
\begin{equation}\label{eq:Doppler_redshift}
  Z \equiv  \frac{z_- - z_+}{2} = \frac{b
  \sqrt{r_e\Phi'(r_e)}}{r_e\sqrt{1-r_e\Phi'(r_e)}} \, .
\end{equation}
Similar derivations of $z_g$ and $Z$ can be found in \cite{Lake:2004,Nucamendi:2000}.

Through \eref{com_energy}, \eref{com_angmom}, \eref{com_energy_emit}
and \eref{com_angmom_emit}, these terms are also related to
$\left.\d t/\d\tau\right|_e$ and $\left.\d\cphi/\d\tau\right|_e$:
\begin{eqnarray}
\label{eq:grav_redshift_coord}
  1+z_g &=& \gamma_e\, e^{-2\Phi(r_e)} = \left[ \frac{\d t}{\d \tau} \right]_e\\
\label{eq:Doppler_redshift_coord}
  Z &=& b\, \frac{\ell_e}{r_e^2 } = b \left[ \frac{\d \cphi}{\d \tau} \right]_e \, .
\end{eqnarray}
From the proper time interval for the worldline of the emitter's circular orbit, we get an expression for $\left.\d t/\d\tau\right|_e$:
\begin{eqnarray}
 \d\tau^2 &=& e^{2\Phi(r_e)}\,\d t^2 - r_e^2\,\d\cphi^2 \\
\nonumber
\Rightarrow\quad \left[ \frac{\d t}{\d \tau} \right]_e &=& e^{-\Phi(r_e)} \sqrt{1+r_e^2\,\left[ \frac{\d \cphi}{\d \tau} \right]_e^2} \\
&\approx&
e^{-\Phi(r_e)} \left\lbrace 1 + \frac{r_e^2}{2}\,\left[ \frac{\d \cphi}{\d \tau}
  \right]_e^2 + \O{\left(r\, \frac{\d \cphi}{\d \tau}\right)_{\!e}^4} \right\rbrace \, .
\end{eqnarray} 
Since the first term of the particle's ``circular velocity'', $r_e\,\left.\d\cphi/\d\tau\right|_e \ll c$, occurs only in the second order in \eref{grav_redshift_coord}, one can already suspect that the gravitational redshift will be negligible compared to the Doppler redshift \eref{Doppler_redshift_coord} in which the ``circular velocity'' occurs in first order. The quotation marks are necessary, as at the moment the different quantities $b$ and $r_e$ are compared. To get a valid comparison, their relationship has to be examined.

\subsubsection{Impact parameter $b$ vs. coordinate radius $r$} \label{sec:obtain_Phi}

For legibility the index $e$ is now dropped and all radii $r$
refer to the radius of the emitter's circular orbit.

Generally it is not possible to measure $Z(r)$ directly. This is
firstly due to the fact that the observer measures the position of
a signal in terms of the apparent radius or impact parameter $b$.
Secondly only the total redshift $z_\pm(b)$ is observable.
Finally, the observer measures many different redshift values from
various emitters along the line of sight, given by $b$. A mapping
$r(b)$ for a corresponding redshift value $z_\pm(b)$ is necessary
to filter out the appropriate $z_\pm(b)$ from all available
values. Thus, a way to obtain the gravitational potential $\Phi(r)$ is
\begin{enumerate}
  \item Measure total redshifts $z_+(b)$ and $z_-(b)$.
  \item From that calculate $Z(b) = \frac{1}{2}\left(z_-(b) - z_+(b)\right)$. \label{it:calcZ}
  \item Obtain $\Phi(b)$ from $Z(b)$.
  \item Interchange variables $b \leftrightarrow r$ to get $\Phi(r)$. \label{it:changeVars}
\end{enumerate}
This procedure is of course idealized and in practice, one faces problems that arise from noisy data and imperfect symmetry between the approaching and receding emitters for the same impact parameter, $z_\pm(b)$, which introduces even more noise in step \ref{it:calcZ}. Other problems that arise due to insufficient modeling are discussed in \sref{redshift_corrections}.

In the given scenario, the observer is looking at redshifts of an edge-on galaxy. It is therefore not obvious which radius belongs to a redshift value that originated from some emitter along the line of sight. Hence, a way of relating a characteristic redshift to the emitter's radius is needed.

I will now establish the relationship between the coordinate radius and the impact parameter, $r(b)$, which is needed to perform step \ref{it:changeVars}. This will also identify which redshift value $z_\pm(b)$ should be picked. To find the corresponding radius for the redshift values of a given impact parameter $b$, I square \eref{Doppler_redshift} to get
\begin{equation}\label{eq:sq_redshift}
  \frac{Z^2}{b^2} = \frac{\Phi'}{r(1-r\Phi')} \, .
\end{equation}
Now assume this quantity to be monotonically decreasing with
increasing $r$, which is synonymous with
\begin{equation}\label{eq:monotone_requirement}
  \Phi' > r\Phi'' + 2r(\Phi')^2 \, .
\end{equation}
The maximum value of $Z^2$ for a fixed impact parameter $b$ is then related to the minimum value of
the radius of the observed null geodesic, $r=r_N=r_{N,\mathrm{min}}
\;\Leftrightarrow\; \d r_N = 0$, see \fref{impact_parameter+radius}.
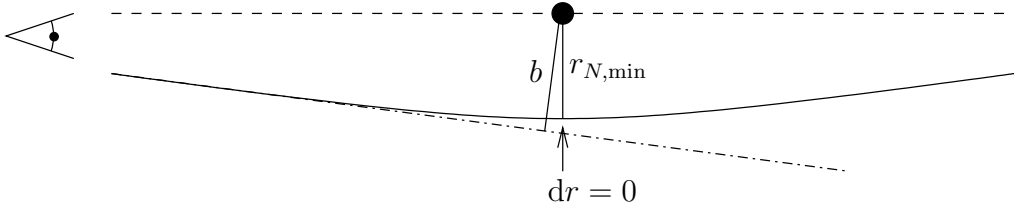
\begin{figure}[htb]
\begin{center}
\input{figures/322_impact_parameter+radius.pstex_t}
\end{center}
\caption[Impact parameter related to the radius of closest approach.]{\label{fig:impact_parameter+radius}
This sketch shows the trajectory (solid line) of all photons that are observed with an impact parameter $b$. The dash-dotted line marks the line of sight corresponding to $b$. The observed photons could have been emitted from any point on the solid line. However, from \eref{sq_redshift}, one can conclude that photons emitted at the point of closest approach, $r_{N,\mathrm{min}}$, to the coordinate origin exhibit the strongest Doppler shift. See also \fref{terminal_velocity_method}.}
\end{figure}

\noindent It then follows from the lightlike line element
\begin{equation}
  \d s^2 = g_{tt}\,\d t_N^2 + g_{\cphi\cphi}\,\d\cphi_N^2 = 0
\end{equation}
that
\begin{equation}\label{eq:lightlike_LE}
  \left. \frac{\d \cphi}{\d t} \right|_{N,\mathrm{min}} = \left. \sqrt{\frac{-g_{tt}}{g_{\cphi\cphi}}} \,\right|_{N,\mathrm{min}} = \pm
  \frac{e^{\Phi(r_{N,\mathrm{min}})}}{r_{N,\mathrm{min}}} \, .
\end{equation}
Using the definition of the impact parameter \eref{impact_param},
one obtains the mapping $r(b)$ from \eref{lightlike_LE} for the
moment of emission at the radius $r=r_{N,\mathrm{min}}$:
\begin{eqnarray}
  \label{eq:r_of_b}
  r &=& b\,e^{\Phi(r)} = b\,e^{\Phi(b)} \\
  \label{eq:dr_of_db}
  \d r &=& e^{\Phi(b)} \left(1 + b \frac{\d\Phi}{\d b} \right) \d b \,.
\end{eqnarray}
Since \eref{sq_redshift} is assumed to be monotonically
decreasing, \eref{r_of_b} and \eref{dr_of_db} are only valid for
the maximum redshift value of $Z$ for a given impact parameter
$b$.

To perform step 3, I want to find a relationship only involving
$Z(b)$, $b$ and $\Phi(b)$. Hence, eliminate $r$ in
\eref{sq_redshift}, using \eref{r_of_b} and \eref{dr_of_db},
so that after careful substitution the result
\begin{equation}\label{eq:Z_of_b}
  Z^2 = b \frac{\d\Phi}{\d b} e^{-2\Phi(b)} = b \frac{\d}{\d b}\left( -\frac{e^{-2\Phi(b)}}{2}
  \right) 
\end{equation}
is obtained, which corresponds nicely to the analogous relation in the
Newtonian limit\footnote{The index $N$ refers here to the Newtonian limit, not the null curve.},
\begin{equation}\label{eq:newton_d_redshift}
  Z^2 = r_N \frac{\d\Phi_N}{\d r_N}
\end{equation}
where $r_N \approx b$ is the observed radius and $e^{-2\Phi(b)}
\approx 1$.

In principle, the potential $\Phi(b)$ can now be obtained from the
measured quantities $Z$ and $b$, by integrating using an appropriate boundary
condition for $\Phi(b)$.
Once $\Phi(b)$ is determined, $\Phi(r)$ is given by the mapping
\eref{r_of_b} between $b \leftrightarrow r$. From here on, I
assume that $\Phi(r)$, and especially $\Phi'(r) = \d\Phi/\d r$,
has been extracted from the data and can be considered as
``known''.

Since the relation between $r$ and $b$ is now available, it is possible to compare the gravitational and the Doppler redshift directly. First I define the circular velocity $v_c\equiv \left[ r\, \d\cphi/\d\tau \right]_e$ and then apply the mapping \eref{r_of_b} to \eref{Doppler_redshift_coord}:
\begin{eqnarray}
\label{eq:grav_redshift_neglect}
  1+z_g &\approx& e^{-\Phi(r)} \left\lbrace 1 + \frac{1}{2}\,v_c^2 + \O{v_c^4} \right\rbrace\\
\label{eq:Doppler_redshift_dominant}
  Z &=& e^{-\Phi(r)}\, v_c \, .
\end{eqnarray}
For $e^{\Phi(r)}\approx 1$ it is now clear that $z_g \ll Z$ since the rotation velocities have typical values around $v_c\approx 250\,\unit{km/s} \ll c=1$ (in geometrical units) and thus, the gravitational redshift can be neglected.

\subsection{Observational techniques}

Astrophysical observations are usually carried out without considering general relativity and as I have shown with \eref{Doppler_redshift_dominant}, this approximation is very good to first order in the rotation velocity $v_c$. The derivation outlined in the previous section, however, considers only edge-on galaxies, which posit only a fraction of the observed spiral galaxies. Due to random orientation in space, most spiral galaxies are somewhat inclined to the line of sight. I will now outline actual observational methods, starting with nearly edge-on observations, followed by methods of observing the rotation curve for the more common case of inclined spiral galaxies.

All direct observations of rotation curves use the Doppler shift of a well known emission or absorption line to determine the emitting particle's velocity component along the line of sight. Typical lines that are utilized are those of \textsc{H$\alpha$}, \textsc{CO}, especially neutral hydrogen \textsc{Hi}, and other optical emission lines \cite{Sofue:2001}. Since there are many emitting particles along any line of sight, and different emitters generally have slightly different velocities, a whole redshift- / velocity-profile is obtained for each observed position. The different methods of data reduction attempt to estimate an appropriate average rotation velocity.

\begin{figure}[htb]
\begin{center}
\input{figures/323_terminal_velocity_method.pstex_t}
\end{center}
\caption[Illustrating the Terminal-Velocity method for determining the redshift.]{\label{fig:terminal_velocity_method}
A wavelength- ($\lambda$) (or analoguously velocity- or also redshift-) profile showing the signal intensity $I$ for a specific impact parameter $b$. The redshift $Z_\mathrm{terminal}$ is determined to correspond to the radius of closest approach for $b$. The other symbols are explained in the text.}
\end{figure}
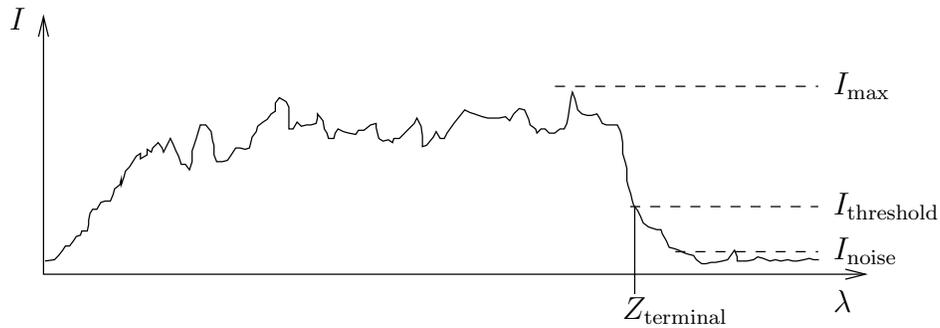

For nearly edge-on galaxies, the Envelope-Tracing method, also called the Terminal-Velocity method, is used. As illustrated in \fref{terminal_velocity_method}, it establishes the maximum or terminal velocity along the major axis of the observed galaxy. To do that, the maximum intensity $I_\mathrm{max}$ and the noise level $I_\mathrm{noise}$ are determined from the velocity profile, and from that a threshold intensity $I_\mathrm{threshold}$ is calculated which corresponds to about 20\% of the maximum intensity. The velocity that occurs with the threshold intensity near the edge of the velocity-profile is defined to be the terminal velocity $Z_\mathrm{terminal}$ and then corrected for systemic velocity, inclination, velocity dispersion of the observed gas and instrumental resolution to obtain the rotation curve \cite{Sofue:2001}. This is essentially the same as picking the maximum velocity as required by the derivation in \sref{measuring_rotcurves} plus observationally necessary corrections. See also \fref{impact_parameter+radius}.

In galaxies with a reasonable inclination, the line of sight does not run through the plane of the disk, but pierces it from top to bottom. The smeared out distribution in the velocity-profile for any observed position then originates from the thickness of the disk and the random and anomalous velocities of the emitters within, see \fref{inclined_galaxy}. A representative rotation velocity is generally obtained by an intensity-weighted velocity average of the entire profile. In the outer parts of the galactic disk, the line profile can be assumed to be rather symmetric around the peak-intensity value and therefore, the intensity-weighted velocity can be approximated by the velocity with the highest intensity \cite{Sofue:2001}.

\begin{figure}[htb]
\begin{center}
\input{figures/323_inclined_galaxy.pstex_t}
\end{center}
\caption[Finite thickness effects of the galactic disk while measuring the redshift.]{\label{fig:inclined_galaxy}
A galaxy (grey ellipse) which is inclined to the line of sight (solid) is shown from the side to illustrate the finite thickness effects of the disk. The wavelength/velocity profile (below) arises from the Doppler shifts of the different velocities of the emitters along that line of sight. The emitters which are closer to the center of the galaxy (right dotted line) have a higher circular velocity. Therefore, in case they are receding from the observer, they appear redder.}
\end{figure}
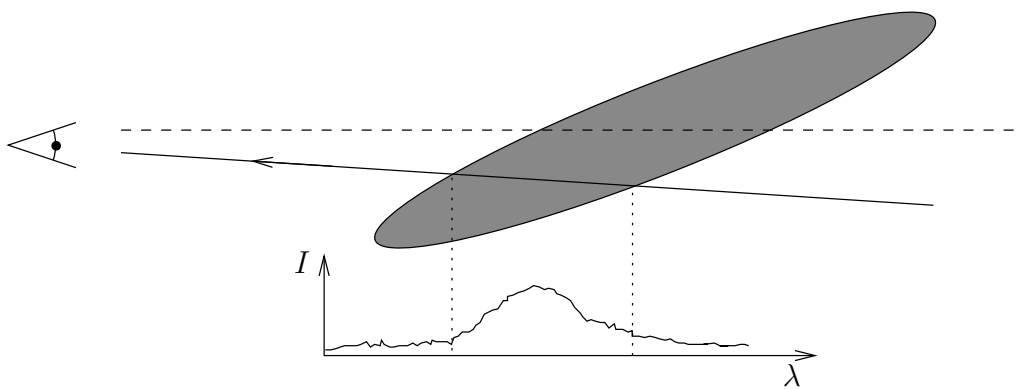

To increase the reliability of the obtained rotation velocities, the initial rotational curve, RC0, can be used to reconstruct the measurable velocity profiles. The difference between these profiles and the actually observed ones yields a corrected rotation curve RC1. This procedure can be repeated until the difference between the observed and corrected velocity-profiles becomes minimal and stable \cite{Sofue:2001}.

If the rotation velocity is obtained for the whole surface of an inclined galaxy, one can apply a tilted ring fitting procedure to obtain a rotation curve that relates only radius and circular velocity \cite{Fraternali:2002}. This averages all obtained rotation velocities of a certain radius and maximises the amount of data that was used to obtain the rotation curve.

Since the actual observations use more data than that which can be obtained using the method in \sref{measuring_rotcurves}, I will now go on to extend the general relativistic model to account for inclined galaxies and systemic velocity.

\subsection{Observationally necessary improvements of the model}
\label{sec:redshift_corrections}

In \sref{obs_light}, I presented the redshift of light emitted by a
source in the geometry \eref{ss_metric}, as it is detected by an observer resting in the galactic plane ($\theta=\pi/2$) at infinity.

Real life galaxies usually show a radial systemic velocity away
from the observer on Earth (or at least the solar system). This is
generally due to the Hubble expansion of the universe and peculiar motion of the observed galaxy, both of which lead to an additional shift of the observed emission/absorption line.

Most observed galaxies are also inclined to the line of sight and
therefore the observer is not within the galactic plane. This affects
the Doppler redshift of the observed light.

In this section, I am going to introduce these further conditions
to the total redshift \eref{total_redshift}. Finally I will argue
that it is generally not necessary to consider general
relativistic effects when obtaining the Doppler redshift from
galaxy observations.

\subsubsection{Inclination correction}

To see how the inclination angle affects the Doppler redshift, I
separate the observer and galaxy coordinate system. The galaxy
coordinate system $(\tilde{x},\tilde{y},\tilde{z})$ is still
aligned with the galactic disc and the coordinates of any particle
therein shall be given by
\begin{eqnarray}
  \label{eq:galaxy_coordinates}
  \tilde{x} &=& \tilde{r}\cos\tilde{\cphi} \\
  \nonumber
  \tilde{y} &=& \tilde{r}\sin\tilde{\cphi} \\
  \nonumber
  \tilde{z} &=& 0 \, .
\end{eqnarray}
In his or her coordinate system, the observer is thought to sit on
the $x$-axis at infinity, i.e.
$(r=\infty,\theta=\pi/2,\cphi=0)$. Thus, the observer
coordinate system $(x,y,z)$ has the same origin as the galaxy
coordinate system, but the relative orientation between the two depends on the galaxy's inclination angle $\psi$ and the position of the observed particle.

To find the appropriate transformation rules between both coordinate systems, two rotations have to be performed: The first rotation about the $y$=$\tilde{y}$-axis accounts for the observed inclination angle $\psi$, see \fref{coord_rot_1}. The second rotation is necessary since for convenience, I want the connecting light ray between the emission event and the observation to lie in the observer's $\theta=\pi/2$ plane. This will be a rotation about the observer $x$-axis (\fref{coord_rot_2}) that makes sure that the observed particle's $z$-coordinate vanishes, i.e. $\theta=\pi/2$.

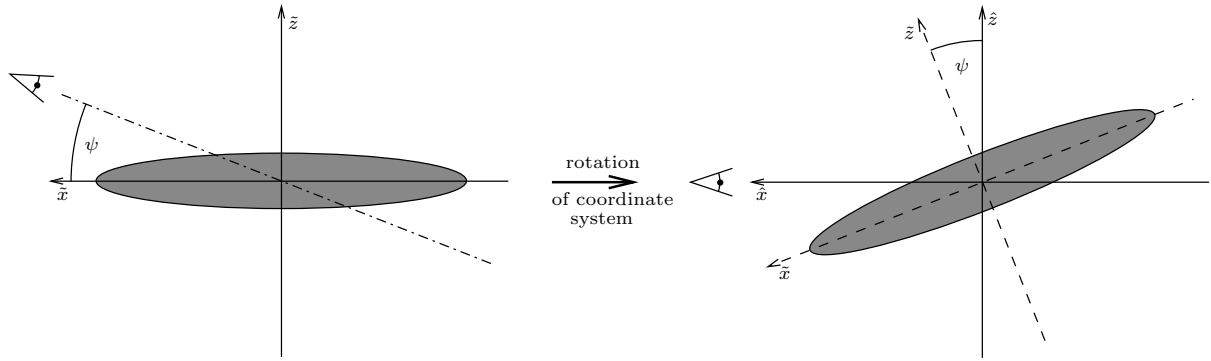
\begin{figure}[htb]
\begin{center}
\input{figures/324_coord_rot_1_shrunk.pstex_t}
\end{center}
\caption[Illustrating the coordinate rotation which represents the inclination of a galaxy.]{\label{fig:coord_rot_1}
The left image shows the galaxy coordinate system $(\tilde{x},\tilde{y},\tilde{z})$ as viewed from the $\tilde{y}$-axis. The galaxy (grey ellipse) is inclined to the line of sight (dash-dotted line) with an angle $\psi$. The right image shows the galaxy with its coordinate system (dashed) in the new intermediate coordinate system $(\hat{x},\hat{y},\hat{z})$ as viewed from the $\hat{y}$-axis. The line of sight now coincides with the $\hat{x}$-axis.}
\end{figure}

The inclination by an angle $\psi$ between the $x$- and $\tilde{x}$-axis is represented by a simple rotation about the coinciding $y$=$\tilde{y}$-axis (\fref{coord_rot_1}):
\begin{equation}\label{eq:coord_trafo}
  \left(%
\begin{array}{c}
  \hat{x} \\
  \hat{y} \\
  \hat{z} \\
\end{array}%
\right)
 =
\left(%
\begin{array}{ccc}
  \cos\psi & 0 & \sin\psi \\
  0 & 1 & 0 \\
  -\sin\psi & 0 & \cos\psi
\end{array}%
\right)
\left(%
\begin{array}{c}
  \tilde{x} \\
  \tilde{y} \\
  \tilde{z} \\
\end{array}%
\right)
 = \tilde{r}
\left(%
\begin{array}{c}
  \cos\psi \,\cos\tilde{\cphi} \\
  \sin\tilde{\cphi} \\
  -\sin\psi \,\cos\tilde{\cphi}
\end{array}%
\right) \, .
\end{equation}
The observer $x$-axis is the line of sight to the center of the observed galaxy. Hence, a rotation about this axis does not change any results obtained by that observation and can be used to simplify the geometric constraints. Such a rotation (\fref{coord_rot_2}) by an arbitrary angle $\zeta$ is given by
\begin{equation}\label{eq:coord_trafo_2}
  \left(%
\begin{array}{c}
  x \\
  y \\
  z \\
\end{array}%
\right)
 =
\left(%
\begin{array}{ccc}
  1 & 0 & 0 \\
  0 & \cos\zeta & -\sin\zeta \\
  0 & \sin\zeta & \cos\zeta 
\end{array}%
\right)
\left(%
\begin{array}{c}
  \hat{x} \\
  \hat{y} \\
  \hat{z} \\
\end{array}%
\right)
 = \tilde{r}
\left(%
\begin{array}{c}
  \cos\tilde{\cphi} \,\cos\psi\\
  \sin\tilde{\cphi} \,\cos\zeta + \cos\tilde{\cphi} \,\sin\zeta \,\sin\psi\\
  \sin\tilde{\cphi} \,\sin\zeta - \cos\tilde{\cphi} \,\cos\zeta \,\sin\psi
\end{array}%
\right) .
\end{equation}
Both angles, $\psi$ and $\zeta$, are Euler angles due the rotation about the $\hat{y}$- and $x$-axis. The third Euler angle is zero, as a rotation about the $\tilde{z}$-axis is not necessary. Since the $z$-component has to vanish, the rotation angle $\zeta$ is given by
\begin{eqnarray}
\nonumber
 0 &=& \sin\tilde{\cphi} \,\sin\zeta - \cos\tilde{\cphi} \,\cos\zeta \,\sin\psi \\
\label{eq:rotang_chi}
\Rightarrow \quad \tan\zeta &=& \,\frac{\sin\psi}{\tan\tilde{\cphi}} \, .
\end{eqnarray} 

\begin{figure}[htb]
\begin{center}
\input{figures/324_coord_rot_2_shrunk.pstex_t}
\end{center}
\caption[Another coordinate rotation that simplifies the final set of coordinates.]{\label{fig:coord_rot_2}
The left image shows the inclined galaxy in the intermediate coordinate system $(\hat{x},\hat{y},\hat{z})$ as seen by the observer. The small solid circle marks the position for which the redshift is to be determined. The right image shows the galaxy with the intermediate coordinate system (dashed) in the final observer coordinate system $(x,y,z)$ as seen from the $x$-axis. The galaxy has been rotated by an angle $\zeta$ so that the observed emitter (circle) lies on the $y$-axis.}
\end{figure}
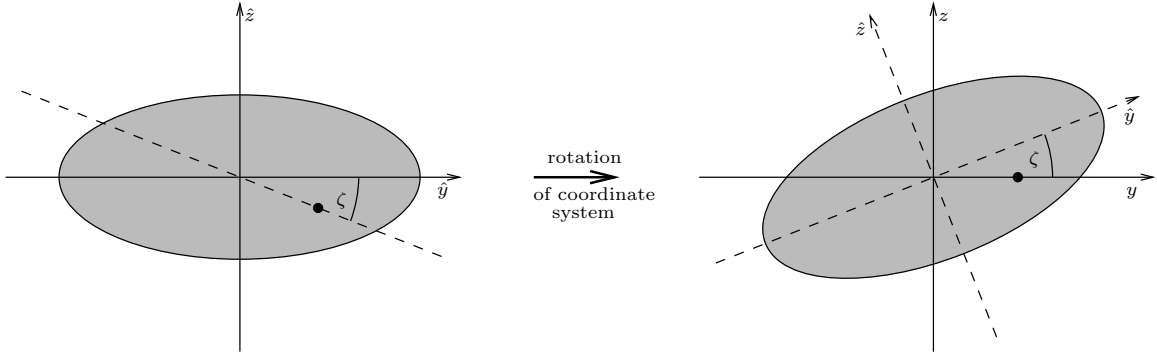

\noindent So $\zeta$ is just the angle between the apparent major axis of the observed galaxy and the connecting line between the galactic center and the observed particle. Finally, a particle in the galactic plane with coordinates \eref{galaxy_coordinates} has the following spherical polar coordinates in the observer system:
\begin{eqnarray}
  \label{eq:sph_obs_coords}
  r &=& \tilde{r} \\
  \nonumber
  \theta &=& \frac{\pi}{2} \\
  \nonumber
  \cphi &=& \arctan\left(\frac{y}{x}\right) = \arctan\left( \frac{\tan\psi}{\sin\zeta} \right) \, ,
\end{eqnarray}
where $\zeta$ is given by $\tilde{\cphi}$ and $\psi$ through \eref{rotang_chi}.

\subsubsection{Corrected redshift relation}
\label{sec:corr_redshift}

Now I apply the additional conditions of inclination $\psi$ and
systemic radial velocity $v_s$ to the redshift relation
\eref{redshift1}. In the observer coordinate system, the
inclination leads to
\begin{equation}\label{eq:obs_coords_phi_dot}
\left[\frac{\d x^\cphi}{\d \tau}\right]_e = \left[\frac{\partial
\cphi}{\partial \tilde{\cphi}}\frac{\d \tilde{\cphi}}{\d
\tau}\right]_e = \frac{\ell_e}{r_e^2} \left[\frac{\partial
\cphi}{\partial \tilde{\cphi}} \right]_e 
\end{equation}
and the radial systemic velocity of the observer, $v_s$, yields
\begin{eqnarray}\label{eq:obs_coords_sys_vel}
  \left[\frac{\d x^t}{\d \tau}\right]_o &=& \gamma_s \\
  \nonumber
  \left[\frac{\d x^r}{\d \tau}\right]_o &=& \gamma_s v_s
\end{eqnarray}
where $\gamma_s \equiv (1-v_s^2)^{-1/2}$. Although the emitter's 4-velocity has a non-zero $\theta$-component, we do not need to know it, since the photon's worldline lies entirely within the $\theta=\pi/2$ plane, which completely suppresses the occurance of $\d\theta/\d\tau|_e$ in the redshift formula.

The term arising from coordinate transformation, $\partial\cphi/\partial\tilde{\cphi}|_e$, can be evaluated, and a tedious, but straightforward calculation yields
\begin{equation}
\left[\frac{\partial \cphi}{\partial \tilde{\cphi}} \right]_e = \cos\psi \,\cos\zeta \, .
\end{equation}
This is a standard result given by the relative orientation between the observer and the observed galaxy \cite{Warner:1973} and thus, is independent of relativistic effects. It is now possible to specify the total redshift of any emitter in an inclined galaxy with systemic velocity $v_s$:
\begin{equation}\label{eq:corrected_redshift1}
  1+z = \frac{ \left[ g_{tt} \sqrt{1+\frac{\ell_e^2}{r_e^2}}\, e^{-\Phi(r_e)} \frac{\d t}{\d \chi} +
  g_{\cphi\cphi}\, \frac{\ell_e}{r_e^2} \frac{\d \cphi}{\d \chi} \cos\psi\,\cos\zeta \right]_e}{
  \left[ g_{tt}\, \gamma_s \frac{\d t}{\d \chi} + g_{rr}\, \gamma_s v_s \frac{\d r}{\d \chi} \right]_o} \, .
\end{equation}
Inserting the affine parameter $\d \chi=g_{tt}\, \d t$ gives
\begin{equation}\label{eq:corrected_redshift2}
  1+z = \frac{ \left[ \sqrt{1+\frac{\ell_e^2}{r_e^2}}\, e^{-\Phi(r_e)} +
  \frac{g_{\cphi\cphi}}{g_{tt}}\, \frac{\ell_e}{r_e^2} \frac{\d \cphi}{\d t} \cos\psi\,\cos\zeta  \right]_e}{
  \left[ \gamma_s + \frac{g_{rr}}{g_{tt}}\, \gamma_s v_s \frac{\d r}{\d t} \right]_o} \, .
\end{equation}
Since the observer is far away from the origin, flat space may be
assumed, so that $\left. \frac{g_{rr}}{g_{tt}} \right|_o \rightarrow -1$ and the
coordinate speed of light $\left. \frac{\d r}{\d t} \right|_o \rightarrow 1$. The
denominator then simplifies to
\begin{equation}\label{eq:corr_redshift2_denom}
  \gamma_s\, (1 - v_s) = \sqrt{\frac{1-v_s}{1+v_s}} \, .
\end{equation}
Finally, analogous to \eref{total_redshift}, the exact result is
\begin{equation}\label{eq:corr_total_redshift}
  (1+z_\pm) \sqrt{\frac{1-v_s}{1+v_s}} = \frac{1}{\sqrt{1-r\Phi'(r)}} \left( \frac{1}{e^{\Phi(r)}}
  - \frac{\pm |b| \sqrt{r\Phi'(r)}}{r} \,\cos\psi\,\cos\zeta \right) .
\end{equation}
Please note that $b$ is only determined by \eref{r_of_b}, when particles on the apparent major axis are observed. In this case, at the moment of emission, $\d r_N=0$ for the light ray, which was the requirement for \eref{r_of_b}. Particles that don't lie on the apparent major axis need a different mapping between $b$ and $r$.

For an exact mapping, the null trajectory of the light ray would have to be known, which in turn requires the metric functions $\Phi(r)$ and $m(r)$ to be known. As a first approximation, however, it is satisfactory to assume that the light ray's trajectory in the galactic plane is only bent marginally by the weak gravitational field. Using this ``near flat space'' approximation, I assume
\begin{equation}
b = e^{-\Phi(r_{N,\mathrm{min}})}\,r_{N,\mathrm{min}} \approx e^{-\Phi(r\sin\cphi)}\,r \sin\cphi \approx r \sin\cphi
\end{equation} 
as one would naturally from the classical definition of the impact parameter (\fref{impact_parameter}).
It follows then that the observed Doppler redshift is
\begin{equation}
Z \,\cos\psi\,\cos\zeta \approx \frac{\sqrt{r\,\Phi'(r)}}{\sqrt{1-r\,\Phi'(r)}}\,\sin\cphi\,\cos\psi\,\cos\zeta \, .
\end{equation} 
One of these three angles, however, is redundant and in general, only two angles are necessary to express the angular dependence. The relations \eref{rotang_chi} and \eref{sph_obs_coords} can be used to express the angular dependence in any of the equivalent forms
\begin{equation}
\sin\cphi\,\cos\psi\,\cos\zeta
= \frac{\pm \sin\psi\,\cos\psi}{\sqrt{\sin^2\psi + \tan^2\zeta}}
= \pm \sqrt{\cos^2\psi - \cos^2\cphi}
= \cos\psi\, \sin\tilde{\cphi} \, .
\end{equation} 
The result is then most conveniently written in terms of the inclination angle $\psi$ and the angular position within the galactic disk $\tilde{\cphi}$,
\begin{equation}
Z \,\cos\psi\,\cos\zeta \approx \frac{\sqrt{r\,\Phi'(r)}}{\sqrt{1-r\,\Phi'(r)}}\,\cos\psi\,\sin\tilde{\cphi} \, .
\end{equation} 
\label{sec:rPhiprime_small} With $Z \ll 1$ being a valid approximation, one can conclude that $r\,\Phi'(r) \ll 1$ and thus
\begin{equation}
\label{eq:Newtonian_redshift}
Z \,\cos\psi\,\cos\zeta \approx \sqrt{r\,\Phi'(r)} \,\cos\psi\,\sin\tilde{\cphi} \equiv v_N(r)\,\cos\psi\,\sin\tilde{\cphi} \, ,
\end{equation} 
where $v_N(r)\equiv \sqrt{r\,\Phi'(r)}$ was defined analogously to the rotation velocity in Newtonian gravity. When applying the approximations $z_\pm \ll 1$ and $v_s \ll 1$ for
a Taylor expansion of the left hand side of \eref{corr_total_redshift}, it can be simplified using definitions \eref{grav_redshift}, \eref{Doppler_redshift} and \eref{Newtonian_redshift}:
\begin{equation}\label{eq:obs_redshift}
  1+z_\pm \;\approx\; 1+z_g \pm v_N(r)\cos\psi\,\sin\tilde{\cphi} + v_s.
\end{equation}
Please note that the distinction between the approaching and receding half of the disk is now absorbed into the position angle $\tilde{\cphi}$ and the $\pm$ in front of the Doppler term is used to distinguish the direction of the disk rotation field, relative to the galaxy coordinate system: $+$ for counterclockwise and $-$ for clockwise rotation of the stars in the galaxy. Apart from the term $z_g$, this equation is identical to the equivalent in
Newton dynamics \cite{Warner:1973}.

The assumption $v_s \ll 1$, however, may not be applicable since there are now observations of rotation curves for high redshift galaxies \cite{Sofue:2001}, up to $z\approx 1$. In these cases, the correct general relativistic calculation must be carried out in an expanding FLRW background metric. This is a task that goes well beyond the scope of this thesis.

When analyzing observational data, $Z$, $v_s$ and $\psi$ are
relatively easy to determine, but direct observations of $z_g$ are
theoretically only possible along the apparent minor axis of an
inclined galactic disc, where $\sin\tilde{\cphi}=0$, i.e. no Doppler shift occurs.
Practically, the gravitational redshift $z_g$ is negligible
compared to the Doppler redshift $Z$, as was shown in \eref{grav_redshift_neglect}, and furthermore it is buried in noise, which generally arises from a finite thickness of the
galactic disc. Typical observed Doppler redshifts are $Z \approx
250 \fracunit{km}{s} / c \ll 1$ and therefore by \eref{grav_redshift_neglect}, we can approximate $z_g \ll Z$, leading to
\begin{equation}\label{eq:obs_redshift2}
  z_\pm \approx \pm\, v_N(r)\cos\psi\,\sin\tilde{\cphi} + v_s.
\end{equation}

This is the same redshift relation as in Newtonian dynamics for
emitters in a galaxy with inclination
angle $\psi$ \cite{Warner:1973}. Generalizing this result, we see that it is safe to
assume that no general relativistic effects need be considered
when determining the Doppler-shift. This is especially easy to understand since I have shown that the gravitational redshift $z_g$ is always negligible compared to the Doppler redshift $Z$, and since a systemic redshift occurring from the expansion of the universe would affect all observed photons in the same way.

Thus a rotation velocity curve
$v(r)$, that was obtained using common procedures in Newtonian
dynamics, is a good approximation for the general relativistic Doppler
redshift $Z(r)$ and the corresponding gravitational potential $\Phi(r)$.

\subsection{Spacetime metric of flat rotation curve region}

As I argued in \sref{rotcurve_shape}, the direct evidence given by observations of individual rotation curves suggests that the rotation velocity in the observed region is approximately constant and hence, implies that the dark matter density falls off as $\rho \propto r^{-2}$:
\begin{equation}
v_c \approx \mathrm{const.} \quad \Rightarrow \quad \rho \propto r^{-2} \, .
\end{equation} 
Furthermore, observations have shown that the rotation velocity is small compared to the speed of light,
\begin{equation}
v_c \approx 250\,\fracunit{km}{s}/c \approx 8\times10^{-4} \ll 1 \, ,
\end{equation}
in geometrical units. From the definition
\begin{equation}
v_c \equiv r\, \frac{\d\cphi}{\d\tau} = \frac{\ell}{r} \, ,
\end{equation} 
it follows that the energy of such a circular orbit is given by \eref{com_drdtau} with $\d r=0$:
\begin{equation}
\gamma^2 = e^{2\Phi} \left( 1 - \frac{\ell^2}{r^2} \right) \approx e^{2\Phi} \, .
\end{equation} 
In this approximation, the geodesic equation for circular motion, \eref{gde_r_circ_orbit}, yields
\begin{equation}
\frac{\d\Phi}{\d r} = \frac{\ell^2\,e^{2\Phi}}{r^3\,\gamma^2} \approx \frac{v_c^2}{r} \, ,
\end{equation} 
which is identical to the corresponding formula arising in the Newtonian case. This can be easily integrated to
\begin{equation}
\Phi(r)-\Phi_0 \approx v_c^2 \int_{r_0}^r \frac{\d\rr}{\rr} = \ln\,\left[\left(\frac{r}{r_0}\right)^{v_c^2}\right] \, ,
\end{equation} 
where $\Phi_0$ and $r_0$ are constants of integration that have to be chosen by appropriate boundary conditions. This gives a good approximation to the metric component $g_{tt}$ for the region of flat rotation curves:
\begin{equation}
g_{tt}(r) = -e^{2\Phi(r)} \approx -e^{2\Phi_0}\left(\frac{r}{r_0}\right)^{2v_c^2} \approx -e^{2\Phi_0}\left(\frac{r}{r_0}\right)^{10^{-6}} \, .
\end{equation}
The very small exponent $10^{-6}$ has the effect that the metric component is almost constant for a very large range of values for $r/r_0$. To see this, let $-g_{tt}\,e^{-2\Phi_0}$ be bounded by
\begin{eqnarray*}
1-\epsilon < & \left(\frac{r}{r_0}\right)^{10^{-6}} & < 1+\epsilon \\
\left(1-\epsilon\right)^{10^6} < & \frac{r}{r_0} & < \left(1+\epsilon\right)^{10^6} \, .
\end{eqnarray*}
Then for e.g. $\epsilon = 10^{-4}$,
\begin{equation}
10^{-43} \;\lesssim\; \frac{r}{r_0} \;\lesssim\; 10^{43} \quad \Rightarrow \quad e^{2\Phi(r)} \approx e^{2\Phi_0} \, .
\end{equation} 
Since it is unlikely that the region of flat rotation curve extends to infinity, and since there have been indications of a decreasing rotation curve for large radii (see \sref{rotcurve_shape}), the integration constants $\Phi_0$ and $r_0$ can be determined from matching an appropriate metric at the ``far-away radius'' $r_0$.

\subsubsection{Matching onto the Schwarzschild exterior metric}

The Schwarzschild exterior metric,
\begin{equation}
\label{eq:Schwarzschild_exterior_metric}
\d s^2 = -\left(1-\frac{2M}{r}\right)\d t^2 + \left(1-\frac{2M}{r}\right)^{-1}\d r^2 + r^2\,\d\Omega^2 \, ,
\end{equation}
represents the asymptotically flat geometry that is induced by a point mass $M$ in an otherwise empty space. Furthermore, by Birkhoff's theorem, all asymptotically flat spherically symmetric vacuum spacetimes must be of this form. 
Thus, the Schwarzschild exterior metric is suitable as a far-away metric for the galaxy metric, and (assuming cosmological effects are negligible) can be matched with the smooth matching conditions 
\begin{eqnarray}
g_{tt}(r_0) &=& g_{tt,\mathrm{Schwarzschild}}(r_0) \\
\frac{\d}{\d r}\,g_{tt}(r_0) &=& \frac{\d}{\d r}\,g_{tt,\mathrm{Schwarzschild}}(r_0) \, ,
\end{eqnarray} 
which yield
\begin{eqnarray}
e^{2\Phi_0} &=& 1-\frac{2M}{r_0} \\
e^{2\Phi_0}\, \frac{2v_c^2}{r_0} &=& \frac{2M}{r_0^2} \, ,
\end{eqnarray} 
so that eliminating $M$ gives
\begin{equation}
e^{2\Phi_0} = \left( 1+2v_c^2 \right)^{-1} \approx e^{-2v_c^2} \approx 1 \, .
\end{equation} 
The result is the $g_{tt}$ component of the galaxy metric for the flat rotation curve region, truncated at $r_0$, merging onto an asymptotically flat exterior Schwarzschild metric:
\begin{equation}
\label{eq:rotcurve_flat_gtt}
g_{tt}(r) \approx -e^{-2v_c^2}\left(\frac{r}{r_0}\right)^{2v_c^2} \, .
\end{equation}
Similar derivations can be found in \cite{Bharadwaj:2003,Nucamendi:2000}.

Depending on the cosmological context, one might wish to consider different metrics as the exterior to the region of flat rotation curves, e.g. the de Sitter metric to represent empty space with a cosmological constant. Please note that the former derivation of observed redshift $z_\pm$ assumed asymptotically flat space and hence, this derivation would not hold in that form when using de Sitter space, which is not asymptotically flat. There might be other metrics of a simple form, that are suitable to model flat rotation curves and/or the asymptotic far-away region.

The Tolman IV solution (see \sref{tolmanIV}), for example, exhibits a density profile that goes as $\rho \approx k r^{-2} + \mathrm{const}$ for very large $r$ while the pressure goes as $p \approx k r^{-2} - \mathrm{const}$. Thus, it seems that it represents flat rotation curves and asymptotic de Sitter behaviour and hence, might be a suitable metric for the outer region of a galaxy in a cosmology with cosmological constant. Unfortunately, the two free parameters of the Tolman IV solution are fixed by the given rotation velocity and density of the cosmological constant. The resulting distance scale $\sim 10^9\,\unit{kpc}= 10^3\,\unit{Gpc}$ is far too large to be useful for galaxies.

\subsection{Interpretation of rotation curve measurements in general relativity} 
\label{sec:rotcurve_interpretation}

While the flat rotation curve implies a metric of approximate form \eref{rotcurve_flat_gtt}, a rotation curve of arbitrary shape determines the gravitational potential $\Phi(r)$ or equivalently, the metric component $g_{tt}(r)$, as was shown in \sref{obtain_Phi}. However, the rotation curve does not by itself yield any information about the metric component $g_{rr}(r)$, i.e. the metric function $m(r)$.

In Newtonian gravity, the gravitational potential $\Phi(r)$ alone is sufficient to determine the gravitational field completely. Furthermore, it is linked to the matter density through
\begin{equation}
\label{eq:Newton_grav_source}
\nabla^2 \Phi(r) = 4\pi\,\rho(r) \, ,
\end{equation}
which then entirely determines the matter distribution.

In general relativity, the metric \eref{ss_metric}, which represents arbitrary, spherically symmetric, non-vacuum spacetimes and was chosen as a model for the galactic halo, is determined by the metric components $g_{tt}(r)$ and $g_{rr}(r)$, or equivalently $\Phi(r)$ and $m(r)$. Hence, knowledge of $\Phi(r)$ alone does not constrain the spacetime completely and thus, the matter distribution cannot be determined uniquely.

This can be remedied by either refining the halo model, i.e. introducing further assumptions, or obtaining more data through a different observational procedure which does include information about $g_{rr}(r)$.

Most commonly, a further assumption is made. The CDM paradigm suggests that the equation of state for dark matter is $p(\rho) \approx 0$ and therefore, the source equation of weak field gravity \eref{pressure_field_eq},
\begin{equation}
\nabla^2 \Phi = 4\pi\,(\rho + 3p)\, ,
\end{equation}
reduces to its Newtonian counterpart \eref{Newton_grav_source}. Then, the matter distribution is again given by the gravitational potential $\Phi(r)$ only. If, however, the equation of state was different, the derived matter distribution would of course be invalid.

The scientifically more conservative approach is to see if there are independent means of measuring the equation of state. For the problem at hand, this can be achieved by observing galactic events that do not exclusively depend on $g_{tt}(r)$, but also on $g_{rr}(r)$. As I will outline in the next section, gravitational lensing is a suitable physical event to measure the dark matter equation of state in combination with rotation curve measurements.


%% file: figures/322_impact_parameter.pstex_t
\begin{picture}(0,0)%
\includegraphics{figures/322_impact_parameter.pstex}%
\end{picture}%
\setlength{\unitlength}{4144sp}%
\begingroup\makeatletter\ifx\SetFigFont\undefined%
\gdef\SetFigFont#1#2#3#4#5{%
  \reset@font\fontsize{#1}{#2pt}%
  \fontfamily{#3}\fontseries{#4}\fontshape{#5}%
  \selectfont}%
\fi\endgroup%
\begin{picture}(5424,1089)(439,-629)
\put(2971,-106){\makebox(0,0)[lb]{\smash{{\SetFigFont{12}{14.4}{\familydefault}{\mddefault}{\updefault}{\color[rgb]{0,0,0}$\cphi$}%
}}}}
\put(3061,-556){\makebox(0,0)[lb]{\smash{{\SetFigFont{12}{14.4}{\familydefault}{\mddefault}{\updefault}{\color[rgb]{0,0,0}$\vec{v}$}%
}}}}
\put(2791, 74){\rotatebox{21.8}{\makebox(0,0)[lb]{\smash{{\SetFigFont{12}{14.4}{\familydefault}{\mddefault}{\updefault}{\color[rgb]{0,0,0}$\vec{r}$}%
}}}}}
\put(3691, 29){\makebox(0,0)[lb]{\smash{{\SetFigFont{12}{14.4}{\familydefault}{\mddefault}{\updefault}{\color[rgb]{0,0,0}$b$}%
}}}}
\end{picture}%

%% file: figures/322_impact_parameter+radius.pstex_t
\begin{picture}(0,0)%
\includegraphics{figures/322_impact_parameter+radius.pstex}%
\end{picture}%
\setlength{\unitlength}{4144sp}%
\begingroup\makeatletter\ifx\SetFigFont\undefined%
\gdef\SetFigFont#1#2#3#4#5{%
  \reset@font\fontsize{#1}{#2pt}%
  \fontfamily{#3}\fontseries{#4}\fontshape{#5}%
  \selectfont}%
\fi\endgroup%
\begin{picture}(6054,1260)(-191,-800)
\put(3189, 34){\makebox(0,0)[lb]{\smash{{\SetFigFont{12}{14.4}{\familydefault}{\mddefault}{\updefault}{\color[rgb]{0,0,0}$r_{N,\mathrm{min}}$}%
}}}}
\put(3061,-736){\makebox(0,0)[lb]{\smash{{\SetFigFont{12}{14.4}{\familydefault}{\mddefault}{\updefault}{\color[rgb]{0,0,0}$\d r=0$}%
}}}}
\put(2950,-21){\makebox(0,0)[lb]{\smash{{\SetFigFont{12}{14.4}{\familydefault}{\mddefault}{\updefault}{\color[rgb]{0,0,0}$b$}%
}}}}
\end{picture}%

%% file: figures/323_terminal_velocity_method.pstex_t
\begin{picture}(0,0)%
\includegraphics{figures/323_terminal_velocity_method.pstex}%
\end{picture}%
\setlength{\unitlength}{4144sp}%
\begingroup\makeatletter\ifx\SetFigFont\undefined%
\gdef\SetFigFont#1#2#3#4#5{%
  \reset@font\fontsize{#1}{#2pt}%
  \fontfamily{#3}\fontseries{#4}\fontshape{#5}%
  \selectfont}%
\fi\endgroup%
\begin{picture}(5180,1955)(233,-1529)
\put(3916,-1456){\makebox(0,0)[lb]{\smash{{\SetFigFont{12}{14.4}{\familydefault}{\mddefault}{\updefault}{\color[rgb]{0,0,0}$Z_\mathrm{terminal}$}%
}}}}
\put(248,279){\makebox(0,0)[lb]{\smash{{\SetFigFont{12}{14.4}{\familydefault}{\mddefault}{\updefault}{\color[rgb]{0,0,0}$I$}%
}}}}
\put(5176,-106){\makebox(0,0)[lb]{\smash{{\SetFigFont{12}{14.4}{\familydefault}{\mddefault}{\updefault}{\color[rgb]{0,0,0}$I_\mathrm{max}$}%
}}}}
\put(5176,-826){\makebox(0,0)[lb]{\smash{{\SetFigFont{12}{14.4}{\familydefault}{\mddefault}{\updefault}{\color[rgb]{0,0,0}$I_\mathrm{threshold}$}%
}}}}
\put(5176,-1096){\makebox(0,0)[lb]{\smash{{\SetFigFont{12}{14.4}{\familydefault}{\mddefault}{\updefault}{\color[rgb]{0,0,0}$I_\mathrm{noise}$}%
}}}}
\put(5176,-1411){\makebox(0,0)[lb]{\smash{{\SetFigFont{12}{14.4}{\familydefault}{\mddefault}{\updefault}{\color[rgb]{0,0,0}$\lambda$}%
}}}}
\end{picture}%

%% file: figures/323_inclined_galaxy.pstex_t
\begin{picture}(0,0)%
\includegraphics{figures/323_inclined_galaxy.pstex}%
\end{picture}%
\setlength{\unitlength}{4144sp}%
\begingroup\makeatletter\ifx\SetFigFont\undefined%
\gdef\SetFigFont#1#2#3#4#5{%
  \reset@font\fontsize{#1}{#2pt}%
  \fontfamily{#3}\fontseries{#4}\fontshape{#5}%
  \selectfont}%
\fi\endgroup%
\begin{picture}(6099,2307)(-236,-1205)
\put(1486,-466){\makebox(0,0)[lb]{\smash{{\SetFigFont{12}{14.4}{\familydefault}{\mddefault}{\updefault}{\color[rgb]{0,0,0}$I$}%
}}}}
\put(4411,-1141){\makebox(0,0)[lb]{\smash{{\SetFigFont{12}{14.4}{\familydefault}{\mddefault}{\updefault}{\color[rgb]{0,0,0}$\lambda$}%
}}}}
\end{picture}%

%% file: figures/324_coord_rot_1_shrunk.pstex_t
\begin{picture}(0,0)%
\includegraphics{figures/324_coord_rot_1_shrunk.pstex}%
\end{picture}%
\setlength{\unitlength}{4144sp}%
\begingroup\makeatletter\ifx\SetFigFont\undefined%
\gdef\SetFigFont#1#2#3#4#5{%
  \reset@font\fontsize{#1}{#2pt}%
  \fontfamily{#3}\fontseries{#4}\fontshape{#5}%
  \selectfont}%
\fi\endgroup%
\begin{picture}(7198,2162)(15,-1333)
\put(5373,590){\makebox(0,0)[lb]{\smash{{\SetFigFont{7}{8.4}{\familydefault}{\mddefault}{\updefault}{\color[rgb]{0,0,0}$\tilde{z}$}%
}}}}
\put(472,-75){\makebox(0,0)[lb]{\smash{{\SetFigFont{7}{8.4}{\familydefault}{\mddefault}{\updefault}{\color[rgb]{0,0,0}$\psi$}%
}}}}
\put(5678,396){\makebox(0,0)[lb]{\smash{{\SetFigFont{7}{8.4}{\familydefault}{\mddefault}{\updefault}{\color[rgb]{0,0,0}$\psi$}%
}}}}
\put(1690,645){\makebox(0,0)[lb]{\smash{{\SetFigFont{7}{8.4}{\familydefault}{\mddefault}{\updefault}{\color[rgb]{0,0,0}$\tilde{z}$}%
}}}}
\put(306,-379){\makebox(0,0)[lb]{\smash{{\SetFigFont{7}{8.4}{\familydefault}{\mddefault}{\updefault}{\color[rgb]{0,0,0}$\tilde{x}$}%
}}}}
\put(3352,-186){\makebox(0,0)[lb]{\smash{{\SetFigFont{7}{8.4}{\familydefault}{\mddefault}{\updefault}{\color[rgb]{0,0,0}rotation}%
}}}}
\put(3269,-407){\makebox(0,0)[lb]{\smash{{\SetFigFont{7}{8.4}{\familydefault}{\mddefault}{\updefault}{\color[rgb]{0,0,0}of coordinate}%
}}}}
\put(3379,-518){\makebox(0,0)[lb]{\smash{{\SetFigFont{7}{8.4}{\familydefault}{\mddefault}{\updefault}{\color[rgb]{0,0,0}system}%
}}}}
\put(4487,-379){\makebox(0,0)[lb]{\smash{{\SetFigFont{7}{8.4}{\familydefault}{\mddefault}{\updefault}{\color[rgb]{0,0,0}$\hat{x}$}%
}}}}
\put(4626,-850){\makebox(0,0)[lb]{\smash{{\SetFigFont{7}{8.4}{\familydefault}{\mddefault}{\updefault}{\color[rgb]{0,0,0}$\tilde{x}$}%
}}}}
\put(5872,673){\makebox(0,0)[lb]{\smash{{\SetFigFont{7}{8.4}{\familydefault}{\mddefault}{\updefault}{\color[rgb]{0,0,0}$\hat{z}$}%
}}}}
\end{picture}%

%% file: figures/324_coord_rot_2_shrunk.pstex_t
\begin{picture}(0,0)%
\includegraphics{figures/324_coord_rot_2_shrunk.pstex}%
\end{picture}%
\setlength{\unitlength}{4144sp}%
\begingroup\makeatletter\ifx\SetFigFont\undefined%
\gdef\SetFigFont#1#2#3#4#5{%
  \reset@font\fontsize{#1}{#2pt}%
  \fontfamily{#3}\fontseries{#4}\fontshape{#5}%
  \selectfont}%
\fi\endgroup%
\begin{picture}(6922,2140)(-11,-1289)
\put(1430,702){\makebox(0,0)[lb]{\smash{{\SetFigFont{7}{8.4}{\familydefault}{\mddefault}{\updefault}{\color[rgb]{0,0,0}$\hat{z}$}%
}}}}
\put(3244,-150){\makebox(0,0)[lb]{\smash{{\SetFigFont{7}{8.4}{\familydefault}{\mddefault}{\updefault}{\color[rgb]{0,0,0}rotation}%
}}}}
\put(3161,-370){\makebox(0,0)[lb]{\smash{{\SetFigFont{7}{8.4}{\familydefault}{\mddefault}{\updefault}{\color[rgb]{0,0,0}of coordinate}%
}}}}
\put(3271,-480){\makebox(0,0)[lb]{\smash{{\SetFigFont{7}{8.4}{\familydefault}{\mddefault}{\updefault}{\color[rgb]{0,0,0}system}%
}}}}
\put(6690,106){\makebox(0,0)[lb]{\smash{{\SetFigFont{7}{8.4}{\familydefault}{\mddefault}{\updefault}{\color[rgb]{0,0,0}$\hat{y}$}%
}}}}
\put(5085,619){\makebox(0,0)[lb]{\smash{{\SetFigFont{7}{8.4}{\familydefault}{\mddefault}{\updefault}{\color[rgb]{0,0,0}$\hat{z}$}%
}}}}
\put(5580,702){\makebox(0,0)[lb]{\smash{{\SetFigFont{7}{8.4}{\familydefault}{\mddefault}{\updefault}{\color[rgb]{0,0,0}$z$}%
}}}}
\put(6706,-343){\makebox(0,0)[lb]{\smash{{\SetFigFont{7}{8.4}{\familydefault}{\mddefault}{\updefault}{\color[rgb]{0,0,0}$y$}%
}}}}
\put(2584,-343){\makebox(0,0)[lb]{\smash{{\SetFigFont{7}{8.4}{\familydefault}{\mddefault}{\updefault}{\color[rgb]{0,0,0}$\hat{y}$}%
}}}}
\put(6124,-156){\makebox(0,0)[lb]{\smash{{\SetFigFont{7}{8.4}{\familydefault}{\mddefault}{\updefault}{\color[rgb]{0,0,0}$\zeta$}%
}}}}
\put(1980,-398){\makebox(0,0)[lb]{\smash{{\SetFigFont{7}{8.4}{\familydefault}{\mddefault}{\updefault}{\color[rgb]{0,0,0}$\zeta$}%
}}}}
\end{picture}%

%% file: ts_galaxy_lensing.tex
\section{Gravitational lensing}
\label{sec:grav_lensing}

Gravitational lensing is another measurable effect of a gravitational field. Eddington first used the gravitational lens of the sun to test Einstein's theory of general relativity during the solar eclipse of the 29$^\mathrm{th}$ May 1919 \cite{Dyson:1920,Dyson:1920a}. Today, \textit{microlensing} events, which temporarily magnify the apparent magnitude of a star, are further cases of gravitational lensing by stellar mass objects.

Stronger gravitational lenses require more mass, and there are indeed observations of strong gravitational lensing by galaxies and galaxy clusters. \textit{Strong lensing} can produce multiple images of the source galaxy (see \sref{strong_lensing}). Lensing by galaxies on a larger distance scale is used to map the gravitational field of whole galaxy clusters, or using statistical methods, to construct a model of the gravitational field of a typical average galaxy. This technique is called \textit{weak lensing}, see \sref{weak_lensing}.

The significant difference between measurements of the gravitational field with rotation curves and gravitational lensing is the speed of the probe particles. For rotation curves, the probe particles are the stars or the gas in the galactic disk which rotate at a clearly subluminal speed $v_c \ll c$. In gravitational lensing, however, the probe particles are the photons of background objects, whose trajectory is bent by the foreground object. Photons travel at the speed of light and therefore, their perception of the gravitational field is quite different, compared to subluminal particles.

In this section, I will first solve the equations of motion for a photon in a weak gravitational field, apply this result to gravitational lensing of point masses and then go on to a fully general relativistic description of lensing by arbitrary geometries using Fermat's principle and an effective refractive index. Finally I will comment on the actual observational techniques that are employed today.

\subsection{Solving the equations of motion directly} 
\label{sec:lens_solve_eq_of_motion}

The simplest case of gravitational lensing is the one where a light ray is bent by the weak gravitational field of a point mass. Such a lensing event is characterized by the observable shift in apparent position on the sky, i.e. the deflection angle $\Delta\cphi$ (see \fref{weinberg_deflection}). One way to find this deflection angle is to solve the equations of motion, as they are given by the geodesic equations for null curves \eref{geodesic_affine_parameter}, directly. 

\begin{figure}[htb]
\begin{center}
\input{figures/331_weinberg_deflection.pstex_t}
\end{center}
\caption[Deflection of a light ray by a point mass.]{\label{fig:weinberg_deflection}
Deflection of a light ray by a point mass. The solid curved line demarks the light ray's trajectory. The angle $\cphi(r)$ is the azimuthal angle for a point on the trajectory at radius $r$. the total deflection angle is given by $\Delta\cphi$. The radius of closest approach is marked by $r_\mathrm{min}$. The light ray's trajectory is symmetric with respect to $r_\mathrm{min}$.}
\end{figure}
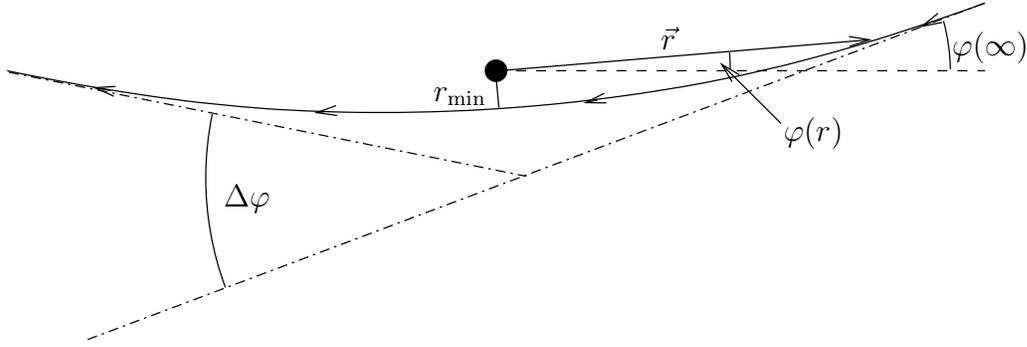

The observer and the photon source are assumed to be far away from the lensing object, such that both can be approximated to rest in flat space at $r=\infty$. Then, in the plane of the light ray's trajectory, the total change of a photon's azimuthal angle $\cphi$ is twice the change from $r=\infty$ to $r=r_\mathrm{min}$ (\fref{weinberg_deflection}). If the trajectory was a straight line, this would be just $\pi$ and the deflection angle would be $\Delta\cphi=0$. Hence, the deflection angle is usually defined as
\begin{equation}
\label{eq:deflection_angle_def}
\Delta\cphi \equiv 2\, \left| \cphi(\infty) - \cphi(r_\mathrm{min}) \right| - \pi \, .
\end{equation} 
The following derivation of $\Delta\cphi$ is similar to the one found in Weinberg \cite[\S 8.5]{weinberg72}. However, a new formula \eref{deflection_angle_weak_field} is given that relates the weak gravitational field of arbitrary geometries specified by $\Phi(r)$ and $m(r)$ to the deflection angle $\Delta\cphi$.

To calculate the angular difference that corresponds to the difference in radial distance between $r=\infty$ and $r=r_\mathrm{min}$, I first combine the definition of a photon's angular momentum \eref{com_angmom_null} with \eref{com_drdchi_null} which yields
\begin{equation}
\label{eq:com_drdcphi_null}
0= -B(r)\,\frac{\tilde{\ell}^2}{r^4} \left( \frac{\d r}{\d\cphi} \right)^2 + \frac{\tilde{\gamma}^2}{A(r)} - \frac{\tilde{\ell}^2}{r^2} \, .
\end{equation}
Before integrating this equation, please note that for the radius of closest approach, $r=r_\mathrm{min}$, where $\d r = 0$, this equation gives
\begin{equation}
\frac{\tilde{\gamma}^2}{\tilde{\ell}^2} = \frac{A(r_\mathrm{min})}{r_\mathrm{min}^2} = \frac{e^{2\Phi(r_\mathrm{min})}}{r_\mathrm{min}^2}\, .
\end{equation} 
The constants of motion vanish upon inserting this into \eref{com_drdcphi_null},
\begin{equation}
\left( \frac{\d r}{\d\cphi} \right)^2 = \frac{r^2}{B(r)} \left[ \left(\frac{r}{r_\mathrm{min}}\right)^2 \, \frac{A(r_\mathrm{min})}{A(r)} - 1  \right] \, ,
\end{equation} 
or in terms of the metric functions $\Phi(r)$ and $m(r)$,
\begin{equation}
\left( \frac{\d r}{\d\cphi} \right)^2 = r^2\,\left( 1-\frac{2m(r)}{r} \right) \left[ \left(\frac{r}{r_\mathrm{min}}\right)^2 \, e^{2\Phi(r_\mathrm{min})-2\Phi(r)} - 1  \right] \, .
\end{equation} 
The differentials can now easily be separated and formally integrated with the upper limit $r=\infty$:
\begin{equation}
\label{eq:exact_deflection_angle_integral_form}
\cphi(\infty) - \cphi(r) = \pm \int_r^\infty \frac{1}{\sqrt{1-2m(\rr)/\rr}}
\frac{1}{\sqrt{(\rr/r_\mathrm{min})^2\,e^{2\Phi(r_\mathrm{min})-2\Phi(\rr)}-1}} \frac{\d\rr}{\rr}
\end{equation} 
Up to this point, no approximations have been made, and the angle of deflection \eref{exact_deflection_angle_integral_form} is valid for any static spherically symmetric spacetime. Assuming that the gravitational field is weak in the relevant region of the galaxy, I can certainly Taylor expand the first square root:
\begin{equation}
\frac{1}{\sqrt{1-2m(\rr)/\rr}} = 1 + \frac{m(\rr)}{\rr} + \O{\left(\frac{m}{r}\right)^2} \, .
\end{equation} 
The second square root needs to be looked at more carefully. Again for weak fields:
\begin{eqnarray}
&~& \frac{1}{\sqrt{(\rr/r_\mathrm{min})^2\,e^{2\Phi(r_\mathrm{min})-2\Phi(\rr)}-1}} = \\  
\nonumber
&~& =\frac{1}{\sqrt{(\rr/r_\mathrm{min})^2-1 + 2[\Phi(r_\mathrm{min})-\Phi(\rr)]\,(\rr/r_\mathrm{min})^2 + \O{\Phi^2}}} \, ,
\end{eqnarray} 
which can be further expanded to yield
\begin{eqnarray}
\nonumber
&~& \frac{1}{\sqrt{(\rr/r_\mathrm{min})^2-1}}\, \left\lbrace 1 - \frac{\Phi(r_\mathrm{min})-\Phi(\rr)}{(\rr/r_\mathrm{min})^2-1}\,\left(\frac{\rr}{r_\mathrm{min}}\right)^2 + \cdots \right. \\
\label{eq:2ndsqrtxxx}
&~& \left. + \O{\Phi^2,  \left(\frac{\Phi(r_\mathrm{min})-\Phi(\rr)}{(\rr/r_\mathrm{min})^2-1}\,\left(\frac{\rr}{r_\mathrm{min}}\right)^2\right)^2} \right\rbrace \, ,
\end{eqnarray} 
where the smallness of the series term still has to be shown. This term can be rearranged to
\begin{equation}
\label{eq:pre_lHopital}
-\frac{\Phi(r_\mathrm{min})-\Phi(\rr)}{(\rr/r_\mathrm{min})^2-1}\,\left(\frac{\rr}{r_\mathrm{min}}\right)^2 =
\frac{\Phi(\rr)-\Phi(r_\mathrm{min})}{1-(r_\mathrm{min}/\rr)^2} \ll 1 \quad \mbox{for} \quad \rr \gg r_\mathrm{min} \, .
\end{equation} 
However, at $\rr=r_\mathrm{min}$ both numerator and denominator vanish. Application of de l'H\^opital's rule gives
\begin{equation}
\lim_{\rr\rightarrow r_\mathrm{min}} \frac{\Phi(\rr)-\Phi(r_\mathrm{min})}{1-(r_\mathrm{min}/\rr)^2} =
\lim_{\rr\rightarrow r_\mathrm{min}} \frac{\Phi'(\rr)}{2(r_\mathrm{min}^2/\rr^3)} =
\frac{r_\mathrm{min}}{2}\, \Phi'(r_\mathrm{min}) \ll 1 \, ,
\end{equation} 
which has been shown to be small for the relevant region of a typical galaxy on page \pageref{sec:rPhiprime_small}. When the gravitational field is reasonably smooth, both numerator and denominator of \eref{pre_lHopital} are monotonically decreasing. Hence, I conclude that
\begin{equation}
\forall \rr \geq r_\mathrm{min}:\: \frac{\Phi(\rr)-\Phi(r_\mathrm{min})}{1-(r_\mathrm{min}/\rr)^2} = \O{r\Phi'} \, .
\end{equation} 
Finally, the second square root of \eref{exact_deflection_angle_integral_form}, that is \eref{2ndsqrtxxx}, can be written as
\begin{equation}
\frac{1}{\sqrt{(\rr/r_\mathrm{min})^2-1}}\, \left\lbrace 1 + \frac{\Phi(\rr)-\Phi(r_\mathrm{min})}{\rr^2-r_\mathrm{min}^2}\,\rr^2 + \O{\Phi^2,  \left(r\Phi'\right)^2} \right\rbrace \, .
\end{equation} 
Therefore, the Taylor series of \eref{exact_deflection_angle_integral_form} for weak gravitational fields is
\begin{eqnarray}
\nonumber
\cphi(\infty) - \cphi(r) &=& \pm \int_r^\infty \frac{1}{\sqrt{(\rr/r_\mathrm{min})^2-1}}\, \left\lbrace 1 + \frac{m(\rr)}{\rr} +  \frac{\Phi(\rr)-\Phi(r_\mathrm{min})}{\rr^2-r_\mathrm{min}^2}\,\rr^2 \right. + \cdots \\
&~& \left. +\O{\left(\frac{m}{r}\right)^2,\Phi^2,  \left(r\Phi'\right)^2} \right\rbrace \frac{\d\rr}{\rr}\, . 
\end{eqnarray}
The first term can be integrated analytically:
\begin{equation}
\int_r^\infty \frac{1}{\sqrt{(\rr/r_\mathrm{min})^2-1}}\frac{\d\rr}{\rr} = \left.\arcsin\left( \frac{r_\mathrm{min}}{\rr} \right)\right|_r^\infty
\end{equation} 
The deflection angle \eref{deflection_angle_def} for gravitational lensing in a weak gravitational field is then given by
\begin{equation}
\Delta\cphi \approx 2 \int_{r_\mathrm{min}}^\infty \frac{1}{\sqrt{(\rr/r_\mathrm{min})^2-1}}\, \left\lbrace  \frac{m(\rr)}{\rr} + \frac{\Phi(\rr)-\Phi(r_\mathrm{min})}{\rr^2-r_\mathrm{min}^2}\,\rr^2  \right\rbrace \frac{\d\rr}{\rr}\, .
\end{equation} 
Since $r_\mathrm{min}$ cannot be observed directly, I want to rewrite this result in terms of the impact parameter \eref{r_of_b},
\begin{equation}
r_\mathrm{min} = b\,e^{\Phi(r_\mathrm{min})} = b + \O{\Phi} \, .
\end{equation} 
To the required order of accuracy, this gives
\begin{equation}
\Delta\cphi \approx 2 \int_b^\infty \frac{1}{\sqrt{(\rr/b)^2-1}}\, \left\lbrace  \frac{m(\rr)}{\rr} + \frac{\Phi(\rr)-\Phi(b)}{\rr^2-b^2}\,\rr^2  \right\rbrace \frac{\d\rr}{\rr} \, ,
\end{equation} 
or equivalently
\begin{equation}
\label{eq:deflection_angle_weak_field}
\Delta\cphi \approx 2\,b \int_b^\infty \frac{\d\rr}{\sqrt{\rr^2-b^2}}\, \left\lbrace  \frac{m(\rr)}{\rr^2} + \frac{\Phi(\rr)-\Phi(b)}{\rr^2-b^2}\,\rr  \right\rbrace \, .
\end{equation} 
This formula is valid for all spherically symmetric spacetimes in which the lensed photons experience only weak gravitational fields. However, to evaluate this integral, one has to make a specific choice for the metric functions $\Phi(r)$ and $m(r)$, as the only quantity that is accessible through lensing oberservations is $\Delta\cphi$.

One can find an analytical solution for the Schwarzschild metric, which represents a ``point'' object with total mass $M$. Similarly, if the halo falls off sufficiently rapidly, the Schwarzschild solution will be a good approximation to the geometry outside of the halo. The metric components of the Schwarzschild solution \eref{Schwarzschild_exterior_metric} are
\begin{equation}
m(r) = M\, ; \qquad \Phi(r) = \frac{1}{2} \ln \left(1 -\frac{2M}{r}\right) \, ,
\end{equation} 
for which the integral \eref{deflection_angle_weak_field} still cannot be solved analytically, but at least an approximate solution can be obtained for weak gravitational fields. Therefore, I assume that the trajectory of the photons passes the point mass far away from event horizon at $r=2M$. Using the smallness of the term 
\begin{equation}
\frac{2M}{r} \leq \frac{2M}{r_\mathrm{min}} \ll 1
\end{equation} 
I can approximate
\begin{equation}
\Phi(r) \approx -\frac{M}{r} \, ,
\end{equation} 
so that
\begin{equation}
\Delta\cphi \approx 2\,b \int_b^\infty \frac{\d\rr}{\sqrt{\rr^2-b^2}}\, \left\lbrace  \frac{M}{\rr^2} + \frac{M\,(\rr-b)}{b\,(\rr^2-b^2)}  \right\rbrace \, .
\end{equation} 
Introducing the dimensionless quantity $x\equiv \rr/b$ yields the solvable integral
\begin{equation}
\Delta\cphi \approx \frac{2\,M\,b}{b^2} \int_1^\infty \frac{\d x}{\sqrt{x^2-1}}\, \left\lbrace  \frac{1}{x^2} + \frac{1}{x+1}  \right\rbrace \, ,
\end{equation} 
which evaluates to
\begin{equation}
\label{eq:tot_deflect_angle_weinberg}
\Delta\cphi \approx \frac{2\,M}{b} \left\lbrace 1 + 1 \right\rbrace = \frac{4\,M}{b} \, .
\end{equation}  

Although this method of deriving the deflection angle is straightforward, the integral involved can usually not be solved analytically for more general metrics. Also, the physical insight gained for the relation between the metric functions, $\Phi(r)$ and $m(r)$, and the deflection angle $\Delta\cphi$ is limited. To understand the influence of the metric functions on the bending of light of spherically symmetric objects better, it is helpful to follow the analogy of an effective refractive index, which relates the bending of light by gravitational fields to familiar problems in classical optics.

\subsection{Lensing with an effective refractive index}

To introduce the notion of an ``effective refractive index'' associated with a gravitational field, I will relate the path of a null curve in a static metric to Fermat's principle of least time or equivalently, least optical length. Then I will show how the path of a light ray in a static spherically symmetric spacetime is described by one scalar function only, namely the effective refractive index $n(r)$. Finally, I emphasise how one can use this refractive index to easily read off the relation between the metric components, $g_{tt}$ and $g_{rr}$, and the deflection angle $\Delta\cphi$.

\subsubsection{Null geodesics in spacetime and the 3-dimensional Fermat metric $\hat{g}_{ab}$}

Following Misner, Thorne \& Wheeler \cite[exercise 40.3]{MTW}, I will show that null geodesics in a static gravitational field can be expressed as geodesics in a Riemannian 3-space with Fermat metric $\hat{g}_{ab}$ and therefore satisfy Fermat's principle of a light ray's least optical length\footnote{Fermat's original formulation required the integral $\int\d t$ to be minimal. This statement was later shown to be physically incorrect and replaced with the requirement that the first variation must vanish locally, i.e. be extremal: $\delta\int\d t=0$. \cite{Kline65}}. A static gravitational field is characterized by $\partial_t\, g_{\alpha\beta} = 0$ and $g_{tb}=g_{at}=0$. Hence for null curves with
\begin{equation}
\d s^2 = g_{\mu\nu}\, \d X^\mu \, \d X^\nu = 0
\end{equation}
follows
\begin{equation}
\label{eq:null_curve_1}
\d t^2 = \left( \d X^t \right)^2 = -\frac{g_{ab}}{g_{tt}} \, \d X^a \, \d X^b \, .
\end{equation}
The non-vanishing Christoffel symbols of the first kind are:
\begin{eqnarray}
\label{eq:null_curve_chris1}
\Gamma_{ttc} &=& \frac{1}{2}\, \partial_c\, g_{tt} \\
\Gamma_{att} &=& -\frac{1}{2}\, \partial_a\, g_{tt} \\
\label{eq:null_curve_chris3}
\Gamma_{abc} &=& \frac{1}{2}\, \left[ \partial_b\, g_{ac} + \partial_c\, g_{ab} - \partial_a\, g_{bc} \right] \, .
\end{eqnarray} 
Splitting the geodesic equations \eref{geodesic_affine_parameter} with affine parameter $\chi$ into time and space yields,
\begin{eqnarray}
0 &=& g_{tt}\,\frac{\d^2 X^t}{\d\chi^2} + 2\,\Gamma_{ttc}\, \frac{\d X^t}{\d\chi}\, \frac{\d X^c}{\d\chi} \, ; \\
\nonumber
0 &=& g_{ad}\,\frac{\d^2 X^d}{\d\chi^2} + \Gamma_{att}\, \frac{\d X^t}{\d\chi}\, \frac{\d X^t}{\d\chi} + \Gamma_{abc}\, \frac{\d X^b}{\d\chi}\, \frac{\d X^c}{\d\chi} \\
  &=& g_{ad}\,\frac{\d^2 X^d}{\d\chi^2} + \left[ \Gamma_{abc} - \Gamma_{att} \, \frac{g_{bc}}{g_{tt}} \right] \, \frac{\d X^b}{\d\chi}\, \frac{\d X^c}{\d\chi} \, ,
\end{eqnarray} 
where \eref{null_curve_1} was used in the last line. Reparametrizing with the time coordinate $t\equiv X^t$ gives
\begin{eqnarray}
g_{tt}\,\frac{\d^2 t}{\d\chi^2} + 2\,\Gamma_{ttc}\, \frac{\d X^c}{\d t}\, \left( \frac{\d t}{\d\chi} \right)^2 &=& 0 \, ; \\
g_{ad}\,\frac{\d^2 X^d}{\d t^2} + \left[ \Gamma_{abc} - \Gamma_{att} \, \frac{g_{bc}}{g_{tt}} \right] \, \frac{\d X^b}{\d t}\, \frac{\d X^c}{\d t} &=& - g_{ad}\,\frac{\d X^d}{\d t} \, \frac{\d^2 t/\d\chi^2}{(\d t/\d\chi)^2} \, .
\end{eqnarray} 
Please note that the time coordinate $t$ is not an affine parameter in the 4-dimensional spacetime $g_{\mu\nu}$. Eliminating the second derivative $\d^2 t/\d\chi^2$ from the last two equations, I find
\begin{equation}
g_{ad}\,\frac{\d^2 X^d}{\d t^2} + \left[ \Gamma_{abc} - \Gamma_{att} \, \frac{g_{bc}}{g_{tt}} - 2\,\Gamma_{ttc}\,\frac{g_{ab}}{g_{tt}} \right] \, \frac{\d X^b}{\d t}\, \frac{\d X^c}{\d t} = 0 \, ,
\end{equation} 
which upon inserting the Christoffel symbols \eref{null_curve_chris1} to \eref{null_curve_chris3} takes the form
\begin{equation}
\label{eq:null_curve_geodesic_1}
g_{ad}\,\frac{\d^2 X^d}{\d t^2} + \left[ \partial_c\, g_{ab} - \frac{1}{2}\, \partial_a\, g_{bc} + \frac{1}{2}\, \frac{g_{bc}}{g_{tt}} \, \partial_a\, g_{tt} - \frac{g_{ab}}{g_{tt}} \, \partial_c\, g_{tt} \right] \, \frac{\d X^b}{\d t}\, \frac{\d X^c}{\d t} = 0 \, .
\end{equation} 
When defining the so called Fermat metric in three space dimensions,
\begin{equation}
\hat{g}_{ab} \equiv \frac{g_{ab}}{-g_{tt}} \, ,
\end{equation}
one can easily identify \eref{null_curve_geodesic_1} as the geodesics of the three dimensional $\hat{g}_{ab}$ with affine parameter $t$:
\begin{equation}
\hat{g}_{ad}\,\frac{\d^2 X^d}{\d t^2} + \left[ \partial_c\, \hat{g}_{ab} - \frac{1}{2}\, \partial_a\, \hat{g}_{bc} \right] \, \frac{\d X^b}{\d t}\, \frac{\d X^c}{\d t} = 0 \, .
\end{equation} 
By the basic notion of geodesics in Riemannian metric spaces (\sref{geodesics_in_metric_spaces}), this corresponds to minimal travelling time $t$ in the Riemannian 3-space or analogously minimal lapsed time coordinate $X^t=t$ in the Lorentzian 4-spacetime:
\begin{equation}
\label{eq:Fermats_principle_conformal_static}
\delta\!\! \int \d t = \delta\!\! \int \sqrt{ \hat{g}_{ab} \, \d X^a \, \d X^b } = 0 \, .
\end{equation} 
This is the exact mathematical formulation of Fermat's principle, which states that the first variation of the lapsed time along the path between two points of a light ray's trajectory must vanish.
An even more general result can be obtained for conformally stationary spacetimes, 
\begin{equation}
\label{eq:conformally_stationary_metric}
\d s^2 = e^{2f(t,X^a)} \left[ - (\d t + \hat\phi_i\,\d X^i)^2 + \hat{g}_{ij}\,\d X^i\,\d X^j \right] \, ,
\end{equation} 
where $\exp [2f(t,X^a)]$ is the conformal factor and $\hat\phi_i(X^a)$ is called the Fermat one-form \cite{Perlick:2004}. Fermat's principle of the least lapsed light travel time then takes the following form:
\begin{equation}
\delta\!\! \int \left[ \sqrt{ \hat{g}_{ab} \, \d X^a \, \d X^b } - \hat\phi_a \, \d X^a \right] = 0 \, ,
\end{equation} 
which reduces to \eref{Fermats_principle_conformal_static} in the conformally static case with $\hat\phi_i = \partial_i h$, where $h(X^a)$ is a scalar function of the space coordinates only.

Since the speed of light in transparent media, $v_n$, is given by the vacuum speed of light, $c$, divided by the refractive index $n$, one can formulate 
Fermat's principle equivalently as the principle of shortest optical length
\begin{equation}
\label{eq:Fermats_principle_3d}
c\;\: {\delta\!\! \int \d t} = \delta\!\! \int n(x,y,z)\,\d\sigma =  0 \, ,
\end{equation} 
where $\sigma$ is Euclidean arclength and
\begin{equation}
\frac{\d\sigma}{\d t} = v_n = \frac{c}{n} \, .
\end{equation} 
If the Fermat metric is conformally Euclidean, i.e. $\hat{g}_{ab}=n^2(x,y,z)\,\delta_{ab}$, equation \eref{Fermats_principle_conformal_static} takes exactly the form \eref{Fermats_principle_3d},
\begin{equation}
\delta\!\! \int n\,\sqrt{ \delta_{ab} \, \d X^a \, \d X^b } = \delta\!\! \int n(x,y,z)\,\d\sigma = 0 
\end{equation} 
and therefore, the scalar effective refractive index $n$ can be conceived from the metric coefficients of conformally static spacetimes $g_{\mu\nu}$ that have a conformally Euclidean space-space part $g_{ab}$.

If the space-space portion of the spacetime is not conformally Euclidean, one can still define an effective refractive index tensor $n_{ab} \equiv \sqrt{\hat{g}_{ab}}$, analogously to a refractive index tensor in anisotropic media like e.g. crystals \cite{Born03}. This is the strong field generalization of the static weak field refractive index tensor presented in Appendix \ref{sec:refindex_static}.

\subsubsection{Non-Euclidean spaces and the propagation of light}


Although I already showed how null curves correspond to Fermat's principle, I want to present the relation between geodesic motion in a non-Euclidean 3-space and the propagation of electromagnetic waves in Euclidean space, as it is shown by Kline \& Kay \cite[\S V.A]{Kline65}.

Consider the non-Euclidean 3-space given by the differential line element
\begin{equation}
\label{eq:fermat_metric_isotropic}
\d\chi^2 \equiv n^2(x,y,z) \left( \d x^2 + \d y^2 + \d z^2 \right) = n^2\, \d\sigma^2 \, ,
\end{equation} 
where $n(x,y,z)$ is the refractive index and $\d\sigma$ is Euclidean distance. From the laws of optics, we know that the propagation of light rays is governed by the relation
\begin{equation}
\label{eq:standard_optics}
\frac{\d}{\d\sigma} \left( n\, \frac{\d\vec{r}}{\d\sigma} \right) = \vec\nabla\, n \, ,
\end{equation}
where $\vec{r} = (x;y;z)$ is the position vector of a point on the light ray's trajectory (for reference see \cite[\S 3.2.1]{Born03} or \cite[\S V.4]{Kline65}). This can be reparametrized in terms of $\chi$,
\begin{eqnarray}
\left| \frac{\d\vec{r}}{\d\sigma} \right| &=& n\, \left| \frac{\d\vec{r}}{\d\chi} \right| =1 \qquad \mathrm{and}\\
\label{eq:standard_optics_chi}
\frac{\d}{\d\sigma} \left( n\, \frac{\d\vec{r}}{\d\sigma} \right) &=&
n\,\frac{\d}{\d\chi} \left( n^2\, \frac{\d\vec{r}}{\d\chi} \right) =
\vec\nabla\, n \, ,
\end{eqnarray} 
where the gradient $\vec\nabla$ is still to be taken in the flat Euclidean space. Applying the product rule to \eref{standard_optics_chi} gives
\begin{equation}
2\,\frac{\d n}{\d\chi}\,\frac{\d\vec{r}}{\d\chi} + n \, \frac{\d^2\vec{r}}{\d\chi^2} = \frac{1}{n^2} \vec\nabla\, n \, ,
\end{equation} 
so that inserting
\begin{equation}
\frac{\d n}{\d\chi} = \frac{\partial n}{\partial x}\,\frac{\d x}{\d\chi} + \frac{\partial n}{\partial y}\,\frac{\d y}{\d\chi} + \frac{\partial n}{\partial z}\,\frac{\d z}{\d\chi} = \frac{\d\vec{r}}{\d\chi} \cdot \vec\nabla\, n 
\end{equation} 
and
\begin{equation}
\frac{1}{n^2} = \left| \frac{\d\vec{r}}{\d\chi} \right| ^2 
\end{equation} 
results in
\begin{equation}
n \, \frac{\d^2\vec{r}}{\d\chi^2} + 2\, \left( \frac{\d\vec{r}}{\d\chi} \cdot \vec\nabla\, n \right) \,\frac{\d\vec{r}}{\d\chi} - \left| \frac{\d\vec{r}}{\d\chi} \right| ^2 \vec\nabla\, n = 0 \, .
\end{equation} 
Rewriting this in the usual index notation with $\vec{r} = X_a$ yields
\begin{equation}
n \, \frac{\d^2 X_a}{\d\chi^2} + 2\, \left( n_{,c} \, \frac{\d X^c}{\d\chi} \right) \,\frac{\d X_a}{\d\chi} - n_{,a}\,\delta_{bc}\, \frac{\d X^b}{\d\chi}\, \frac{\d X^c}{\d\chi}   = 0 \, ,
\end{equation} 
which when put into the form
\begin{equation}
\label{eq:light_ray_3d_geodesic}
n^2 \, \frac{\d^2 X_a}{\d\chi^2} + 2n \left\lbrace n_{,c}\,\delta_{ab} - \frac{1}{2}\,n_{,a}\,\delta_{bc} \right\rbrace \, \frac{\d X^b}{\d\chi}\, \frac{\d X^c}{\d\chi}   = 0 \, ,
\end{equation} 
is easily identified as the geodesic equation \eref{geodesic_eqns_1st_christoffel} of $\hat{g}_{ab}=n^2\,\delta_{ab}$ with the affine connexion
\begin{equation}
\Gamma_{abc} = \frac{1}{2} \left\lbrace \hat{g}_{ab,c} + \hat{g}_{ac,b} - \hat{g}_{bc,a} \right\rbrace \, .
\end{equation} 
Therefore, light travels along a trajectory with
\begin{equation}
\delta\!\! \int \d\chi = \delta\!\! \int n\,\d\sigma = c\; \delta\!\! \int \d t = 0\, , 
\end{equation} 
which has been shown in \sref{geodesics_in_metric_spaces}. Since $\d t \equiv \d\sigma / v_n$ is the light's travel time, Fermat's principle is satisfied.

In the next section I will show how this geometric formulation of the refractive index can be used to relate the metric components $g_{tt}$ and $g_{rr}$ of a static spherically symmetric spacetime to the bending of light.

\subsubsection{Refractive index of a static spherically symmetric spacetime}

I will now derive an explicit formula for the exact refractive index of a spherically symmetric spacetime. If a conformally static spacetime is spherically symmetric, it can be expressed in isotropic coordinates where all information about a light ray's trajectory is contained in one scalar function, the effective refractive index $n(r)$. I follow the approach of Perlick \cite{Perlick:2004} for the exact analytic formulation and then introduce a weak field approximation that casts $n(r)$ in a simple and familiar form that allows straightforward interpretation of the relation between the metric components and the bending of light.

The spacetime \eref{ss_metric} can always be written in conformal static form,
\begin{eqnarray}
\nonumber
\d s^2 &=& e^{2\Phi(r)} \left[ -\d t^2 + \frac{e^{-2\Phi(r)}}{1-2m(r)/r}\, \d r^{2} + r^{2}\,e^{-2\Phi(r)}\, \d\Omega^{2} \right] \\
\label{eq:ss_metric_isotropic_coords}
&\equiv& e^{2\Phi(r)} \left[ -\d t^2 + S^2(r)\, \d r^{2} + R^2(r)\, \d\Omega^{2} \right]
\, .
\end{eqnarray}
To perform a coordinate transformation to isotropic coordinates,
\begin{equation}
\label{eq:ss_metric_isotropic_coords_refindex}
\d s^2 = e^{2\Phi(r)} \left[ -\d t^2 + n^2(\rr)\, \left( \d \rr^{2} + \rr^{2}\, \d\Omega^{2} \right) \right] \, ,
\end{equation}
one finds the conditions
\begin{equation}
n(\rr)\, \d\rr = S(r)\, \d r \, ; \qquad n(\rr)\, \rr = R(r) \, ,
\end{equation} 
so that eliminating $n(r)$ gives the transformation for the $r$-coordinate in integral form
\begin{equation}
\label{eq:isotropic_coord_condition}
\int \frac{\d\rr}{\rr} = \int \frac{S(r)}{R(r)} \,\d r \quad \Leftrightarrow \quad \ln \frac{\rr}{\rr_C} = \int \frac{\d r}{r\,\sqrt{1-2m(r)/r}} \, ,
\end{equation} 
where $\rr_C$ is a constant of integration. The refractive index is then given exactly by
\begin{equation}
\label{eq:ref_index_strong_gravity}
n(\rr) = \frac{R(r)}{\rr} = \frac{r}{\rr}\,e^{-\Phi(r)} \, ,
\end{equation} 
where $r=r(\rr)$ still has to be determined from integration of \eref{isotropic_coord_condition}. Although this is a simple and straightforward relation, the coordinate $\rr$ and the refractive index can only be obtained analytically if the function $m(r)$ is known. Before I go on to introduce the approximation of weak gravitational fields, I note that the change to isotropic coordinates allows the use of one refractive index $n(r)$ only. For arbitrary coordinates, two refractive indices are necessary in spherical symmetry, as shown by Atkinson \cite{Atkinson:1965}, and with less symmetry one generally needs an entire $3\times 3$ tensor of refractive indices, see Appendix \ref{sec:refindex}.

For weak gravitational fields, one can approximate $2m(r)/r \ll 1$ and therefore, the integral in \eref{isotropic_coord_condition} simplifies tremendously
\begin{eqnarray}
\nonumber
\ln \frac{\rr}{\rr_C} &=&  \int \frac{\d r}{r}\,\left\lbrace 1 + \frac{m(r)}{r} + \O{\left( \frac{2m}{r} \right)^2} \right\rbrace \\
&=& \ln \frac{r}{r_C} + \int \frac{m(r)}{r^2}\, \d r + \O{\left( \frac{2m}{r} \right)^2} \, .
\end{eqnarray} 
Once one chooses the constants of integration to satisfy $\rr_C = r_C$, one gains the relation
\begin{equation}
\label{eq:isotropic_radius_curvature_radius}
\rr = r\, \exp\left\lbrace \int \frac{m(r)}{r^2}\, \d r + \O{\left( \frac{2m}{r} \right)^2} \right\rbrace \, .
\end{equation} 
Since $\rr = r + \O{2m/r}$, the refractive index to second order in $2m/r$ takes the form
\begin{eqnarray}
n(\rr\approx r) &=& \exp\left\lbrace - \Phi(r) - \int \frac{m(r)}{r^2}\, \d r  + \O{\left( \frac{2m}{r} \right)^2,\, \frac{2m}{r}\,\Phi} \right\rbrace \\
\label{eq:ref_index_weak_gravity}
&=& 1 - \Phi(r) - \int \frac{m(r)}{r^2}\, \d r + \O{\left( \frac{2m}{r} \right)^2,\, \frac{2m}{r}\,\Phi,\, \Phi^2} \, .
\end{eqnarray} 
The physical interpretation of this last line is simple and enlightening:
\begin{enumerate}
\item The trajectory of a null curve in a static spherically symmetric spacetime is entirely determined by one scalar function $n(r)$ only. This is especially also true for strong gravitational fields, where $n(r)$ is given by \eref{ref_index_strong_gravity}. The requirement of two refractive indices in other coordinate representations apart from isotropic coordinates is a mere reflection of coordinate artefacts. The physically relevant information about bending of light in static spherically symmetric spacetimes is encoded in a single refractive index.
\item The refractive index can be used to calculate photon trajectories in Euclidean 3-space. The resulting trajectories are null curves in the isotropic coordinate representation \eref{ss_metric_isotropic_coords} of that spacetime. Especially for asymptotically flat spacetimes this means that asymptotic observables, that were calculated using the refractive index $n(r)$, have the same value for the Euclidean 3-space and for the spacetime. For example, the asymptotically observed  deflection angle $\Delta\cphi$ can directly be calculated from $n(r)$.
\item The metric functions $\Phi(r)$ and $m(r)$ \textit{both} contribute to the refractive index and therefore to the bending of light. This means, that observation of gravitational lensing reflects contributions from both the metric components $g_{tt}$ and $g_{rr}$, unlike observations of e.g. rotation curves that yield information about $g_{tt}$ only. Furthermore, if one uses the Newtonian approximation $\Phi'(r) \approx m(r)/r^2$, it is evident from \eref{ref_index_weak_gravity} that both metric components contribute in equal amounts to the bending of light in weak gravitational fields. \label{it:both_components}
\end{enumerate}

It is now evident that combined observations of rotation curves and gravitational lensing in principle completely determine the geometry, if it can be modelled by a static spherically symmetric spacetime. Before I show how this can be quantified, let me illustrate how to calculate the deflection angle of a lensing event of a point mass and comment on current observational techniques.

\label{sec:refindex_spherical_symmetric}

\subsubsection{Derivation of $\Delta\cphi$ using the effective refractive index $n(r)$}

Finding the deflection angle for arbitrary spacetimes can only be achieved numerically, no matter which approach of calculation one persues: solving the equations of motion directly or utilizing the effective refractive index $n(r)$. However, if one expects the deflection angle to be small, one can approximate the deflection angle analytically. 

Let $\vec{e} \equiv \d\vec{r}/\d\sigma$ be a unit length tangent vector of a light ray in Euclidean 3-space. The (small) total angle of deflection is then given by the difference between the asymptotic tangent vectors $\vec{e}_\mathrm{in}$ and $\vec{e}_\mathrm{out}$ of the incoming and outgoing light rays,
\begin{equation}
\label{eq:def_deflection_vector}
\vec{\alpha} \equiv \vec{e}_\mathrm{out} - \vec{e}_\mathrm{in}\, ,
\end{equation} 
so that for small angles $\alpha$,
\begin{equation}
\Delta\cphi \approx |\vec{\alpha}| = \sqrt{2}\,\sqrt{1-\cos\alpha} \approx \alpha \, .
\end{equation} 
Equation \eref{def_deflection_vector} can also be written in integral form:
\begin{equation}
\vec{\alpha} = \int \frac{\d\vec{e}}{\d\sigma}\, \d\sigma \, ,
\end{equation} 
with the integral to be evaluated along the photon's trajectory. The geodesic equation \eref{light_ray_3d_geodesic}, or equivalently \eref{standard_optics}, yields
\begin{equation}
n\, \frac{\d\vec{e}}{\d\sigma} = \vec\nabla n - \vec{e} \, \left(\vec{e} \cdot \vec\nabla n \right) \equiv \vec\nabla_{\!\!\perp} n \, ,
\end{equation} 
where $\vec\nabla_{\!\!\perp}$ is defined to give the gradient only in the directions perpendicular to $\vec{e}$. Hence the total angle of deflection is given by \cite{Schneider92}
\begin{equation}
\label{eq:tot_alpha}
\vec\alpha = \int \frac{\vec\nabla_{\!\!\perp} n}{n}\, \d\sigma \approx \int \vec\nabla_{\!\!\perp} n\: \d\sigma \, ,
\end{equation} 
where the variation of $n$ was approximated to be small for small deflection angles:
\begin{equation}
n(r) \approx 1 + \epsilon(r)\; ; \quad \mathrm{with}~\epsilon(r) \ll 1 \, .
\end{equation}
The integral in \eref{tot_alpha} is to be taken along the light ray's trajectory. For weak gravitational fields, e.g. the ones described by the refractive index \eref{ref_index_weak_gravity}, the deflection angle will be very small and hence, the path of integration can be approximated by a straight line, instead of following the exact and unknown trajectory of the light ray. 

Let the $z$-coordinate be pointing into the direction of this straight line of integration and let $b = |(b_{x}, b_{y})| = b\,|\hat{b}|$ be the impact parameter whose unit vector $\hat{b}$ lies in the $x$-$y$ plane by definition. Then, both $\vec\nabla_{\!\!\perp} n$ and $\vec\alpha$ will approximately lie in the $x$-$y$ plane as well and by symmetry point into the $\hat{b}$ direction. 
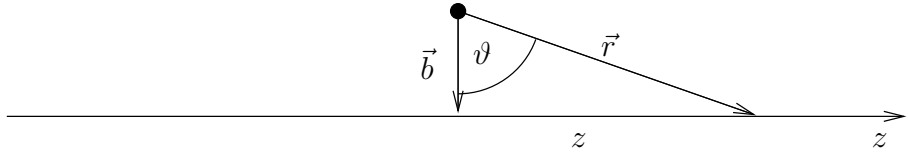
\begin{figure}[htb]
\begin{center}
\input{figures/332_refindex_trick.pstex_t}
\end{center}
\caption[Geometry of the integration path to calculate the deflection angle.]{\label{fig:refindex_trick}
Geometry of the integration path to calculate the deflection angle. The $z$-axis forms a perpendicular triangle with the vector of the impact parameter $\vec{b}=b\,\hat{b}$ and the position vector $\vec{r} = r\,\hat{r}$~; $|\hat{b}|=|\hat{r}|=1$.}
\end{figure}
Therefore, I can write for a spherically symmetric refractive index $n(r)$,
\begin{equation}
\Delta\cphi \approx |\vec\alpha| = \hat{b} \cdot \vec\alpha \approx \int_{-\infty}^{\infty} \hat{b} \cdot \vec\nabla_{\!\!\perp} n(r)\: \d z \, .
\end{equation}
Then,
\begin{equation}
\hat{b} \cdot \vec\nabla_{\!\!\perp} n(r) = \hat{b} \cdot \vec\nabla n(r) = \hat{b} \cdot \hat{r}\, \frac{\d n}{\d r} = \frac{b}{r}\, n'(r) \, ,
\end{equation}
where I used the perpendicular triangle (\fref{refindex_trick})
\begin{equation}
\label{eq:perp_triangle}
r^{2} = b^{2} + z^{2}
\end{equation}
to deduce
\begin{equation}
\hat{b} \cdot \hat{r} = \cos\vartheta = \frac{b}{r} \, .
\end{equation}
From \eref{perp_triangle} also follows that
\begin{equation}
r\, \d r = z\, \d z
\end{equation}
so that
\begin{equation}
\Delta\cphi \approx 2\int_{\infty}^b \frac{b}{z}\, n'(r)\, \d r = -2\,b \int_b^{\infty} \frac{n'(r)}{\sqrt{r^2-b^2}}\, \d r \, .
\end{equation}
Inserting the derivative of the refractive index \eref{ref_index_weak_gravity},
\begin{equation}
n'(r) \approx  - \Phi'(r) - \frac{m(r)}{r^2} \, ,
\end{equation}
gives
\begin{equation}
\label{eq:deflection_angle_weak_field_refindex}
\Delta\cphi \approx 2\,b \int_b^{\infty} \frac{\d r}{\sqrt{r^2-b^2}}\, \left\lbrace \Phi'(r) + \frac{m(r)}{r^2} \right\rbrace  \, .
\end{equation}
This formula looks similar to the one obtained from solving the equations of motion directly \eref{deflection_angle_weak_field}, and indeed, the identity of both formulae can be shown by an integration by parts:
\begin{equation}
\int_b^{\infty} \frac{\Phi'(r)\,\d r}{\sqrt{r^2-b^2}} = \int_b^{\infty} \frac{[\Phi(r)-\Phi(b)]'\,\d r}{\sqrt{r^2-b^2}} = \left. \frac{\Phi(r)-\Phi(b)}{\sqrt{r^2-b^2}} \right|_b^\infty + \int_b^{\infty} \frac{[\Phi(r)-\Phi(b)]\,r\,\d r}{(r^2-b^2)^{3/2}} \, .
\end{equation}
Noting that the surface term vanishes after applying de l'H\^optial's rule for $r=b$, I have shown that the deflection angles calculated from both independent approaches, solving the equations of motion directly \eref{deflection_angle_weak_field} and using the refractive index \eref{deflection_angle_weak_field_refindex}, are consistent with each other.

Therefore, it is no surprise that \eref{deflection_angle_weak_field_refindex} yields the familiar solution when inserting the metric functions of the Schwarzschild solution \eref{Schwarzschild_exterior_metric}:
\begin{equation}
\Phi'(r) \approx \frac{M}{r^2} \, ; \qquad m(r) = M \, ,
\end{equation} 
into the formula for the deflection angle \eref{deflection_angle_weak_field_refindex},
\begin{equation}
\label{eq:tot_deflect_angle_refindex}
\Delta\cphi \approx 2\,b \int_b^{\infty} \frac{\d r}{\sqrt{r^2-b^2}}\, \left\lbrace \frac{M}{r^2} + \frac{M}{r^2} \right\rbrace = \frac{4\,M\,b}{b^2} \int_1^{\infty} \frac{\d x}{x^2\,\sqrt{x^2-1}} = \frac{4\,M}{b}\, .
\end{equation}
Please note that half the deflection angle originates in the $g_{tt}$ component of the Schwarzschild metric, while the other half is contributed by the $g_{rr}$ component. This is also the case for the deflection angle derived via the equations of motion \eref{deflection_angle_weak_field} and more generally for the effective refractive index of the Schwarzschild exterior geometry:
\begin{equation}
\label{eq:schwarzschild_refindex}
n(r) \approx 1 + \frac{M}{r} - \int \frac{M}{r^2}\,\d r = 1 + \frac{2M}{r} \, .
\end{equation} 
This ``50:50'' behaviour is a special feature of the Schwarzschild solution and not typical for more general spherically symmetric geometries. Especially when gravitational lensing occurs within a galaxy, where the point mass model ceases to be an appropriate approximation, it is necessary to realise that there are two distinct contributions to the deflection angle or equivalently to the refractive index. In the remainder of this section, I will outline how to interpret observations of lensing by extended objects.

\subsubsection{Advantages of the notion of the effective refractive index}

One might ask, what is the advantage of the derivation of the deflection angle using the refractive index? After all, it was necessary to introduce several approximations, and then the only result that could be obtained analytically is the deflection angle for the lensing of the weak gravitational field of a point mass $M$, which can be achieved more easily by solving the equations of motion directly, as in \sref{lens_solve_eq_of_motion}.

However, while deriving this result using the refractive index, we learned how the metric components contribute to the lensing. Roughly speaking, for point masses, the effects are split up 50:50 between the $g_{tt}$ and $g_{rr}$ component. We also learned that the refractive index $n(r)$ is the \textit{only} observable function that can be measured by observing gravitational lensing events. Hence, only if one \textit{assumes} a-priori the quasi-Newtonian refractive index
\begin{equation}
\label{eq:refindex_Newtonian_assumption}
n(r) \equiv 1 - 2\Phi_\mathrm{lens}(r)
\end{equation}
will one measure the potential $\Phi_\mathrm{lens}(r)$, which is generally \textit{interpreted} as the gravitational potential $\Phi(r)$. From \eref{ref_index_weak_gravity}, we now know that actually
\begin{equation}
\label{eq:lensing_potential}
2\Phi_\mathrm{lens}(r) \approx \Phi(r) + \int \frac{m(r)}{r^2}\,\d r \, .
\end{equation} 
Since the only observable is $n(r)$, it is not possible to separate the contributions from $\Phi(r)$ and $m(r)$ by observing gravitational lensing events only. A different means of obtaining information about $\Phi(r)$ and/or $m(r)$ is necessary, and as I have previously outlined, rotation curve measurements are suitable to determine $\Phi(r)$ only.

Before I quantify the interpretation of combined observations of rotation curves and gravitational lensing, I will give an overview of the current observational situation of strong and weak gravitational lensing.


\subsection{Observational techniques of gravitational lensing}

The notation in this chapter differs slightly from the previous notation to match the formulae as they are usually employed in the observational gravitational lensing literature. All vectors and gradients are two dimensional and live in the ``lens plane'' which will be defined later on.

\subsubsection{Lensing formalism as employed in the astrophysics community}

The standard formalism \cite{Blandford:1986,Schneider:1985,Schneider92}, which is actually used in virtually all current observations, is based on the calculation of the deflection angle $\Delta\cphi$ using the Schwarzschild metric of a point mass $M$, as derived in \eref{tot_deflect_angle_weinberg} or equivalently \eref{tot_deflect_angle_refindex}:
\begin{equation}
\vec{\hat\alpha} \approx \frac{4M}{\xi}\,\frac{\vec{\xi}}{\xi} \, ,
\end{equation} 
where $\vec{\hat\alpha}$ is now the total angle of deflection (previously $\Delta\cphi$) and $\vec{\xi}$ is the 2-dimensional position vector of the lensed image in the lens plane. Since the distance between the source, lens and the observer each are all larger than the region in which most of the lensing takes place ($\sim |\vec\xi| = b$), a thin screen approximation is generally employed. This means that all matter density, which is responsible for gravitational lensing, is projected into a 2-dimensional plane perpendicular to the line of sight. If one abandons the notion of a point mass and allows for a finite matter distribution, the surface density is given by the summation of all mass along a line of sight:
\begin{equation}
\Sigma(\vec\xi) \equiv \int \rho(\vec\xi, z)\, \d z \, ,
\end{equation} 
where $\Sigma(\vec\xi)$ is the surface density and $z$ is the space coordinate along the line of sight. The deflection angle induced by a small mass $\d M$ in a small area $\d^2\xi$ is then given by
\begin{equation}
\d\vec{\hat\alpha} \approx \frac{4\, \Sigma(\vec\xi)\,\d^2\xi}{\xi}\,\frac{\vec{\xi}}{\xi} \, ,
\end{equation} 
so that the total deflection angle of an image at position $\vec\xi$ is taken to be the superposition of all small deflections as they would occur by the isolated small masses $\d M$:
\begin{equation}
\label{eq:tot_deflect_angle_spread_matter}
\vec{\hat\alpha} \approx 4 \int \frac{(\vec{\xi}-\vec{\bar\xi})\,\Sigma (\vec{\bar\xi})}{ | \vec\xi - \vec{\bar\xi} |^2} \,\d^2\bar\xi \, .
\end{equation} 
This is the central formula, on which the whole formalism is based. Note that this formula is not restricted to spherical symmetry, but allows arbitrary matter distributions in the lens plane. \textit{Basically all inferences that have been made from observations of gravitational lensing events assume that this formula accurately describes gravitational lensing for weak gravitational fields.} Since superposition is a valid tool in linearized weak field gravity, there is no objection against this approach and many consistent observations seem to support the validity of \eref{tot_deflect_angle_spread_matter}. However, to interpret gravitational lensing results in a first post-Newtonian context, one needs to be aware that the use of \eref{tot_deflect_angle_spread_matter} implicitly assumes for a point mass
\begin{equation}
\Phi_\mathrm{lens}(r) = \Phi(r) = \int \frac{M}{r^2}\, \d r \, ,
\end{equation} 
which can be concluded from comparison of \eref{schwarzschild_refindex}, \eref{refindex_Newtonian_assumption} and \eref{lensing_potential}. When applying the previously introduced superposition, it is straightforward to see that the spherically symmetric Poisson equation for Newtonian gravity implies
\begin{equation}
\Delta\Phi_\mathrm{lens}(r) = 4\pi\,\rho_\mathrm{lens}(r) \quad \Rightarrow \quad
\Phi_\mathrm{lens}(r) = \int \frac{m_\mathrm{lens}(r)}{r^2}\, \d r \, .
\end{equation}
As I argued previously in \sref{rotcurve_interpretation}, this is equivalent to assuming that the pressure of the matter which generates the gravitational field vanishes or is negligible ($|p|\ll\rho$). If this assumption is not imposed a-priori, gravitational lensing oberservations will only yield information about $\Phi_\mathrm{lens}(r)$. The actual mass distribution $m(r)$ can then only be obtained from \eref{lensing_potential} when $\Phi(r)$ is known from a different kind of observation, e.g. rotation curves. Hence, $m_\mathrm{lens}(r)$ and $\rho_\mathrm{lens}(r)$ are the spherically symmetric matter and density distributions under the assumption that pressure does not contribute significantly to gravitational lensing.

Without derivation, I note the three basic equations of gravitational lensing, following from \eref{tot_deflect_angle_spread_matter}, that relate the source and image properties \cite{Blandford:1986,Kneib:2001,Schneider:1985}:
\begin{enumerate}
\item Arrival time:
\begin{equation}
\label{eq:lensing_timedelay}
\tau = \frac{1}{2}\, \left(\vec\xi_I - \vec\xi_S \right)^2 - \psi(\vec\xi_I) \, .
\end{equation} 
The appropriately scaled arrival time $\tau$ of an image at position $\vec\xi_I$ is defined by this equation. $\vec\xi_S$ is the position of the source of the image. The time delay between two images is given by the difference of their arrival time $\tau$. The lensing potential $\psi(\vec\xi_I)$ is the appropriately scaled 2-dimensional equivalent of the 3-d gravitational lensing potential:
\begin{equation}
\psi(\vec\xi_I) \propto \int \Phi_\mathrm{lens}(\vec\xi_I,z)\, \d z \propto 2 \int \Sigma(\vec{\bar\xi})\, \ln\left|\vec\xi_I - \vec{\bar\xi}\right|\, \d^2\bar\xi \, .
\end{equation}
\item Lens equation: 
\begin{equation}
\label{eq:lensing_lensequation}
\vec\xi_I = \vec\xi_S + \vec\nabla\psi(\vec\xi_I) \, .
\end{equation} 
The lens equation is equivalent to Fermat's principle of least arrival time, $\vec\nabla\tau=0$, and gives the position of the source $\vec\xi_S$ for one of its images at $\vec\xi_I$ if the lensing potential $\psi$ is known.
\item Magnification tensor:
\begin{equation}
\label{eq:lensing_magtensor}
M^{-1}_{ij} = \frac{\partial^2 \tau}{\partial\xi_{Ii}\, \partial\xi_{Ij}} = \delta_{ij} - \frac{\partial^2 \psi}{\partial\xi_{Ii}\, \partial\xi_{Ij}} = \frac{\partial\xi_{Si}}{\partial\xi_{Ij}} = \left[ \begin{array}{cc}
\kappa + \mu & 0 \\
0 & \kappa - \mu
\end{array} \right]
\end{equation} 
This equation defines the inverse of the magnification tensor $M_{ij}$ at the image position $\vec\xi_I$. The components of the magnification matrix are given in suitably rotated coordinates of the lens plane which diagonalize the matrix. The total magnification is given by $M = \left( \det M^{-1}_{ij} \right)^{-1} = \left(\kappa^2-\mu^2 \right)^{-1}$. $\kappa$ denotes the expansion of the image and $\mu$ denotes its shear.
\end{enumerate}

The arrival time $\tau$ and the lensing potential $\psi$ are suitably scaled to include all relevant cosmological information about the distances and expansion between the observer, lens plane and the source \cite{Schneider:1985}. The lensing potential is defined by the lens equation to satisfy
\begin{equation}
\vec\nabla\psi \equiv \vec{\hat\alpha} \, .
\end{equation} 
Therefore, a spherically symmetric mass distribution yields by \eref{tot_deflect_angle_spread_matter} \cite{Kneib:2001,Schneider92}
\begin{equation}
\partial_r\, \psi(r) = \frac{4 \, m_\mathrm{lens}(r)}{r} \, ,
\end{equation} 
where $m_\mathrm{lens}(r)$ is the inferred mass within a sphere of radius $r$ under the assumption that pressure in the matter which causes the bending of light is insignificant.

Before I give a brief overview over the use of these formulae in observational data analysis, I want to point out that these three equations depend on the lensing potential $\psi$ or its first and second derivatives, all of which are related directly to $\Phi_\mathrm{lens}$ and its first and second derivatives. 


\subsubsection{Strong Lensing}
\label{sec:strong_lensing}

The term \textit{strong lensing} is generally used to describe gravitational lensing that happens within $1$-$10\,\unit{kpc}$ of the lensing object, which usually leads to the occurrence of multiple images. Despite the name ``strong'' lensing, the gravitational fields involved are weak enough to allow the use of the linearized formalism that was introduced earlier. At present there about $70$ multiple image systems of galaxies or galaxy clusters known\footnote{For an up to date list see \url{http://www.cfa.harvard.edu/glensdata/}.} \cite{CASTLE:2005}, with more data being collected and therefore increasing this number.
Due to the small number of images, typically $2$-$4$, the number of observable parameters -- image positions, amplification and time delay -- is also very limited. Furthermore, conclusions about the lensing potential $\psi(\vec\xi_I)$ are only possible at the few image positions. Hence, lensing systems that have their multiple images at different radii are especially suitable for probing the potential $\Phi_\mathrm{lens}(r)$ or the mass profile $m_\mathrm{lens}(r)$. \cite{Kneib:2001}

The data from a strong lensing system is mostly fitted with some promising parametric model for the mass distribution, such as e.g. the singular isothermal sphere (which corresponds to a flat rotation curve for all radii), or the NFW halo \cite{Navarro:1996}. This procedure allows a reasonable mass estimate, despite the limited number of data points, but it assumes the shape of the mass profile $m_\mathrm{lens}(r)$ a-priori \cite{Kneib:2001}.

Recently, with the upcoming availability of high-resolution observations of lensing systems, non-parametric methods are being developed and employed \cite{Diego:2005a,Diego:2005b,Saha:2004,Williams:2004}. These methods allow the reconstruction of a ``pixelated'' density map of the surface density $\Sigma$ from the lens equation \eref{lensing_lensequation} and hence, are suitable for actually measuring the mass profile $m_\mathrm{lens}(r)$ without any prior assumptions about its shape.

While the time delay \eref{lensing_timedelay} between the lensed images was initially thought to be a suitable tool for measuring the Hubble parameter $H_0$ \cite{Kneib:2001,Kochanek:2004}, it turned out that the time delays are very sensitive to the mass distribution \cite{Oguri:2002} and estimated values for the Hubble parameter tend to be too low \cite{Kochanek:2004} when compared to cosmological measurements of the Hubble parameter (see \sref{comological_parameters}). Conversely however, if an accepted value for the Hubble parameter is assumed, time delay measurments yield important constraints on the mass distribution. Unfortunately, time delay mesurements are still very rare \cite{Kochanek:2004}, since extensive monitoring of the image flux is necessary to determine the time delay between the images which ranges from a few days to years.


\subsubsection{Weak Lensing}
\label{sec:weak_lensing}

\textit{Weak lensing} involves even weaker gravitational fields than strong lensing. Generally the images of source galaxies are lensed by the weak gravitational field of a foreground galaxy, at a distance of $\sim 10$-$1000\,\unit{kpc}$. Because the distance between lens and image is so much bigger than for strong lensing, the number of source/lens galaxy pairs for weak lensing is very large. On the downside this also means that there are no multiple images or significant distortion and hence, the signal of a single weak lensing event is practically not detectable. However, due to the abundant occurence of weak lensing events one can constrain the halo potential of an ``average galaxy'' by combining a statistically large sample \cite{Brainerd:2004}.

Since there are no multiple images, time delay is not accessible for weak lensing and the position of the image is also not measurably different from the source position. Hence, the only way to obtain a signal from weak lensing is by analyzing the small distortion of the images that turns spherical sources into ellipses that are tangentially aligned with respect to the vector that connects the lensed image with the center of the lensing potential well. A large number of lensed images then should draw up a coherent pattern of the gravitational potential in a galaxy field \cite{Brainerd:2004}. Of course the majority of lensed galaxies are not spherical in their unperturbed appearance, but already elliptical. The complex image ellipticity $\varepsilon$, often also called the ``image polarization'', contains the magnitude of the ellipticity of an image and its orientation. For a completely random distribution of unperturbed images of source galaxies, one would expect the ensemble average to vanish: $\left\langle \varepsilon_\mathrm{unper} \right\rangle = 0$. The weak lensing signal hides in the fact that the observed average ellipticity does not vanish: $\left\langle \varepsilon_\mathrm{obs} \right\rangle \neq 0$. Therefore, the extraction of the weak lensing signal from a large sample involves rigorous statictics which are explained in detail in e.g. \cite{Guzik:2002,Hoekstra:2004,Kleinheinrich:2003,Kleinheinrich:2005}.

Straightfoward averaging over all lensed images produces a picture of an ``average galaxy'' that is not necessarily characteristic of any individual galaxy. To understand the gravitational potential of galaxies in their diversity, one has to apply appropriate scaling laws in the weak lensing data analysis \cite{Kneib:2001}. The first critical scaling is the distance of the lens and the source galaxies. This means that ideally the data contains the redshift of all involved galaxies. If redshift is not available, distance statistics have to be applied which reduces the quality of the data \cite{Brainerd:2004}. The second scaling is to assume a universal density profile that is followed by all galaxies. This is again a parametric approach to determine the mass profile $m_\mathrm{lens}(r)$, as was already discussed for strong lensing \cite{Kneib:2001}. Hence, weak lensing does not yield an unbiased mass profile that is free of prior assumptions.

The last issue of weak lensing that I want to mention in this context is that weak lensing can only use \eref{lensing_magtensor} to determine the second derivative of the lensing potential $\psi$. Therefore, only the first derivative of the mass distribution can be inferred. The constant of integration that is necessary to obtain an absolute value for the mass profile $m_\mathrm{lens}(r)$ is not always unambiguous, and hence the mass profile can generally not be uniquely determined \cite{Kneib:2001}. Recent efforts to break this so called \textit{mass-sheet degeneracy} combine both weak and strong lensing techniques \cite{Bradac:2005}.


\subsection{Interpretation of gravitational lensing observations}

As I have outlined in \sref{refindex_spherical_symmetric}, gravitational lensing observations of a spherically symmetric weak gravitational field described by \eref{ss_metric} are entirely determined by the scalar refractive index
\begin{equation}
n(r) = 1 - 2 \Phi_\mathrm{lens}(r) \, .
\end{equation} 
Different types of lensing observations yield information about $\Phi_\mathrm{lens}(r)$ and/or its first and second derivative. $\Phi_\mathrm{lens}(r)$ is usually interpreted as the 
Newtonian gravitational potential which leads to the conclusion that the density and mass of a spherically symmetric object are given by
\begin{equation}
\rho_\mathrm{lens}(r) = \frac{1}{4\pi}\, \Delta\Phi_\mathrm{lens}(r) \, ; \qquad m_\mathrm{lens}(r) = r^2\, \frac{\d \Phi_\mathrm{lens}}{\d r} \, .
\end{equation} 
However, I already showed that if one does not make this assumption about $\Phi_\mathrm{lens}(r)$, but instead interprets it in terms of the metric functions $\Phi(r)$ and $m(r)$ for spherically symmetric weak gravitational fields, one finds the relation \eref{lensing_potential},
\begin{equation}
2\,\Phi_\mathrm{lens}(r) \approx \Phi(r) + \int \frac{m(r)}{r^2}\,\d r \, ,
\end{equation} 
or equivalently,
\begin{equation}
2\,m_\mathrm{lens}(r) \approx r^2\,\frac{\d \Phi}{\d r} + m(r) \, .
\end{equation}  
This relation, which handles non-negligible pressure in a static spherically symmetric weak gravitational field correctly, is the key equation for combining measurements of rotation curves with gravitational lensing.

%% file: figures/331_weinberg_deflection.pstex_t
\begin{picture}(0,0)%
\includegraphics{figures/331_weinberg_deflection.pstex}%
\end{picture}%
\setlength{\unitlength}{4144sp}%
\begingroup\makeatletter\ifx\SetFigFont\undefined%
\gdef\SetFigFont#1#2#3#4#5{%
  \reset@font\fontsize{#1}{#2pt}%
  \fontfamily{#3}\fontseries{#4}\fontshape{#5}%
  \selectfont}%
\fi\endgroup%
\begin{picture}(5874,2046)(-11,-1240)
\put(1298,-425){\makebox(0,0)[lb]{\smash{{\SetFigFont{12}{14.4}{\familydefault}{\mddefault}{\updefault}{\color[rgb]{0,0,0}$\Delta\cphi$}%
}}}}
\put(3913,517){\makebox(0,0)[lb]{\smash{{\SetFigFont{12}{14.4}{\familydefault}{\mddefault}{\updefault}{\color[rgb]{0,0,0}$\vec{r}$}%
}}}}
\put(4646,-30){\makebox(0,0)[lb]{\smash{{\SetFigFont{12}{14.4}{\familydefault}{\mddefault}{\updefault}{\color[rgb]{0,0,0}$\cphi(r)$}%
}}}}
\put(2551,209){\makebox(0,0)[lb]{\smash{{\SetFigFont{12}{14.4}{\familydefault}{\mddefault}{\updefault}{\color[rgb]{0,0,0}$r_\mathrm{min}$}%
}}}}
\put(5656,481){\makebox(0,0)[lb]{\smash{{\SetFigFont{12}{14.4}{\familydefault}{\mddefault}{\updefault}{\color[rgb]{0,0,0}$\cphi(\infty)$}%
}}}}
\end{picture}%

%% file: figures/332_refindex_trick.pstex_t
\begin{picture}(0,0)%
\includegraphics{figures/332_refindex_trick.pstex}%
\end{picture}%
\setlength{\unitlength}{4144sp}%
\begingroup\makeatletter\ifx\SetFigFont\undefined%
\gdef\SetFigFont#1#2#3#4#5{%
  \reset@font\fontsize{#1}{#2pt}%
  \fontfamily{#3}\fontseries{#4}\fontshape{#5}%
  \selectfont}%
\fi\endgroup%
\begin{picture}(5424,927)(-11,-305)
\put(3556,299){\makebox(0,0)[lb]{\smash{{\SetFigFont{12}{14.4}{\familydefault}{\mddefault}{\updefault}{\color[rgb]{0,0,0}$\vec{r}$}%
}}}}
\put(3376,-241){\makebox(0,0)[lb]{\smash{{\SetFigFont{12}{14.4}{\familydefault}{\mddefault}{\updefault}{\color[rgb]{0,0,0}$z$}%
}}}}
\put(2791,254){\makebox(0,0)[lb]{\smash{{\SetFigFont{12}{14.4}{\familydefault}{\mddefault}{\updefault}{\color[rgb]{0,0,0}$\vartheta$}%
}}}}
\put(2476,164){\makebox(0,0)[lb]{\smash{{\SetFigFont{12}{14.4}{\familydefault}{\mddefault}{\updefault}{\color[rgb]{0,0,0}$\vec{b}$}%
}}}}
\put(5176,-241){\makebox(0,0)[lb]{\smash{{\SetFigFont{12}{14.4}{\familydefault}{\mddefault}{\updefault}{\color[rgb]{0,0,0}$z$}%
}}}}
\end{picture}%

%% file: ts_galaxy_rotcurves+lensing.tex
\section[Combining rotation curves \& gravitational lensing]{Combined measurements of rotation curves and gravitational lensing}

The simplest metric that can be used as a realistic model for a galaxy is a static spherically symmetric spacetime as it is given by \eref{ss_metric}:
\begin{equation}
\label{eq:ss_metric_2}
\d s^{2}= 
-e^{2\Phi(r)}\, \d t^{2}
+\frac{\d r^{2}}{1-2m(r)/r}
+r^{2}\, \d\Omega^{2} \, .
\end{equation}
Naturally, this metric stops to be an appropriate description if the considered galaxy is too aspherical or not dominated by a spherical dark matter halo. For all other cases, the system is completely described by the two metric functions $\Phi(r)$ and $m(r)$.

As I have shown in \sref{rotcurve} and \sref{grav_lensing}, neither rotation curve nor gravitational lensing observations on their own can determine both functions uniquely. Moreover, in both cases it is necessary to make an assumption about the equation of state, $p=p(\rho)$, of the matter which generates the gravitational field. Without such an assumption, the data cannot be interpreted. The commonplace assumption is $p=0$ for which the interpretation of both kinds of observations takes the familiar Newtonian form of weak gravitational fields.

Along the way, I indicated that combined observations of rotation curves \textit{and} gravitational lensing eliminate the need for such an a priori assumption and furthermore, allow to deduct the equation of state -- or depending on the quality of the data, at least limits on it. I will now quantify how one can interpret measurements of combined rotation curve and gravitational lensing observations.

\subsection{Formalism combining rotation curves and lensing}

The relation between the gravitational potentials that are obtained by rotation curve or lensing observations, $\Phi_\mathrm{RC}$ and $\Phi_\mathrm{lens}$, and the metric functions $\Phi(r)$ and $m(r)$ has been derived previously in \sref{corr_redshift} and \eref{lensing_potential}:
\begin{eqnarray}
\label{eq:Phi_rotcurve}
\Phi_\mathrm{RC}(r) &=& \Phi(r) \, , \\
\label{eq:Phi_lens}
\Phi_\mathrm{lens}(r) &=& \frac{1}{2}\,\Phi(r) + \frac{1}{2}\, \int \frac{m(r)}{r^2}\, \d r \, .
\end{eqnarray}
In general, these potentials do not agree. Only in the Newtonian limit, where the pressure is negligible, are they identical: $\Phi_\mathrm{RC} = \Phi_\mathrm{lens} = \Phi_\mathrm{N}$.
Since this is the standard assumption for interpreting rotation curve and lensing data,
most observations are processed to report the mass profile instead of the gravitational potential -- which also avoids the problem of finding a suitable constant of integration for the potential. Therefore, it is useful to discuss the combined measurements of rotation curves and gravitational lensing in terms of the inferred mass distributions directly.
For both kinds of measurements, the usual assumption is that the Newtonian relation 
\begin{equation}
\label{eq:Newtonian_assumption}
m_\mathrm{RC/lens}(r) \equiv r^2\,\frac{\d\Phi_\mathrm{RC/lens}(r)}{\d r} = r^2\,\Phi'_\mathrm{RC/lens}(r)
\end{equation} 
defines the mass distribution in terms of the obtained potential. 
Under this assumption, the relations between the obtained ``pseudo-mass'' distributions and the metric functions are:
\begin{eqnarray}
\label{eq:mRC_Phi}\label{eq:m_rotcurve}
m_\mathrm{RC}(r) &=& r^2\,\Phi'(r) \\
\label{eq:mlens_Phi_m}\label{eq:m_lens}
m_\mathrm{lens}(r) &=& \frac{1}{2}\,r^2\,\Phi'(r) + \frac{1}{2}\,m(r) \, .
\end{eqnarray} 
These ``masses'' describe the observations equivalently to the potentials \eref{Phi_rotcurve} and \eref{Phi_lens}. The quotation marks around ``masses'' indicate that these profiles no not generally describe the total mass-energy within a sphere of radius $r$. Recalling \eref{pressure_field_eq},
\begin{equation}
\nabla^2 \Phi_\mathrm{RC}(r) = \nabla^2 \Phi(r) \approx 4\pi\,(\rho + p_r + 2p_t) \, ,
\end{equation} 
and the $tt$-Einstein equation \eref{Gtt},
\begin{equation}
\frac{m(r)}{r^2} = 4\pi\,\rho(r) \, ,
\end{equation} 
one can easily deduce the Poisson equations
\begin{eqnarray}
\nabla^2 \Phi_\mathrm{RC}(r) &\approx& 4\pi\,(\rho + p_r + 2p_t) \\
\label{eq:lens_potential_field_equation}
\nabla^2 \Phi_\mathrm{lens}(r) &\approx& 4\pi\,\left(\rho + \frac{1}{2}\,[p_r + 2p_t] \right) \, .
\end{eqnarray} 
The physical significance of the measureable ``mass'' distributions $m_\mathrm{RC}$ and $m_\mathrm{lens}$ becomes now clear by integrating these Poisson equations:
\begin{eqnarray}
\label{eq:pseudomass_rotcurve}
m_\mathrm{RC}(r) &\approx& 4\pi \int_0^r \left[ \rho(\rr) + p_r(\rr) + 2p_t(\rr) \right]\,\rr^2\,\d\rr \\
\label{eq:pseudomass_lens}
m_\mathrm{lens}(r) &\approx& 4\pi \int_0^r \left[ \rho(\rr) + \frac{1}{2}\,p_r(\rr) + p_t(\rr) \right]\,\rr^2\,\d\rr \, .
\end{eqnarray} 
It is now evident that the Newtonian assumption \eref{Newtonian_assumption} only leads to its intended physical interpretation as total mass-energy within a sphere of radius $r$ when the sum of the principal pressures vanishes:
\begin{equation}
m(r) \approx m_\mathrm{RC}(r) \approx m_\mathrm{lens}(r) \quad \Longleftrightarrow \quad p_r(r)+2\,p_t(r) \approx 0 \, .
\end{equation} 
Therefore, I shall refer to these ``masses'', as they are determined by the established rotation curve and lensing formalisms, as \emph{pseudo-masses}.
Inverting \eref{mRC_Phi} and \eref{mlens_Phi_m} determines the metric functions in terms of these observed pseudo-masses:
\begin{eqnarray}
\label{eq:Phi_mRC}
\Phi'(r) &=& \frac{m_\mathrm{RC}(r)}{r^2} \\
\label{eq:m_mlens_mRC}
m(r) &=& 2\,m_\mathrm{lens}(r) - m_\mathrm{RC}(r) \, .
\end{eqnarray} 
The gravitational potential $\Phi(r)$ can be obtained by integration under suitable boundary conditions. However, only the derivatives of $\Phi(r)$ contribute to the field equations and hence to the density and pressure profiles. The exact Einstein equations \eref{Gtt}-\eref{Gthetaphi} for a spherically symmetric stress-energy tensor of the form \eref{set_ss_fluid} are:
\begin{eqnarray}
\label{eq:Einstein_tt}
8\pi\,\rho(r) &=& \frac{2\, m'(r)}{r^2} \\
8\pi\,p_r(r) &=& - \frac{2}{r^2}\, \left[ \frac{m(r)}{r} - r\,\Phi'(r)\left( 1-\frac{2\,m(r)}{r} \right)  \right]  \\
\nonumber
8\pi\,p_t(r) &=& - \frac{1}{r^3}\,\left[ m'(r)\,r-m(r) \right] \left[1+r\,\Phi'(r)\right] + \\
\label{eq:Einstein_transverse}
&~&\qquad + \left( 1-\frac{2\,m(r)}{r} \right) \left[ \frac{\Phi'(r)}{r} + \Phi'(r)^2 +\Phi''(r) \right]  \, .
\end{eqnarray}
Inserting \eref{Phi_mRC} and \eref{m_mlens_mRC} and their derivatives gives the density and pressure profiles of the galaxy in terms of the observed pseudo-mass functions $m_\mathrm{RC}(r)$ and $m_\mathrm{lens}(r)$:
\begin{eqnarray}
\label{eq:Einstein_approx_tt}
4\pi\,r^2 \rho(r) &=& 2\,m_\mathrm{lens}'(r) - m_\mathrm{RC}'(r) \, , \\
4\pi\,r^2 p_r(r) &=& 2\, \frac{m_\mathrm{RC}(r)-m_\mathrm{lens}(r)}{r} + \O{\left( \frac{2\,m}{r} \right)^2} \, , \\
\label{eq:Einstein_approx_transverse}
4\pi\,r^2 p_t(r) &=& r\,\left[ \frac{m_\mathrm{RC}(r)-m_\mathrm{lens}(r)}{r} \right]' 
 + \O{\left( \frac{2\,m}{r} \right)^2} \\
&=& \frac{r}{2}\,\left[4\pi\,r^2 p_r(r)\right]'  + \O{\left( \frac{2\,m}{r} \right)^2} \, .
\end{eqnarray}
As consistency checks, I note that:
\begin{itemize}
\item The Einstein equations in curvature coordinates, \eref{Einstein_approx_tt}--\eref{Einstein_approx_transverse}, agree to the given order of $2m/r$ with the Einstein equations of the metric in isotropic coordinates \eref{ss_metric_isotropic_coords_refindex}, and therefore the approximation $\rr \approx r$ as of \eref{isotropic_radius_curvature_radius} is valid;

\item From the equations \eref{Einstein_approx_tt}--\eref{Einstein_approx_transverse} follows that
\begin{equation}
4\pi\,r^2 \left[ \rho(r) + p_r(r) + 2p_t(r) \right] = m_\mathrm{RC}'(r) + \O{\left( \frac{2\,m}{r} \right)^2} \, ,
\end{equation} 
and
\begin{equation}
4\pi\,r^2 \left[ \rho(r) + \frac{1}{2} \left(p_r(r) + 2p_t(r)\right) \right] = m_\mathrm{lens}'(r) + \O{\left( \frac{2\,m}{r} \right)^2} \, ,
\end{equation} 
and thus these results are consistent with the weak field approximation of the field equations \eref{pressure_field_eq}, and the interpretations of the pseudo-masses \eref{pseudomass_rotcurve} and \eref{pseudomass_lens};

\item Should the observed mass profiles agree with each other, i.e. $m = m_\mathrm{RC} = m_\mathrm{lens}$, the density and pressure profiles yield the desired result of the Newtonian limit:
\begin{eqnarray}
4\pi\,r^2 \rho(r) &=& m'(r) \, , \\
4\pi\,r^2 p_r(r) &=&  \O{\left( \frac{2\,m}{r} \right)^2} \, ,  \\
4\pi\,r^2 p_t(r) &=&  \O{\left( \frac{2\,m}{r} \right)^2} \, .
\end{eqnarray}
\end{itemize}
I conclude that the currently existing formalisms for analysing data from rotation curve and gravitational lensing observations can be used to obtain the pseudo-masses $m_\mathrm{RC}(r)$ and $m_\mathrm{lens}(r)$, which by \eref{Einstein_approx_tt}--\eref{Einstein_approx_transverse} yield the density and pressure profiles in a first post-Newtonian approximation. Furthermore, from the combination
\begin{equation}
4\pi\,r^2\,(p_r+2p_t) \approx 2\,(m_\mathrm{RC}'-m_\mathrm{lens}') \, ,
\end{equation} 
one can immediately infer that the observed system is Newtonian, in the sense of negligible pressure content, if and only if $m_\mathrm{RC}'(r) \approx m_\mathrm{lens}'(r)$. Furthermore, defining the dimensionless quantity
\begin{equation}
w(r) = \frac{p_r(r)+2p_t(r)}{3\rho(r)} \approx \frac{2}{3}\,\frac{m_\mathrm{RC}'(r)-m_\mathrm{lens}'(r)}{2\,m_\mathrm{lens}'(r)-m_\mathrm{RC}'(r)}
\end{equation} 
gives a convenient parameter that determines a ``measure'' of the equation of state.

\subsection{Observational situation}
\label{sec:rotcurve+lensing_observations}

The post-Newtonian formalism I have outlined requires the simultaneous measurement of (pseudo-)density profiles from rotation curve and gravitational lensing observations.

While in principle these profiles do not have to be of the same galaxy, they must be comparable in the sense that they accurately describe ``similar'' galaxies. For example, weak lensing measurements (\sref{weak_lensing}) can be used to statistically infer the (pseudo-)density profile of an ``average'' galaxy \cite{Brainerd:2004}. At the same time, analysing the dynamics of satellite galaxies gives the rotation curve and thus, the corresponding pseudo-density profile, of another ``average'' galaxy \cite{Brainerd:2004}. Whether these two ``average'' galaxies are comparable or not depends on many factors, such as e.g. the distribution of galaxy morphologies in both samples, the statistical noise, the employed models for the (pseudo-)density distribution, etc. These statistical issues render the fast-growing collection of weak lensing data problematic for applying the formalism presented.

On the other hand, combined simultaneous measurements of rotation curves and lensing of individual galaxies are extremely well suited for our formalism. However, while there is a large number of individual rotation curves available ($> 100,000$; \cite{Sofue:2001}), the number of individual ``strong'' lensing systems with multiple images is rather limited ($\sim 70$; \cite{CASTLE:2005}\footnote{For an up to date list see \url{http://www.cfa.harvard.edu/glensdata/}.}). Combined observations are further aggravated by the differing distance scales: Most high quality rotation curves are naturally available for galaxies with a low to intermediate distance, corresponding to a redshift of up to $z\sim 0.4$ \cite{Sofue:2001}, while gravitational lenses are easier to detect at intermediate to high redshifts of $z \gtrsim 0.4$ \cite{CASTLE:2005}, since the image separation scales increasingly with the redshift of the lensing galaxy \cite{Schneider:1985,Kochanek:2004}. Therefore, even for nearby galaxies with existing combined measurements of kinematics and lensing (e.g. 2237+0305 at $z\approx 0.039$ and ESO 325-G004 at $z\approx 0.035$), the lensing data is resticted to the core region, while the rotation curve is only described by few data points in the outer region of the lens galaxy \cite{Barnes:1999,Smith:2005,Trott:2002}. Consequently, the inferred pseudo-mass profiles are available for the same galaxy, but unfortunately at different radii and therefore not comparable.

Although the observational situation makes it currently difficult to employ the formalism presented, the situation is likely to improve in the future when observations with a higher resolution will be carried out -- preferably with an emphasis on obtaining high-resolution rotation curves for lensing galaxies that exhibit lensed images at different radii.

For instance, detailed observation of the recently discovered closest known strong lensing galaxy ESO 325-G004 \cite{Smith:2005} could provide satisfactory data to allow the decomposition of density and pressure of the galactic fluid, as outlined in this article. The system consists of an isolated lensing galaxy at redshift $z\approx 0.035$ with an effective radius of $R_\mathrm{eff}=12''5$ and arc-shaped images of the background object at $R\approx 3''$, and possible arc candidates at $R\approx 9''$. Smith \emph{et al.} intend to collect more detailed data \cite{Smith:2005} that hopefully will include extended stellar dymanics and hence, allow for a direct comparison of the rotation curve and lensing data, if the arc candidates at $R\approx 9''$ turn out to contribute to the measurements.

\subsection{Conclusions}
\label{sec:rotcurve+lensing_conclusions}

In this chapter, I have argued that the standard formalism of rotation curve measurements and gravitational lensing make an \textit{a priori} Newtonian assumption that is based on the CDM paradigm. I introduced a post-Newtonian formalism that does not rely on such an assumption, and furthermore allows one to deduce the density- and pressure-profiles in a general relativistic framework. In this framework, rotation curve measurements provide a pseudo-mass profile $m_\mathrm{RC}(r)$ and gravitational lensing observations yield a different pseudo-mass profile $m_\mathrm{lens}(r)$. Combining both pseudo-masses allows one to draw conclusions about the density- and pressure profiles\footnote{These formulae are given in SI units, hence the factor of $c^2$.} in the lensing galaxy,
\begin{eqnarray}
\rho(r) &=& \frac{1}{4\pi\,r^2} \, \left[ 2\,m_\mathrm{lens}'(r) - m_\mathrm{RC}'(r) \right] \, , \\
p_r(r)+2p_t(r) &\approx& \frac{2\,c^2}{4\pi\,r^2} \, \left[m_\mathrm{RC}'(r)-m_\mathrm{lens}'(r) \right] \, . 
\end{eqnarray}
In the case of absent or negligible pressure, this could be used to observationally confirm the CDM paradigm of a pressureless galactic fluid. Conversely, if significant pressure is detected, a decomposition of the galaxy morphology would allow new insight into the equation of state of dark matter.

Since the formalism presented is based on a first-order weak field approximation, I suggest that to confirm the findings, one should re-insert the obtained density and pressure profiles into the metric \eref{ss_metric_2}. The actual observed quantities can then be extracted numerically for comparison from the exact field equations \eref{Einstein_tt}--\eref{Einstein_transverse} and the geodesic equations.

Even though data might not yet be available to constrain the dark matter equation of state noticably, one should note that the possibility of non-negligible pressure in the galactic fluid introduces a new free parameter into the analysis of combined rotation curve and lensing observations. Therefore, the approach presented here might also help to shed some light on prevailing problems that arise when combining rotation curve and lensing observations. For example, an unresolved issue exists when measuring the Hubble constant from the time delay between gravitationally lensed images: Using the standard models for matter distribution in the lens galaxy, the resulting Hubble constant is either too low compared to its value from other observations, or the dark matter halo must be excluded from the galaxy model to obtain the commonly accepted value of $H_0$ \cite{Kochanek:2004}. 
A possible explanation of this trend might lie in a disregarded pressure component of the dark matter halo.

Finally, I want to emphasize that the key point of this new formalism is that in general relativity, density and pressure \textit{both} contribute to generating the gravitational field. Furthermore, the perception of this gravitational field depends on the velocity of probe particles. These effects become especially important when one compares rotation curve and gravitational lensing measurements, where the probe particles are fundamentally different: interstellar gas or stars at subluminal velocities for rotation curves, and photons which travel at the speed of light for lensing measurements. The presented formalism accounts for these crucial differences between the probe particles, and relates observations of both kinds to the the density and pressure profile of the host galaxy. Although only static spherically symmetric galaxies are considered in this thesis in a first post-Newtonian approximation, the basic concept is fundamental and can be extended to more general systems with less symmetry. A suitable framework for considering most exotic weak gravity scenarios is provided by the effective refractive index tensor, as introduced by Boonserm \emph{et al.} \cite{Boonserm:2005}, see appendix \ref{sec:refindex}.

%% file: tc_compact_objects.tex
\chapter{Compact spherical objects}
\label{sec:compact_objects}

After using the spherically symmetric spacetime \eref{ss_metric} to investigate general relativistic effects in galaxies, I now want to focus on compact objects with an extremely high density. While the necessity of general relativistic effects for galaxies depends on the pressure content in the galactic fluid, it is certain that the strong gravitational fields of compact objects cannot be described adequately by Newtonian gravity.

\section{Compact stars and black holes}

Today, basically three types of compact objects that require general relativity for their description are well established in the astrophysics community \cite[\S 24]{MTW}:
\begin{description}
\item[White dwarfs] are remnants of not so massive stars that used up their nuclear fuel. They no longer have a source of energy and radiate away the remaining thermal energy. The pressure gradient that holds up a white dwarf arises from the degeneracy pressure of electrons. Their maximum mass is given by the Chandrasekhar limit of $\sim 1.4\,\Msun$ \cite{Chandra:1939}. General relativity does not have a significant influence on the overall stellar structure, but is necessary to describe the pulsation frequencies accurately.
\item[Neutron stars] are remnants of more massive stars that are held up by the degeneracy pressure of neutrons. Therefore they have approximately the density of atomic nuclei. General relativity is necessary for the appropriate description of the stellar structure, due to the strong gravitational fields inside the neutron star.
\item[Black holes] are the only possible configuration in general relativity for an object composed of ordinary matter, once it becomes more compact than the Buchdahl-Bondi bound \cite{Martin:2003za} of $2M/R < 8/9$. They exhibit a curvature sigularity at the center and an event horizon at the Schwarzschild radius of $R_S=2M$. Rotating and charged black holes have a few more peculiar properties that shall not be discussed in this work. Black holes do not exist in Newtonian gravity and hence, general relativity is indispensible for their description.
\end{description}

While white dwarfs and neutron stars only require general relativistic corrections, I now want to concentrate on black holes which are only conceivable in general relativity. Static black holes (without electric charge) are generally described by the Schwarzschild metric
\begin{equation}
\d s^2 = - \left(1-\frac{2M}{r}\right)\,\d t^2 + \frac{\d r^2}{1-2M/r} + r^2\,\d\Omega^2 \, ,
\end{equation} 
which was the first solution of Einstein's field equations \eref{Einstein_eqns} to be discovered by Karl Schwarzschild in 1916 \cite{Schwarzschild:1916}. Its mathematical simplicity is also its physical curse: The solution exhibits a curvature singularity at the center where $r=0$, which is considered to be a physically pathologic feature that hints that general relativity alone might not be enough to describe the physics of such strong gravitational fields in such a small region. Maybe a unified theory of gravity and quantum mechanics will provide the necessary solution. Until such a solution is found, physicists are happy to accept that the pathological singularity is ``hidden'' behind the event horizon at $R_S=2M$, which acts as a ``one-way membrane'' from which not even light can escape, and anything that passes this horizon is lost in the black hole without hope of recovery\footnote{Attempts to combine black hole physics and quantum theory have led to models in which black holes emit ``Hawking radiation'' \cite{Birrell82,Hawking:1974} and maybe even information that was previously thought to be lost forever \cite{Hawking:2005}.}.

Other than the central singularity, which cannot be detected in principle, the event horizon is the only directly detectable feature of a black hole. The indirect observational evidence for black holes -- through the gravitational fields surrounding it -- has recently become very convincing, as I already outlined in \sref{galaxy_anatomy}. However, until there is direct evidence of the event horizon, the indirect observation only tells us that there is a very compact object, and the only widely accepted object that fits this category is a black hole.

In this chapter, I will explore basic properties of mathematical alternatives to black holes. The use of the word ``mathematical'' indicates that the presented models satisfy all mathematical requirements of forming a compact object with a radius of $R \approx R_S$, while the physical reality of the underlying assumptions is currently as observationally unaccessible as the physical implications of the inside of a black hole.

\section{Gravastars}
\label{sec:gravastar}
One of the small number of serious challenges to the usual concept of black holes is the ``gravastar'' (\textit{gra}vitational \textit{va}cuum \textit{star}) model that was recently developed by Mazur and Mottola \cite{Mazur:2004fk}. In the gravastar picture, or the very closely related quantum phase transition picture developed by Laughlin \emph{et al.}~\cite{Chapline:2000,Chapline:2002}, the quantum vacuum undergoes a phase transition at or near $R_S$ where the event horizon would have been expected to form. 

In the Mazur--Mottola model, a suitable segment of de Sitter space\footnote{De Sitter space has an equation of state $\rho = -p>0$, which is also identified as the equation of state of the cosmological constant and hence, the interior of a Mazur--Mottola gravastar is sometimes said to be filled with ``dark energy''.} is chosen for the interior of the compact object while the outer region of the gravastar consists of a (relatively thin) finite-thickness shell of stiff matter ($p=\rho$) that is in turn surrounded by Schwarzschild vacuum ($p=\rho=0$).
Apart from these three explicitly mentioned layers, the Mazur--Mottola model requires two additional infinitesimally-thin shells with surface densities $\sigma_\pm$, and surface tensions $\vartheta_\pm$, that compensate the discontinuities in the pressure profile and stabilize this 5-layer onion-like construction, effectively by introducing delta-function anisotropic pressures \cite{Mazur:2004fk}. Since infinitesimally thin shells are a mathematical abstraction, for physical reasons it is useful to minimize the use of thin shells, for example by attempting to replace the thin shells completely with a continuous layer of finite thickness, like in \cite{Cattoen:2005he}.

This section is a brief summary of the results that were found by Catto\"en, Visser and myself \cite{Cattoen:2005he}. The full article can be reviewed in appendix \ref{sec:appgravastar}.

\subsection{Generalized model without thin shells}

In the article \cite{Cattoen:2005he} we demonstrate that the pressure anisotropy implicit in the Mazur--Mottola infinitesimally thin shell is not an accident, but instead a \emph{necessity} for all gravastar-like objects. That is, attempting to build a gravastar completely out of perfect fluid will always fail. (Either the gravastar swells up to infinite size, or a horizon will form despite one's best efforts, or worse a naked singularity will manifest itself.)  We derive this result by working with configurations where pressure is assumed continuous and differentiable, and analyzing the resulting static geometry using first the \emph{isotropic} TOV equation \eref{TOV_iso}, and then the \emph{anisotropic} TOV equation \eref{TOV_aniso}.


To have a useful model, we should retain as much of standard physics as possible,  while introducing a minimum of ``new physics''. In the spirit of Mazur and Mottola \cite{Mazur:2004fk}, and Laughlin \emph{et al.}~\cite{Chapline:2000,Chapline:2002}, we will keep the density positive throughout the configuration but permit the pressure to become negative in the gravastar interior. To avoid infinitesimally thin shells one must then demand that the radial pressure $p_r$ is continuous (though the density need not be continuous, and typically is not continuous at the surface of the gravastar). Qualitatively the radial pressure is taken to be that of \fref{grava_pressure}.

\begin{figure}[htb]
\begin{center}
\input{figures/gravastar_like_object3.pstex_t}
\end{center}
\caption[Qualitative sketch of a gravastar, explicitly labelling the ``core'', ``crust'', and ``atmosphere''.]{\label{fig:grava_pressure}
Qualitative sketch of a gravastar, explicitly labelling the ``core'', ``crust'', and ``atmosphere''. $r_g$ refers to the radius at which the local gravitational acceleration $g=\frac{4\pi}{3}\,r\, (\rhoavg+3p_r)/(1-2m/r)$ vanishes; for details see \sref{gravastar_metric}.}
\end{figure}
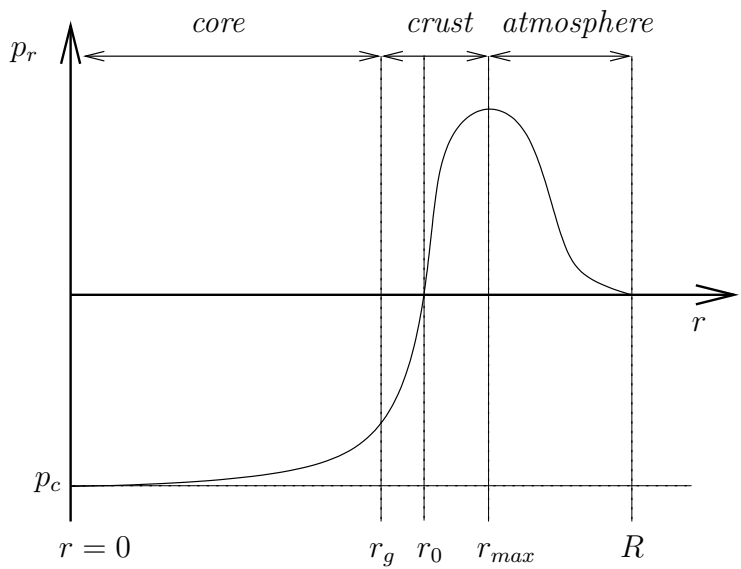

That is, to ``smooth out''  the infinitesimally thin shells of the Mazur--Mottola gravastar model, we shall consider static spherically symmetric geometries such that:
\begin{itemize}
\item Inside the gravastar, $ r < R$, the density is everywhere positive and finite.

\item The  central pressure is negative, $p_c<0$, and in fact $p_c = - \rho_c$.
\\
(We do \emph{not} demand $\rho=-p_r=-p_t$ except at the centre.)

\item The spacetime is assumed to \emph{not} possess an event horizon. \\
This implies that $\forall r$ we have $2m(r)<r$.

\end{itemize}
These three features, positive density, negative central pressure, and the absence of horizons, are the three most important features characterizing a gravastar. They also ensure that a gravastar does not exhibit the pathological central sinuglarity and the event horizon of a black hole. Instead, the spacetime metric of a gravastar is well-defined everywhere.

\subsection{Bounds on the pressure anisotropy}

It is straightforward to demonstrate that a gravastar-like object with negative central pressure cannot consist entirely of a perfect fluid. First, I rewrite the anisotropic TOV equation \eref{TOV_aniso} as
\begin{equation}
\label{eq:grava_TOV2}
\frac{\d p_r}{\d r}  = -\frac{4\pi\,r}{3}\, \frac{(\rho+p_r)\;(\rhoavg+3p_r)}{1-2m(r)/r} 
+{2\;\rho\;\Delta\over r} \, ,
\end{equation}
where I made use of the average density $\rhoavg(r)\equiv m(r)/(\frac{4}{3}\pi r^3)$.
The anisotropy parameter $\Delta$ is defined as
\begin{equation}
\Delta \equiv {p_t - p_r \over \rho} \, ,
\end{equation}
and hence, for $\Delta=0$, \eref{grava_TOV2} represents the isotropic TOV equation.

To realize the failure of a perfect fluid in a gravastar, note that the maximum pressure in an ordinary spherical compact object occurs at $r=0$, where $\d p_r/\d r=0$ naturally for $\Delta=0$. In a gravastar, this extremum is a pressure minimum and the maximum pressure occurs at a certain radius $r_\mathrm{max} \neq 0$, see \fref{grava_pressure}. Therefore, at $r_\mathrm{max}$, the first term of the right hand side of \eref{grava_TOV2} is negative for a gravastar, while the left hand side vanishes by the definition of a pressure maximum. Consequently, $\Delta$ must be positive, i.e. anisotropic pressure is required to allow a pressure maximum for $r>0$. A more detailed discussion of the region in which perfect fluids fail to deliver a valid solution can be found in appendix \ref{sec:iso_fail}.

Once we accept that perfect fluid spheres are not what we are looking for to model gravastars, one might wonder what happens to the Buchdahl--Bondi bound for isotropic fluid spheres. It has been shown that for $\rho' < 0$ and $p_t \leq p_r$ the $8/9$ bound still holds. However if the transverse stress is allowed to exceed the radial stress, $p_t > p_r$, then the upper limit shifts to $2M/R < \kappa \leq 1$, where $\kappa$ depends on the magnitude of the maximal stress anisotropy \cite{Guven:1999wm}.
In the gravastar picture, I already showed that $\Delta>0$ and hence, $p_t > p_r$. Therefore, the compactness of a gravastar is not limited by the Buchdahl--Bondi bound, but only by the magnitude of the maximal pressure anisotropy and the regularity of the metric, i.e. $2m(r)/r < 1$. Let us now make these qualitative considerations more quantitative.
To do this, let us
rewrite \eref{grava_TOV2} in the form
\begin{equation}
\Delta \equiv \frac{p_t - p_r}{\rho} = \frac{r}{2} \left[ \frac{p_r'}{\rho} +
\left(1+\frac{p_r}{\rho} \right)\, \frac{m+4\pi p_r\, r^3}{r^2
\left[ 1-2m/r \right] } \right]
\, .
\end{equation}
In appendix \ref{sec:iso_fail} we have determined the smallest interval in radii for which
anisotropic pressure is necessary to be $(r_g,r_\mathrm{max}]$. We can now ask for explicit bounds on
$\Delta$ in that interval. 

For the interval $r\in [r_0,r_\mathrm{max}]$,  by making use of $p_r' \geq 0$ and $p_r \geq 0$, we find the simple lower bound
\begin{equation}
\Delta \geq \frac{1}{4}\, \frac{2m/r}{1-2m/r} > 0.
\end{equation}
Now in the region $[r_0,r_\mathrm{max}]$ we have $p_t>p_r\geq0$, consequently if the dominant energy condition [DEC; $\rho\geq0$ and $|p_i|\leq \rho$] is to be satisfied we must at the very least have $\Delta \leq 1$.  But this is guaranteed to be violated whenever $2m/r>4/5$. 

That is: If the gravastar is sufficiently close to forming a horizon, in the sense that $2m/r > 4/5$ somewhere in the range $[r_0,r_\mathrm{max}]$, then the DEC must also be violated at this point.\footnote{And even if we were to discard the entire positive-pressure region $(r_0,R)$, we can nevertheless still apply this bound at $r_0$ itself: If $2m(r_0)/r_0>4/5$ then the DEC is violated at $r_0$.} Consequently, any gravastar that is sufficiently close to forming a horizon will violate the DEC in its ``crust''.

For the interval
$r\in (r_g,r_0)$ we find the considerably weaker bounds
\begin{equation}
0 \leq \frac{r\,p_r'}{2\,\rho} < \Delta < \frac{r\,p_r'}{2\,\rho}
+ \frac{1}{4}\, \frac{2m/r}{1-2m/r}
\end{equation}
where we have used $p_r' \geq 0$, $p_r < 0$, and the null energy condition [NEC; $\rho+p_i\geq 0$].

\subsection{Minimizing the anisotropic region}

Let us now attempt to minimize the region over which anisotropy is present. It is easy to see that at $r_g$ we have
\begin{equation}
\Delta(r_g) = \frac{r_g\,p_r'}{2\,\rho} \geq 0 \, .
\end{equation}
Therefore, in the case where the anisotropy is confined to the
smallest interval possible, we want $p_r'(r_g) = 0$, corresponding to an inflexion point for the radial pressure. At the point $r_0$ of zero
radial pressure, the anisotropy \emph{cannot} vanish:
\begin{equation}
\Delta(r_0) = \frac{r_0\,p_r'}{2\,\rho} + \frac{1}{4}\,
\frac{2m/r_0}{1-2m/r_0} > 0 \, .
\end{equation}
At the point of maximal radial pressure, the anisotropy also has
to be non-zero, at least if we take the limit from below:
\begin{equation}
\Delta(r_\mathrm{max}^-) = \frac{1}{4}\, \frac{2m/r_\mathrm{max}}{1-2m/r_\mathrm{max}}
\left(1+\frac{p_r}{\rho} \right)\, \left(1+\frac{4\pi p_r\, r_\mathrm{max}^3}{m
} \right) \, > 0 \, .
\end{equation}
Beyond the peak $\forall r > r_\mathrm{max}$,  it is possible to arrange $\Delta = 0$, though at a price:  If
we wish to confine the anisotropy to the smallest interval
possible, we have to set $\Delta(r \rightarrow r_\mathrm{max}^+) = 0$
which leads to a discontinuity in $p_r'$ and $\Delta$ as well as a
``kink'' in the pressure profile $p_r$ at $r_\mathrm{max}$. (However, $p_r$ itself is still continuous, as is the density $\rho$.) Indeed we then have
\begin{equation}
p_r'(r_\mathrm{max}^+) = 
{2 \rho_\mathrm{max}\over r_\mathrm{max}} \; \Delta(r_\mathrm{max}^-).
\end{equation}
The
implications of confining the pressure anisotropy to the smallest
interval possible are shown in \fref{gravastar_miniiso}, where the anisotropy is confined to the region $r\in(r_g,r_\mathrm{max}]$.

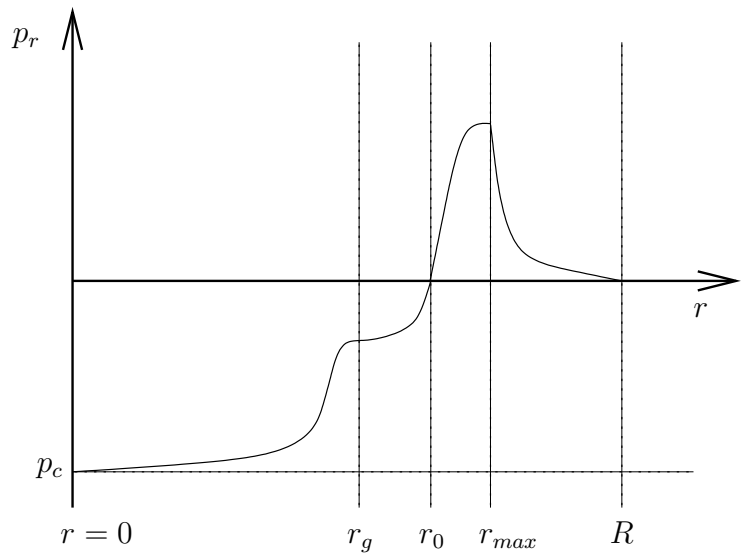
\begin{figure}[htb]
\begin{center}
\input{figures/gravastar_like_object2.pstex_t}
\end{center}
\caption[Qualitative sketch of radial pressure as a function of $r$ for a gravastar that is as near as possible to a perfect fluid.]{\label{fig:gravastar_miniiso}
Qualitative sketch of radial pressure as a function of $r$ for a gravastar that is as near as possible to a perfect fluid. Note the inflexion point at $r_g$ and the kink at $r_\mathrm{max}$.}
\end{figure}

The Mazur--Mottola model is now recovered as the limiting case where
$r_g\to r_0\leftarrow r_\mathrm{max}$, and so all the important anisotropy is confined in their inner
thin shell. The anisotropy
$\Delta\to\infty$, because $p'\to\infty$. Effectively $\Delta$ is replaced by 
choosing an appropriate finite (in
this case negative) surface tension $\vartheta$ and surface energy
density $\sigma$ which is given by the Israel--Lanczos--Sen 
junction conditions \cite{Barrabes:1991,Israel:1966,Lanczos:1924,Sen:1924,visser04}. 

The second, outer thin shell which is present in the Mazur--Mottola model
is \emph{not} a physical necessity, but is merely a convenient
way to avoid an infinitely diffuse atmosphere that would
arise otherwise from the equation of state $p=\rho$. A finite
surface radius $R$ can be modelled by altering the equation of
state slightly to include a finite surface density $\rho_S$ which is
reached for vanishing pressure: $\rho(p)=\rho_S+p$. Then, the
outer thin shell can be omitted when joining the gravastar metric
onto the Schwarzschild exterior metric.

\subsection{Gravastars vs. black holes}

The article \cite{Cattoen:2005he} discusses the results outlined above in more detail. We explored the generalization of the Mazur--Mottola gravastar and found that any object that binds all its mass within a radius of $R\approx R_S = 2M$ with $2M/R > 8/9$ must have anisotropic pressure in certain regions, i.e. $p_t>p_r$. Therefore, the Buchdahl-Bondi bound does not apply, and the compactness is only bound by the absence of an event horizon, i.e. $2m(r)/r < 1$.

Consequently, a black hole and a gravastar cannot be distinguished by their gravitational effects only. Only a direct measurement of the gravastar surface or equivalently, the black hole's event horizon, would provide evidence for one model or the other. While the event horizon is in principle detectable through the effects of being a ``one-way membrane'', i.e. the absence of bremsstrahlung from infalling particles, a prediction for measureable surface effects of a gravastar is not available as yet.

The gravastar model exists in a mathematical context, but there is no generally accepted physical explanation for a spherical object that exhibits the characteristic pressure profile of a gravastar.

\input{ts_anti_gravastar}

%% file: figures/gravastar_like_object3.pstex_t
\begin{picture}(0,0)%
\includegraphics{figures/gravastar_like_object3.pstex}%
\end{picture}%
\setlength{\unitlength}{3947sp}%
\begingroup\makeatletter\ifx\SetFigFont\undefined%
\gdef\SetFigFont#1#2#3#4#5{%
  \reset@font\fontsize{#1}{#2pt}%
  \fontfamily{#3}\fontseries{#4}\fontshape{#5}%
  \selectfont}%
\fi\endgroup%
\begin{picture}(4597,3502)(3226,-6869)
\put(6301,-3511){\makebox(0,0)[lb]{\smash{{\SetFigFont{12}{14.4}{\rmdefault}{\mddefault}{\itdefault}{\color[rgb]{0,0,0}atmosphere}%
}}}}
\put(7501,-5386){\makebox(0,0)[lb]{\smash{{\SetFigFont{12}{14.4}{\familydefault}{\mddefault}{\updefault}{\color[rgb]{0,0,0}$r$}%
}}}}
\put(3526,-6811){\makebox(0,0)[lb]{\smash{{\SetFigFont{12}{14.4}{\familydefault}{\mddefault}{\updefault}{\color[rgb]{0,0,0}$r=0$}%
}}}}
\put(6151,-6811){\makebox(0,0)[lb]{\smash{{\SetFigFont{12}{14.4}{\familydefault}{\mddefault}{\updefault}{\color[rgb]{0,0,0}$r_{max}$}%
}}}}
\put(7051,-6811){\makebox(0,0)[lb]{\smash{{\SetFigFont{12}{14.4}{\familydefault}{\mddefault}{\updefault}{\color[rgb]{0,0,0}$R$}%
}}}}
\put(3376,-6361){\makebox(0,0)[lb]{\smash{{\SetFigFont{12}{14.4}{\familydefault}{\mddefault}{\updefault}{\color[rgb]{0,0,0}$p_c$}%
}}}}
\put(3226,-3661){\makebox(0,0)[lb]{\smash{{\SetFigFont{12}{14.4}{\familydefault}{\mddefault}{\updefault}{\color[rgb]{0,0,0}$p_r$}%
}}}}
\put(5776,-6811){\makebox(0,0)[lb]{\smash{{\SetFigFont{12}{14.4}{\familydefault}{\mddefault}{\updefault}{\color[rgb]{0,0,0}$r_0$}%
}}}}
\put(5476,-6811){\makebox(0,0)[lb]{\smash{{\SetFigFont{12}{14.4}{\familydefault}{\mddefault}{\updefault}{\color[rgb]{0,0,0}$r_g$}%
}}}}
\put(4351,-3511){\makebox(0,0)[lb]{\smash{{\SetFigFont{12}{14.4}{\rmdefault}{\mddefault}{\itdefault}{\color[rgb]{0,0,0}core}%
}}}}
\put(5701,-3511){\makebox(0,0)[lb]{\smash{{\SetFigFont{12}{14.4}{\rmdefault}{\mddefault}{\itdefault}{\color[rgb]{0,0,0}crust}%
}}}}
\end{picture}%

%% file: figures/gravastar_like_object2.pstex_t
\begin{picture}(0,0)%
\includegraphics{figures/gravastar_like_object2.pstex}%
\end{picture}%
\setlength{\unitlength}{3947sp}%
\begingroup\makeatletter\ifx\SetFigFont\undefined%
\gdef\SetFigFont#1#2#3#4#5{%
  \reset@font\fontsize{#1}{#2pt}%
  \fontfamily{#3}\fontseries{#4}\fontshape{#5}%
  \selectfont}%
\fi\endgroup%
\begin{picture}(4597,3455)(3226,-6869)
\put(7501,-5386){\makebox(0,0)[lb]{\smash{{\SetFigFont{12}{14.4}{\familydefault}{\mddefault}{\updefault}{\color[rgb]{0,0,0}$r$}%
}}}}
\put(3526,-6811){\makebox(0,0)[lb]{\smash{{\SetFigFont{12}{14.4}{\familydefault}{\mddefault}{\updefault}{\color[rgb]{0,0,0}$r=0$}%
}}}}
\put(6151,-6811){\makebox(0,0)[lb]{\smash{{\SetFigFont{12}{14.4}{\familydefault}{\mddefault}{\updefault}{\color[rgb]{0,0,0}$r_{max}$}%
}}}}
\put(3376,-6361){\makebox(0,0)[lb]{\smash{{\SetFigFont{12}{14.4}{\familydefault}{\mddefault}{\updefault}{\color[rgb]{0,0,0}$p_c$}%
}}}}
\put(3226,-3661){\makebox(0,0)[lb]{\smash{{\SetFigFont{12}{14.4}{\familydefault}{\mddefault}{\updefault}{\color[rgb]{0,0,0}$p_r$}%
}}}}
\put(5776,-6811){\makebox(0,0)[lb]{\smash{{\SetFigFont{12}{14.4}{\familydefault}{\mddefault}{\updefault}{\color[rgb]{0,0,0}$r_0$}%
}}}}
\put(6976,-6811){\makebox(0,0)[lb]{\smash{{\SetFigFont{12}{14.4}{\familydefault}{\mddefault}{\updefault}{\color[rgb]{0,0,0}$R$}%
}}}}
\put(5326,-6811){\makebox(0,0)[lb]{\smash{{\SetFigFont{12}{14.4}{\familydefault}{\mddefault}{\updefault}{\color[rgb]{0,0,0}$r_{g}$}%
}}}}
\end{picture}%

%% file: ts_anti_gravastar.tex
\section{Anti-Gravastars}
\label{sec:anti-gravastar}

Following our findings that gravastars with negative pressure near the center must have a crust that consists of some sort of anisotopic fluid \cite{Cattoen:2005he}, I now investigate another possible configuration that departs from the conventional assumption that the density and pressure in a stellar object are both positive. I call an object of finite extent with positive pressure, $p>0$,  everywhere and negative mass-energy density at the center, $\rho_c<0$, an \emph{anti-gravastar}.

Such an object will, of course, have a negative total mass $M$ if the energy density $\rho$ is negative throughout the body of the anti-gravastar. Should objects with a net negative mass exist, they would be observable through an unusual microlensing-profile \cite{Cramer:1994qj} or a self-accelerating ``$+M$/$-M$-rocket'' \cite{Bondi57} - both of which have not yet been observed, and probably never will be. Since there is no observational evidence that such negative mass objects exist, we consider this type of anti-gravastar only in a mathematical fashion. 

There is, however, another type of anti-gravastar with a positive total mass $M>0$. It consists of a negative density core with a mass of $M_{core}=-M_-<0$ and a positive density crust with $M_{crust}=M_-+M>0$ (see \fref{anti_grav_pos_mass}). When matched onto a Schwarzschild exterior metric, these anti-gravastars could in theory mimic the gravitational field of a usual spherical stellar object.

Unlike the gravastar, we shall see that an anti-gravastar can consist entirely of a perfect fluid. To see that, I assume the anti-gravastar to consist of a perfect fluid and then find that the isotropic TOV equation satisfies the conditions that constitute an anti-gravastar.

\subsection{Anti-gravastar model}
\label{sec:antigravastar_defs}

The most important properties of a static, spherically symmetric anti-gravastar are:
\begin{itemize}
\item Inside the anti-gravastar, $r<R$, the pressure is positive and finite everywhere.
\item The central density is negative, $\rho_c<0$.
\item The induced spacetime does not have an event horizon, i.e. $\forall r: 2m(r)<r$.
\end{itemize}

To close the set of differential equations \eref{stat_spher_DES_iso} that determine a static spherically symmetric spacetime for a perfect fluid,
\begin{eqnarray}\label{eq:stat_spher_DES_iso}
m' &=& 4\pi \rho\, r^2 \, ; \\ \nonumber
p' &=& -\frac{(\rho+p)(m+4\pi p\, r^3)}{r^2 [1-2m(r)/r]} \, ,
\end{eqnarray} 
two initial conditions and a third equation that fixes the relation between $\rho$ and $p$ are necessary. For the initial conditions, I choose naturally
\begin{enumerate}
\item $m(0)=0$ since I want the anti-gravastar to have a regular origin and
\item $p(R)=0$ which defines the finite radius of the surface $R$ where I want to match a Schwarzschild exterior metric with $M=m(R)$. Equivalently, one can choose $p(0)=p_c>0$ and define the surface radius $R$ by the condition $p(R)=0$. If no surface can be defined this way, one needs to adjust the assumptions made.
\end{enumerate}
In the following sections, I will present two analytical geometries that represent a perfect fluid anti-gravastar and therefore prove their possible (mathematical) existence. These are the constant density Schwarzschild interior geometry and the less known Tolman IV solution \cite{Delgaty:1998uy}. Both geometries constitute anti-gravastars with a negative total mass. 

Although I wasn't able to find an analytical example of an anti-gravastar with a positive total mass, I will present the basic features of an equation of state that should lead to a perfect fluid anti-gravastar with a positive total mass. Such an object could in principle mimic the gravitational field of a black hole, just like the gravastar presented earlier.

\subsection{Schwarzschild interior anti-gravastar}

The Schwarzschild interior solution represents an object whose density is constant throughout its entire body. Due to its simple formulation, it is often used to illustrate basic physical concepts, although a constant density object is generally not thought to be an accurate description of astrophysical bodies.

In the context of anti-gravastars, I highlight the properties of the Schwarzschild interior solution, if one assumes negative constant density. The metric for the Schwarzschild interior solution is given by \cite{Delgaty:1998uy}
\begin{equation}
\d s^2 = - \left( A - B \sqrt{1-\frac{r^2}{C^2}} \right)^2 \d t^2 + \left( 1 - \frac{r^2}{C^2} \right)^{-1} \d r^2 + r^2 \d \Omega^2 \, .
\end{equation} 
Calculating the $tt$-component of the Einstein tensor, this leads to
\begin{equation}
\rho_* = \frac{3}{8\pi G_N C^2} > 0 \, .
\end{equation} 
Hence, to consider negative density I substitute the constant $C^2 \rightarrow -C^2$ which obviously will force negative density $\rho_*$. The resulting metric is:
\begin{equation}\label{eq:SSint_mod_metric}
\d s^2 = - \left( A - B \sqrt{1+\frac{r^2}{C^2}} \right)^2 \d t^2 + \left( 1 + \frac{r^2}{C^2} \right)^{-1} \d r^2 + r^2 \d \Omega^2
\end{equation} 
with a negative constant density
\begin{equation}
\rho_* = -\frac{3}{8\pi G_N C^2} < 0 \, .
\end{equation} 
The $rr$-component of the Einstein tensor yields the well-known density profile
\begin{equation}
\label{eq:SSint_pressure_profile}
p(r) = \left| \rho_* \right| \frac{\sqrt{1-2M/R}-\sqrt{1-2m(r)/r}}{3\sqrt{1-2M/R}-\sqrt{1-2m(r)/r}}
\end{equation} 
where I found that $A=3B\sqrt{1-2M/R}$, $B$ is arbitrary and $C=\pm R \sqrt{-R/2M}$.

Since all masses in \eref{SSint_pressure_profile} are negative, all square-roots exist. From $m(r)/r \propto r^2$, I conclude that
\begin{equation}
\forall r<R: \left| \frac{2m(r)}{r} \right| < \left| \frac{2M}{R} \right| \, .
\end{equation}
Therefore, both numerator and denominator of \eref{SSint_pressure_profile} are positive. The negative density $\rho_*$ appears in \eref{SSint_pressure_profile} with an accompanying minus-sign which has been replaced by the absolute value for clarity. Thus, it is ensured that the pressure profile is positive right up to the surface $R$. Hence, the modified Schwarzschild interior metric \eref{SSint_mod_metric} represents an anti-gravastar.

One should note that a brief calculation shows that
\begin{equation}
\forall r<R: \sqrt{1-2M/R} > \sqrt{1-2m(r)/r} \quad \Rightarrow \quad \rho_*+p(r)<0 \, .
\end{equation} 
Thus, the NEC is always violated.

\subsection{Properties of the Tolman IV solution}
\label{sec:tolmanIV}

The Tolman IV solution is another analytical geometry representing a spherically symmetric perfect fluid object. In its standard form, it is given by the metric \cite{Delgaty:1998uy}
\begin{equation}\label{eq:TolIV_metric}
\d s^2 = - B^2 \left( 1 + \frac{r^2}{A^2} \right)^2\, \d t^2 + \frac{1 + 2\, r^2/A^2}{\left( 1 - r^2/C^2 \right)\left( 1 + r^2/A^2 \right)}\, \d r^2 + r^2\, \d \Omega^2 \, .
\end{equation} 
The pressure and density profiles follow from the Einstein equations: 
\begin{eqnarray}\label{eq:TolIV_pressure}
p(r) &=& \frac{C^2-A^2 - 3r^2}{8\pi\, G_N\, C^2\, (A^2 + 2r^2)} \, , \\
\label{eq:TolIV_density}
\rho(r) &=& \frac{3A^2(A^2+C^2) + (7A^2+2C^2)r^2 + 6r^4}{8\pi\, G_N\, C^2\, (A^2 + 2r^2)^2} \, .
\end{eqnarray} 
If we assume the constants $A^2$ and $C^2$ to be positive, we immediately see that $\forall r: \rho>0$. From \eref{TolIV_pressure} we find that, if we define the surface by $p(R)=0$, the surface radius is given by  $R^2=\frac{1}{3}(C^2-A^2)$ and thus, if $R^2>0$, both pressure and density are positive for $r<R$. The parameter $B$ is irrelevant for the whole discussion -- it can easily be absorbed into a modified time-coordinate -- and hence, will be neglected from now on.

The expressions for the pressure and density simplify tremendously at the center for $r=0$,
\begin{equation}
p_c = \frac{C^2-A^2}{8\pi\, G_N\, C^2\, A^2} \, ; \quad \rho_c = \frac{3(A^2+C^2)}{8\pi\, G_N\, C^2\, A^2} \, ,
\end{equation}
and therefore it is possible to express the constants $A^2$ and $C^2$ in terms of the central pressure $p_c$ and the central density $\rho_c$:
\begin{equation}
A^2 = \frac{3}{4\pi\, G_N} \frac{1}{\rho_c+3p_c} \, ; \quad C^2 = \frac{3}{4\pi\, G_N} \frac{1}{\rho_c-3p_c} \, .
\end{equation} 
It is generally assumed that the parameters $A^2$ and $C^2$ are positive -- hence the square-notation -- but in general, both parameters can have negative values. To keep the notation sensible, this is realised by inverting the accompanying sign of the parameter in the metric and all other derived formulae. Therefore, I distinguish four cases:
\begin{enumerate}
\item $A^2>0, C^2>0 \quad \Leftrightarrow \quad \rho_c>3|p_c|>0$  \label{item:A+C+}
\item $A^2<0, C^2>0 \quad \Leftrightarrow \quad 3p_c<-|\rho_c|<0$ \label{item:A-C+} 
\item $A^2>0, C^2<0 \quad \Leftrightarrow \quad 3p_c>|\rho_c|>0$  \label{item:A+C-}
\item $A^2<0, C^2<0 \quad \Leftrightarrow \quad \rho_c<-3|p_c|<0$ \label{item:A-C-}
\end{enumerate}
In the following, I will discuss the properties of the Tolman IV geometry in all four cases and then show which one is suitable as an anti-gravastar.

\paragraph{Naked singularities.}

Probably the most obvious conclusion from \eref{TolIV_pressure} and \eref{TolIV_density} is that when $A^2<0$ there will be a pole in the pressure and density at $r^2=|A^2/2|$ which implies a naked curvature sigularity (see e.g. \sref{GS_ss_interior}) and therefore geometries of that kind should only be utilized in regions where $r^2 \neq |A^2/2|$. In short, the Tolman IV solution with $A^2<0$ cannot cover the entire spacetime without invoking a naked curvature singularity.

\paragraph{Existence of surface.}

Since the surface (if it exists) is located at $R^2=\frac{1}{3}(C^2-A^2)$, we can conclude immediately that there cannot be a well defined surface (in the sense of $p(R)=0$) for case \ref{item:A+C-} and that there always will be a surface for case \ref{item:A-C+}. Note that the latter will also have a naked singularity either below, above or even at the surface.

To see what happens to the other cases, it is helpful to rewrite the surface radius in terms of the central pressure and density:
\begin{equation}
R^2 = \frac{3}{2\pi\, G_N} \frac{p_c}{\rho_c^2-(3p_c)^2} \, .
\end{equation} 
Thus, cases \ref{item:A+C+} and \ref{item:A-C-} will only have a well-defined surface iff $p_c>0$.

\paragraph{Equation of state.}

Due to the simpicity of the pressure and density profiles \eref{TolIV_pressure} and \eref{TolIV_density}, which are simple rational functions, it is possible to eliminate $r^2$ and gain the equation of state that is inherent to the Tolman IV solution:
\begin{equation}
\rho(p) = \frac{8A^2\left(4\pi\,G_N\,p\,C^2\right)^2 + \left(13A^2+2C^2\right)\left(4\pi\,G_N\,p\,C^2\right) + 3\left(2A^2+C^2\right)^2}{4\pi\,G_N\,C^2\left(A^2+2C^2\right)} \, .
\end{equation} 
This equation of state is quadratic in the pressure $p$ and takes a relatively simple form when expressed in terms of the central pressure and density:
\begin{eqnarray}
\rho(p) &=& \frac{8}{\rho_c+p_c}\, p^2 + \frac{5\rho_c-11p_c}{\rho_c+p_c}\, p + \frac{(\rho_c-p_c)(\rho_c-3p_c)}{\rho_c+p_c} \\ \nonumber
&=& \frac{(16\,p^2+5\rho_c-11p_c)^2}{32(\rho_c+p_c)} + 7\rho_c-25p_c \, .
\end{eqnarray}

\paragraph{Asymptotic behaviour.}

The asymptotic behaviour of the Tolman IV solution is obtained by Taylor-expanding \eref{TolIV_pressure} and \eref{TolIV_density} about $r\rightarrow\infty$:
\begin{eqnarray}
\label{eq:TolIV_asymptote_density}
\left. \rho(r)\right|_{r\approx\infty} &=& \frac{3}{16\pi\,G_N\,C^2} + \frac{A^2+2C^2}{32\pi\,G_N\,C^2}\frac{1}{r^2} + \O{\frac{1}{r^4}} \, , \\
\left. p(r)\right|_{r\approx\infty} &=& - \frac{3}{16\pi\,G_N\,C^2} + \frac{A^2+2C^2}{32\pi\,G_N\,C^2}\frac{1}{r^2} + \O{\frac{1}{r^4}} \, .
\end{eqnarray} 
That is, both pressure and density approach a constant value,
\begin{equation}
\label{eq:TolIV_rho_infinity}
\rho_\infty = - p_\infty = \frac{3}{16\pi\,G_N\,C^2} \, ,
\end{equation} 
and both profiles exhibit a $r^{-2}$-falloff close to infinity which implies that
\begin{equation}
m(r\rightarrow\infty) \quad\propto\quad r + \mbox{``cosmological~background~$\propto r^3$''} \, .
\end{equation}
In a universe with a cosmological constant, this could in principle represent a dark matter halo of a galaxy with a flat rotation curve. However, the density of the cosmological constant, the rotation velocity of halo objects and the size of a typical galaxy give three constraints that would have to fit the two free parameters $A^2$ and $C^2$ of the Tolman IV geometry:
\begin{itemize}
\item The density of the cosmological constant can be worked out from the values given in \tref{cosmo_densities}: $\rho_\infty \approx 0.77 \times 10^{-30}\,\unit{kg/m^3}$. Thus, it follows from \eref{TolIV_rho_infinity} that
\begin{equation}
C^2 \approx 1.2 \times 10^{17}\,\unit{kpc^2} \approx (350\,\unit{Gpc})^2\, .
\end{equation}

\item The mass distribution in a flat rotation curve region is typically
\begin{equation}
\frac{m(r)}{r} \approx v_\mathrm{rot}^2 \approx \left(\frac{250\,\unit{km/s}}{c}\right)^2 \approx 7 \times 10^{-7} \approx \mbox{const.} \, ,
\end{equation}
so that the density in the region of flat rotation curves is given by
\begin{equation}
\rho = \frac{m'(r)}{4\pi\,r^2} \approx \frac{7 \times 10^{-7}}{4\pi}\,\frac{1}{r^2} \, .
\end{equation}
Then, by comparison with \eref{TolIV_asymptote_density}, I conclude that
\begin{equation}
\frac{A^2+2C^2}{8C^2} \approx 7 \times 10^{-7} \quad \Rightarrow \quad A^2 \approx - 2 C^2 \approx 2.4 \times 10^{17}\,\unit{kpc^2} \, ,
\end{equation}
meaning that if the Tolman IV geometry is to describe a galaxy with flat rotation curves exhibiting a rotation velocity of $v_\mathrm{rot}\approx 250\,\unit{km/s}$, embedded in a universe with a cosmological constant that has a density of $\rho_\infty \approx 0.77 \times 10^{-30}\,\unit{kg/m^3}$, the parameter $A^2$ will be negative and the parameter $C^2$ will be positive. This corresponds to case 2.

\item Since $A^2 < 0$, the Tolman IV solution exhibits a naked singularity at $r^2=|A^2/2|$. This means that the asymptotic behaviour, as it is illustrated by the Taylor expansion in \eref{TolIV_asymptote_density}, is only valid for $r \gg|A|/\sqrt{2}$. That is, the Tolman IV geometry with an asymptotic density of $\rho_\infty \approx 0.77 \times 10^{-30}\,\unit{kg/m^3}$ exhibits the flat rotation curve behaviour (with the given velocity) only for
\begin{equation}
r \gg 3.5 \times 10^8\,\unit{kpc} \, ,
\end{equation}
which is of course a distance scale that is far too large for typical galaxies.
\end{itemize}
Summarizing, the Tolman IV solution matches asymptotically onto either de Sitter space (for $C^2>0$, cases \ref{item:A+C+} and \ref{item:A-C+}) or anti-de Sitter space (for $C^2<0$, cases \ref{item:A+C-} and \ref{item:A-C-}). Therefore, it could prove to be a useful geometry if one wishes to embed a spherically symmetric object into a cosmology with a cosmological constant. However, the two free parameters constrain the geometry severely, e.g. I have shown that the Tolman IV geometry cannot represent the flat rotation curve region of a typical galaxy. 

As a last remark about the asymptotic behaviour, I note that the density and pressure of Tolman IV at infinity are related to their central values through $C^2$:
\begin{equation}
\rho_\infty = - p_\infty = \frac{\rho_c-3p_c}{4} \, .
\end{equation}

\paragraph{Derivatives of the profiles.}

The derivatives, with respect to $r$, of the pressure and density profiles are
\begin{eqnarray}
p'    &=& - \frac{(A^2+2C^2)\, r}{4\pi\, G_N\, C^2\, (A^2 + 2r^2)^2} \, , \\
\rho' &=& - \frac{(A^2+2C^2)(5A^2+2r^2)\, r}{4\pi\, G_N\, C^2\, (A^2 + 2r^2)^3} \, .
\end{eqnarray} 
From that we see that $p'$ never changes sign and is zero only for $r=0$ and $r\rightarrow\infty$. Whether $p'$ is increasing or decreasing is determined by the constants $A^2$ and $C^2$: $\sign p'=-\sign[C^2(A^2+2C^2)]$.

For $A^2>0$, the sign of $\rho'$ is the same as that of $p'$. If $A^2<0$ (cases \ref{item:A-C+} and \ref{item:A-C-}), the sign of $\rho'$ first changes at the curvature singularity at $r^2=-A^2/2$ and then there is an extremum at $r^2=-5/2\, A^2$ where the sign also changes. Thus $\sign \rho' = \sign p'$ in the central region and near infinity, and $\sign \rho' = - \sign p'$ for $-A^2/2 < r^2 < -5/2\, A^2$.

\paragraph{Total mass.}

The density profile can easily be integrated to yield the mass $m(r)$ that is enclosed in a sphere with radius $r$ ($G_N=1$ for convenience):
\begin{equation}
m(r) = \frac{(A^2+C^2+r^2)\,r^3}{2 C^2\, (A^2 + 2r^2)} \, .
\end{equation}
Thus, the mass grows as $r^3$ for $r\rightarrow\infty$ as one would expect from a ``cosmological background density''. Subtracting this asymptotic mass gives
\begin{equation}
m(r) - \frac{r^3}{4C^2} = \frac{(A^2+2C^2)\,r^3}{4 C^2\, (A^2 + 2r^2)} \, ,
\end{equation} 
which grows as $r$ for $r\rightarrow\infty$ as I already mentioned in the paragraph about asymptotic behaviour:
\begin{equation}
m(r) - \frac{r^3}{4C^2} = \frac{A^2+2C^2}{8C^2}\,r 
+ \O{\frac{1}{r}} \, .
\end{equation}
Although this behaviour is typical for the halo of flat rotation curve galaxies, I have already shown that the Tolman IV geometry cannot be parametrized suitably to match the densities and distance scales of a realistic galaxy.

If the configuration has a well defined surface where $p(R)=0$, the total mass $M=m(R)$ is given by
\begin{equation}
M = \frac{\sqrt{3} \sqrt{C^2-A^2}^3}{9C^2} = \frac{R^3}{3C^2} \, ,
\end{equation}
and thus we see that the total mass will be negative if $C^2$ is negative and if there is a well defined surface radius.

\subsubsection{Anti-Gravastar configurations of the Tolman IV metric}

We have seen that the Tolman IV solution can represent quite different geometries, depending on the choice of the sign of the parameters $A^2$ and $C^2$. It can exhibit nasty features like a naked curvature singularity or have a well-defined surface with vanishing pressure. In any case, the asymptotic behaviour of the Tolman IV solution is either de Sitter or anti-de Sitter space.

A quite surprising finding was that the Tolman IV solution has a simple, analytical equation of state. It represents a two-parameter ($A$, $C$) sub-class of fluid spheres that have a quadratic three-parameter ($\alpha$, $\beta$, $\gamma$) equation of state of the form
\begin{equation}
\rho(p) = \alpha(A,C)\, p^2 + \beta(A,C)\, p + \gamma(A,C)\, \, .
\end{equation}
The three parameters $\alpha$, $\beta$ and $\gamma$ can be expressed through either $A^2$ and $C^2$ or $\rho_c$ and $p_c$. Thus, if one is considering a quadratic equation of state for a spherical object, one might want to check whether it falls into the Tolman IV category.

The usual object that will be considered is likely to have all parameters positive, which is case \ref{item:A+C+} with $p_c>0$. This will result in a ``normal'' compact object with a radius $R$ and a total mass $M=R^3/3C^2$.

If we are looking for configurations that qualify as an anti-gravastar, we are immediately restricted to cases \ref{item:A+C-} and \ref{item:A-C-} with $p_c>0$ and $\rho_c<0$. Since case \ref{item:A+C-} does not have a surface, we can rule it out already. For case \ref{item:A-C-} we know $r^2<R^2<A^2/2$ and thus, the naked singularity will occur outside the surface and therefore shall not be of concern. The derivatives of both, pressure and density, are negative for $r<R$ and thus, both will be decreasing, i.e. the density will always be negative. Thus, the total mass $M$ will be negative, which also can be deduced from $C^2<0$.

Finally, we find that the modified Tolman IV solution for a negative mass anti-gravastar is
\begin{equation}
\d s^2 = - B^2 \left( 1 - \frac{r^2}{A^2} \right)^2 \d t^2 + \frac{1 - 2\, r^2/A^2}{\left( 1 + r^2/C^2 \right)\left( 1 - r^2/A^2 \right)} \d r^2 + r^2 \d \Omega^2
\end{equation}
with $A^2>C^2$, i.e. $p_c>0$, and $A^2>0$, $C^2>0$.  Example plots of the pressure and density profiles are shown in \fref{anti_grav_TolIV_profiles}. 
\begin{figure}[hbt]
\begin{center}
\includegraphics[width=7cm]{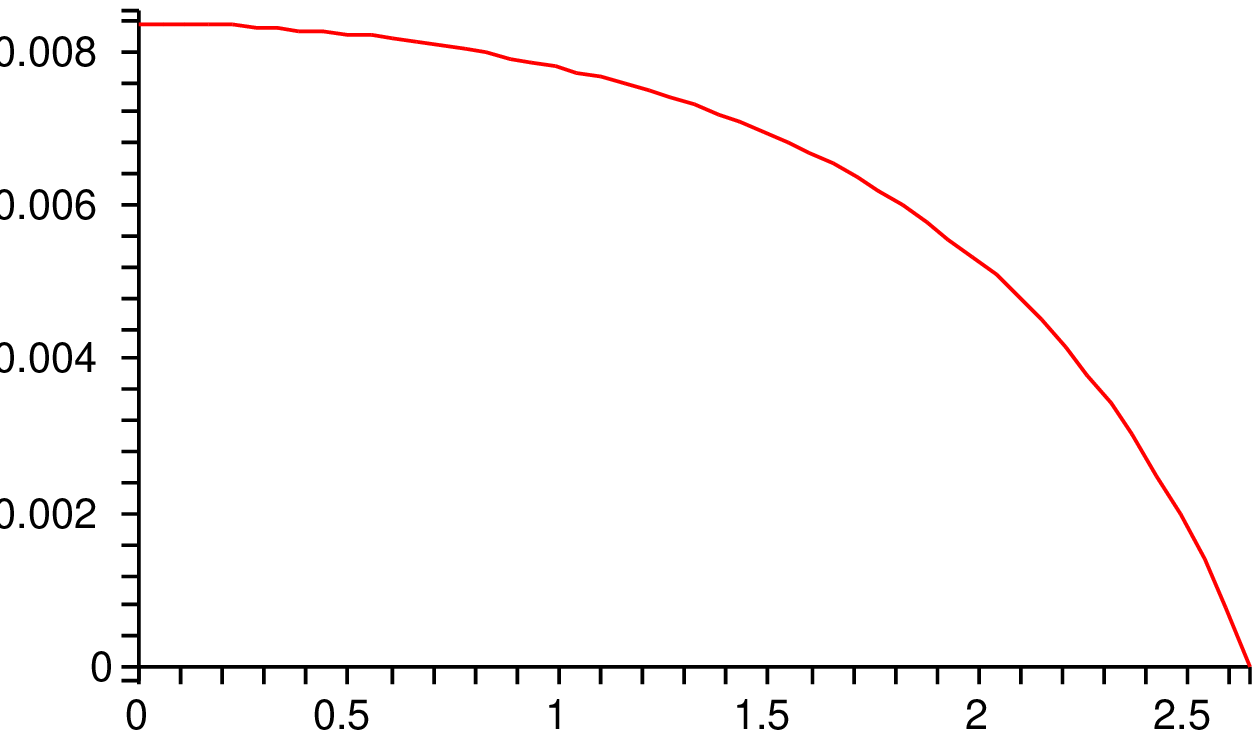}
\includegraphics[width=7cm]{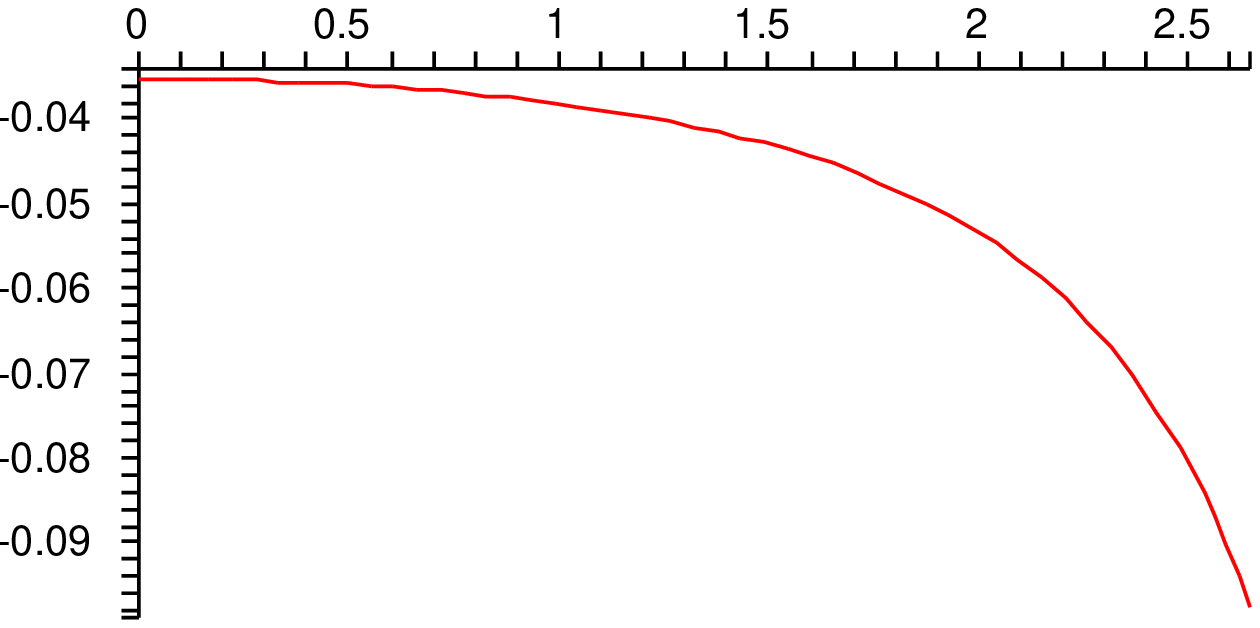}
\end{center}
\caption[Pressure and density profiles of an anti-gravastar represented by the Tolman IV geometry.]{Pressure profile (left) and density profile (right) in dimensionless units for $A=5$, $C=2$ and thus $R=\sqrt{7}$ ($G_N=1$, $c=1$). Note that $\rho(R)= -\frac{27}{88\pi\,G_N}$ is finite.}
\label{fig:anti_grav_TolIV_profiles}
\end{figure}

From the negative density at the surface, we can immediately conclude that the NEC will be violated there. In fact, it turns out that
\begin{equation}
\rho+p = \frac{(A^2+2C^2)(r^2-A^2)}{4\pi\,G_N\,C^2\,(A^2-2r^2)^2} < 0
\end{equation} 
since $r^2<R^2<A^2/2<A^2$. Thus, the NEC is violated everywhere.

\subsection{Equation of state for an anti-gravastar with positive total mass}
\label{sec:antigravastar_positive_mass}

Since we already have seen that perfect fluid anti-gravastars with negative total mass can exist, one might ask whether perfect fluid anti-gravastars that have a positive total mass can exist in theory as well. I will show that such objects can exist when one chooses the equation of state carefully.

An anti-gravastar can only have a positive total mass if there is ``more'' positive density matter than negative density matter. ``More'' in this context means the absolute value of the positive mass is greater than the absolute value of the negative mass, $|M_+| > |M_-|$.

The simplest configuration that can have a positive total mass is a negative density core with a positive density crust (\fref{anti_grav_pos_mass}). This satisfies our requirement that the central density must be negative. The positive density matter in the crust shifts the problem of converting infalling ``normal'' matter into negative density matter from the surface to inside the anti-gravastar. This conversion process still has to be justified in a physically relevant model, but for the time being, we just explore the mathematical implications. From that point of view it is more convenient to model the density-transition with a continuous equation of state and match a positive mass Schwarzschild exterior metric at the surface $R$.
\begin{figure}[hbt]
\begin{center}
\input{figures/anti-gravastar_pos_mass.pstex_t}
\end{center}
\caption{Pressure and density for an anti-gravastar with positive total mass.}
\label{fig:anti_grav_pos_mass}
\end{figure}
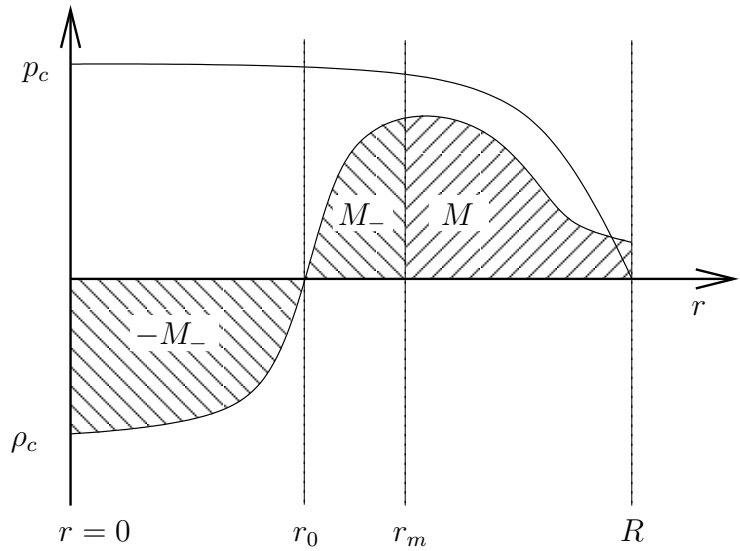

A simple equation of state that can in principle represent all necessary features is
\begin{equation}\label{eq:anti_grav_eos_tanh}
\rho(p) = p \tanh\frac{p_0-p}{D} + \rho_S
\end{equation} 
which is plotted in \fref{anti_grav_tanh_eos}. The parameter $p_0>0$ marks the pressure for which the density changes its sign, $D>0$ is a measure for the width of the transition region and $\rho_S>0$ denotes the surface density, which has to be non-zero if one wants to avoid an infinitely far streching atmosphere.

It is easy to check that this choice of equation of state satisfies the NEC:
\begin{equation}
\rho+p = p\, \left( 1 + \tanh\frac{p_0-p}{D} \right) + \rho_S > \rho_S \geq 0 \, .
\end{equation}
For $p_c \gg p_0$ and $p_c \gg \rho_S$ I also find
\begin{equation}
\label{eq:tanheos_central_pressure+density}
\rho_c = p_c\, \tanh\frac{p_0-p_c}{D} + \rho_S \approx -p_c + \rho_S \approx -p_c
\end{equation}
and thus, this equation of state corresponds approximately to anti-de Sitter space near the center.
\begin{figure}[hbt]
\begin{center}
\input{figures/anti-gravastar_tanh_eos.pstex}
\end{center}
\caption[Equation of state \eref{anti_grav_eos_tanh} for an anti-gravastar with positive total mass.]{Equation of state \eref{anti_grav_eos_tanh} with $p_0=2$, $D=0.1$ and $\rho_S=0.1$.}
\label{fig:anti_grav_tanh_eos}
\end{figure}

To clarify the following discussion, I use the average density $\rhoavg$ and the local gravitational acceleration $g$ as they were defined in \sref{gravastar}.
Additionally, I note that
\begin{equation}\label{eq:signpprime}
\sign p' = -\sign \left[ (\rho+p)\, g \right] = -\sign \left[ (\rho+p)(\rhoavg+3p) \right] \, .
\end{equation} 

The simplest anti-gravastar will have $p'<0$ throughout the interior. Then, for $p_c>0$, the pressure is guaranteed to be positive inside the surface radius $R$, which is defined by $p(R)=0$. Hence, the first property of an anti-gravastar  in \sref{antigravastar_defs} is satisfied. If $p_c > p_0$, the equation of state \eref{anti_grav_eos_tanh} makes sure that the central density is negative, which was the second property of an anti-gravastar. The last important property of an anti-gravastar is the absence of event horizons, i.e. $2m(r)/r < 1$. This property can generally only be evaluated numerically. Thus, to construct an anti-gravastar with the equation of state \eref{anti_grav_eos_tanh}, it is sufficient to show that $p'<0$ everywhere when $p_c>p_0$. However, the absence of horizons still needs to be verified numerically.

By \eref{signpprime} and knowing that the NEC is satisfied, I conclude that 
\begin{equation}
\label{eq:tanheos_g_positive}
p'<0 \quad \Leftrightarrow \quad g \propto \rhoavg+3p>0 \quad \Leftrightarrow \quad p(r)>-\frac{1}{3}\,\rhoavg(r) \, ,
\end{equation}
i.e. the local gravitational acceleration needs to point inwards everywhere. Since the equation of state \eref{anti_grav_eos_tanh} induces a pressure and density profile with the qualitative shape of \fref{anti_grav_pos_mass}, it is safe to assume that
\begin{equation}
\rho_c \leq \rhoavg(r)
\end{equation}
everywhere. Furthermore, from \eref{tanheos_central_pressure+density}, one can easily conclude that
\begin{equation}
-\rho_c = |\rho_c| \leq p_c
\end{equation}
for $p_c>p_0$. Combining these last three inequalities yields
\begin{equation}
\label{eq:tanheos_g_positive_inner}
p(r) > \frac{1}{3}\,p_c \geq -\frac{1}{3}\,\rho_c \geq -\frac{1}{3}\,\rhoavg(r) \quad \Rightarrow \quad g>0 \, .
\end{equation}
This inequality guarantees positive $g$ everywhere. However, it is too restrictive in the outer regions of an anti-gravastar. But since $p>0$ everywhere, it is sufficient to apply the less strict bound
\begin{equation}
\label{eq:tanheos_g_positive_outer}
\rhoavg > 0 \quad \Rightarrow \quad g>0
\end{equation}
for the outer regions of the anti-gravastar.
I conclude that if the parameters of \eref{anti_grav_eos_tanh} are chosen carefully enough to
\begin{enumerate}
\item ensure that $\rho_c<0$, i.e. $p_c > p_0$;
\item have gravitational attraction everywhere, i.e. $g>0$ which is generally established by the inequality \eref{tanheos_g_positive} and specifically by either \eref{tanheos_g_positive_inner} or \eref{tanheos_g_positive_outer};
\item avoid the formation of an event horizon for all $r<R$, i.e. $2m(r)/r<1$,
\end{enumerate}
the equation of state \eref{anti_grav_eos_tanh} will lead to a perfect fluid gravastar. This equation of state can potentially yield an anti-gravastar with a positive total mass. To ensure this, the parameters of \eref{anti_grav_eos_tanh} must be chosen to also
\begin{enumerate}
\item[4.] make sure that the positive density crust dominates the negative density core, i.e. $|M_{crust}| > |M_{core}|$.
\end{enumerate}

Finding an actual set of suitable parameters, and therefore proving that anti-gravastars with a positive total mass are mathematically possible, requires one to solve the TOV equation numerically. This task was too time consuming to be realized within this thesis and therefore, I leave it to be carried out in the future.

One remarkable fact that arises from this configuration, is that the density in the crust can be much higher than the density in an object with the same total compactness $2M/R$. Since the local compactness $2m(r)/r$ gets a ``negative headstart'' from the core, the crust actually has to contain more positive density matter than another object with the same total mass $M$ and radius $R$.

\subsection{Summary}

I have found two analytical solutions that can represent anti-gravastars with a negative total mass: the Schwarzschild interior solution and the Tolman IV geometry. Both solutions consist of a spherically symmetric perfect fluid. Among the known spherical perfect fluid solutions \cite{Delgaty:1998uy}, there may be more that actually can represent an anti-gravastar. However, a systematic search has not yet been carried out.

Unfortunately I was not able to spot an analytical solution that constitutes an anti-gravastar with a positive total mass, although there is no indication that such a solution should not exist. To emphasise the possibility of a positive mass anti-gravastar, I used the equation of state
\begin{equation}
\label{eq:anti_grav_eos_tanh_2}
\rho(p) = p \tanh\frac{p_0-p}{D} + \rho_S
\end{equation} 
to illustrate the necessary conditions for the existence of such an object. However, the ultimate proof for positive mass anti-gravastars, using numerical methods, is yet to be carried out.

While the definition of an anti-gravastar already violates the WEC and DEC near the center where $\rho<0$, there also seems to be a correlation between the NEC and the sign of the total mass of an anti-gravastar. Both negative total mass solutions, the Schwarzschild interior and Tolman IV, violated the NEC everywhere whereas the example for a possible positive mass anti-gravastar satisfied the NEC everywhere.

The version of the positive mass theorem obtained by Penrose, Sorkin \& Woolgar \cite{Penrose:1993ud} states that if the NEC is satisfied everywhere, the total mass must be positive. Applied to the example in \sref{antigravastar_positive_mass}, this theorem implies that if an anti-gravastars with an equation of state \eref{anti_grav_eos_tanh_2} exists, it must have a positive total mass.

Conversely, one can conclude that an anti-gravastar with a negative total mass must violate the NEC at least locally in some region. However, it is an open question whether a local (or also global) violation of the NEC is permittable for a positive total mass or whether violation of the NEC immediately leads to a negative total mass.

Analogously to gravastars, if anti-gravastars with a positive total mass and a sufficient compactness of $2M/R \approx 1$ existed physically, they would exhibit an external gravitational field that is indistinguishable from that of a black hole. Although the model of an anti-gravastar is not favoured by me personally as an alternative for black holes, it demonstrates -- just like the gravastar -- that the current observational evidence for compact objects in the universe does not automatically lead to the conclusion that it must be a black hole --- there is still a little wriggle room left.



%% file: figures/anti-gravastar_pos_mass.pstex_t
\begin{picture}(0,0)%
\includegraphics{figures/anti-gravastar_pos_mass.pstex}%
\end{picture}%
\setlength{\unitlength}{3947sp}%
\begingroup\makeatletter\ifx\SetFigFont\undefined%
\gdef\SetFigFont#1#2#3#4#5{%
  \reset@font\fontsize{#1}{#2pt}%
  \fontfamily{#3}\fontseries{#4}\fontshape{#5}%
  \selectfont}%
\fi\endgroup%
\begin{picture}(4597,3455)(3226,-6869)
\put(7501,-5386){\makebox(0,0)[lb]{\smash{{\SetFigFont{12}{14.4}{\familydefault}{\mddefault}{\updefault}{\color[rgb]{0,0,0}$r$}%
}}}}
\put(3526,-6811){\makebox(0,0)[lb]{\smash{{\SetFigFont{12}{14.4}{\familydefault}{\mddefault}{\updefault}{\color[rgb]{0,0,0}$r=0$}%
}}}}
\put(7051,-6811){\makebox(0,0)[lb]{\smash{{\SetFigFont{12}{14.4}{\familydefault}{\mddefault}{\updefault}{\color[rgb]{0,0,0}$R$}%
}}}}
\put(5626,-6811){\makebox(0,0)[lb]{\smash{{\SetFigFont{12}{14.4}{\familydefault}{\mddefault}{\updefault}{\color[rgb]{0,0,0}$r_m$}%
}}}}
\put(4998,-6811){\makebox(0,0)[lb]{\smash{{\SetFigFont{12}{14.4}{\familydefault}{\mddefault}{\updefault}{\color[rgb]{0,0,0}$r_0$}%
}}}}
\put(3301,-3886){\makebox(0,0)[lb]{\smash{{\SetFigFont{12}{14.4}{\familydefault}{\mddefault}{\updefault}{\color[rgb]{0,0,0}$p_c$}%
}}}}
\put(3226,-6211){\makebox(0,0)[lb]{\smash{{\SetFigFont{12}{14.4}{\familydefault}{\mddefault}{\updefault}{\color[rgb]{0,0,0}$\rho_c$}%
}}}}
\put(5926,-4820){\makebox(0,0)[lb]{\smash{{\SetFigFont{12}{14.4}{\familydefault}{\mddefault}{\updefault}{\color[rgb]{0,0,0}$M$}%
}}}}
\put(4004,-5570){\makebox(0,0)[lb]{\smash{{\SetFigFont{12}{14.4}{\familydefault}{\mddefault}{\updefault}{\color[rgb]{0,0,0}$-M_-$}%
}}}}
\put(5279,-4820){\makebox(0,0)[lb]{\smash{{\SetFigFont{12}{14.4}{\familydefault}{\mddefault}{\updefault}{\color[rgb]{0,0,0}$M_-$}%
}}}}
\end{picture}%

%% file: tc_conclusions.tex
\chapter{Summary and conclusions}
\label{sec:conclusions}

The discussions in this thesis have shown that pressure plays a special role in general relativity as opposed to Newtonian gravity. In general relativity pressure is a source of the gravitational field, and the perception of that gravitational field by passive particles depends on the particle speed (\sref{galactic_halo}). Since gravity and fluid dynamics are distinct theories in the Newtonian picture, these ``relativistic'' aspects of pressure cannot be adequately described in a quasi-Newtonian approximation. 

\section{Rotation curves, gravitational lensing and the equation of state of the galactic fluid}

After giving an overview of typical galactic structure and possible dark matter candidates in \sref{galaxy}, I went on to discuss how to interpret rotation curve measurements in general relativity in \sref{rotcurve}.

It has been shown previously by Lake \cite{Lake:2004}, (and also by Nucamendi, Salgado and Sudarsky \cite{Nucamendi:2000}), that the observable rotation curve data of edge-on galaxies contain contributions from gravitational redshift, as well as the Doppler shift which arises from motion of the tracer particle towards or away from the observer (\sref{obs_light}). However, applying this result in an observational context, I found that the gravitational redshift \eref{grav_redshift_neglect}, as opposed to the Doppler shift \eref{Doppler_redshift_dominant}, contributes at a higher order of the small rotation velocity to the observable frequency shift. Therefore, the gravitational redshift is negligible when processing the observational data. Furthermore, I extended the derivation of the general relativistically derived total redshift to be applicable to arbitrarily inclined galaxies, and found that under the permittable weak field approximations, the resulting total redshift \eref{obs_redshift2} is identical to that found in Newtonian gravity. Consequently, the established data reduction procedures for rotation curve measurements yield the general relativistically correct redshift for weak gravitational fields. However, as I argued in \sref{rotcurve_interpretation}, the \emph{interpretation} of that observed redshift in general relativity is different from the widely adapted interpretation in Newtonian gravity: The potential $\Phi_\mathrm{RC}(r)=\Phi(r)$, which can be determined from the redshift, is generally \emph{not} equal to the Newtonian gravitational potential $\Phi_N(r)$, given by the field equation \eref{Newton_grav_source},
\begin{equation}
\nabla^2 \Phi_N(r) = 4\pi\,\rho(r) \, .
\end{equation} 
Instead, the appropriate weak field equation in general relativity is\footnote{To keep the conclusion as simple as possible, I use $p$ in the sense of $p=(p_r+2p_t)/3$.} \eref{pressure_field_eq},
\begin{equation}
\label{eq:conc_field_eq_RC}
\nabla^2 \Phi_\mathrm{RC}(r) = 4\pi\,[\rho(r) + 3\,p(r)]\, .
\end{equation}
The Newtonian limit is recovered iff $|p| \ll \rho$.

In \sref{grav_lensing} I pointed out that the currently employed formalism for gravitational lensing is based on a superposition of the deflection angle of a point mass. Since this superposition excludes the notion of pressure, it cannot possibly model gravitational lensing appropriately when the lensing galaxy is made of some form of matter with a non-negligible pressure content. To find a description that allows for possible pressure in the active gravitational matter, I argued that the effective refractive index is the only possible observable quantity of gravitational lensing by a spherically symmetric galaxy. In isotropic coordinates \eref{ref_index_weak_gravity}, the effective refractive index is a scalar, while other coordinate systems require a refractive index tensor with at least two independent components (see \sref{refindex_sphersymm}). This is still true for aspherical configurations as long as the 3-dimensional space part of the metric is conformally Euclidean. By comparison of the refractive index \eref{ref_index_weak_gravity},
\begin{equation}
n(r) \equiv 1 - 2\Phi_\mathrm{lens}(r) \approx 1 - \Phi(r) - \int \frac{m(r)}{r^2}\, \d r \, ,
\end{equation} 
with its commonly used quasi-Newtonian counterpart,
\begin{equation}
n(r) \equiv 1 - 2\Phi_\mathrm{lens}(r) \approx 1 - \Phi_N(r) \, ,
\end{equation} 
I conclude that the lensing potential $\Phi_\mathrm{lens}(r)$, as obtained from the established data analysis methodology, has to be interpreted in terms of the metric functions $\Phi(r)$ and $m(r)$ in the following manner:
\begin{equation}
\Phi_\mathrm{lens}(r) = \frac{1}{2}\, \Phi(r) + \frac{1}{2}\, \int \frac{m(r)}{r^2}\, \d r \, .
\end{equation} 
Therefore, the corresponding field equation for the lensing potential \eref{lens_potential_field_equation} is
\begin{equation}
\label{eq:conc_field_eq_lens}
\nabla^2 \Phi_\mathrm{lens}(r) = 4\pi\,\left[\rho(r) + \frac{3}{2}\,p(r) \right] \, .
\end{equation} 
Again, the Newtonian limit is recovered iff $|p| \ll \rho$.

In the case of non-negligible pressure, \eref{conc_field_eq_RC} and \eref{conc_field_eq_lens} can easily be inverted to yield the density $\rho$ and pressure $p$ of the galactic fluid when sufficient data is available from independent rotation curve and lensing observations of the same galaxy. However, the critical word in the previous sentence is ``sufficient'', as it is not easy to simultaneously obtain quality data for rotation curves and gravitational lensing for the same object. A more detailed discussion of the problems involved is given in \sref{rotcurve+lensing_observations}. The basic problem is that good rotation curve data at all radii can be obtained for nearby galaxies, but for these galaxies the lensing data tends to be concentrated towards the center. Since the recent past has shown us that the observational situation is only getting better, I am confident that in the near future there will be enough quality data available to employ the formalism presented and discussed in \sref{rotcurve+lensing_conclusions}.

It will then be possible to infer the equation of state of the galactic fluid directly from observations, which might confirm the Cold Dark Matter paradigm observationally or, if significant pressure is detected, even shed new light on the nature of dark matter.

\section[Black holes vs.~gravastars]{Black holes vs.~gravastars: Mathematical simplicity or unresolved physics?}

Black holes are widely accepted in the astrophysics community because they are a mathematically simple solution in the strong gravity regime of compact objects, that also fits the latest observational evidence for a very dense matter concentration at the center of the Milky Way and other galaxies. However, the simplicity comes at the price of a pathological curvature singularity at the center of a black hole, which indicates that general relativity indeed fails to deliver an appropriate description of physics at this singular point. A unified theory of gravity and quantum theory might resolve this issue, but it is not necessary to appeal to the ``holy grail'' of the great unified theory to find a solution that provides us with compact objects similar to black holes but without mathematical pathologies.

The gravastar picture, as developed by Mazur and Mottola \cite{Mazur:2004fk}, provides a mathematical alternative to black holes: A massive core of matter with the equation of state of the cosmological constant, $\rho=-p>0$, is covered in a thin but finite thickness shell of stiff matter with an equation of state $\rho=p$. Implicit in this model are two infinitesimally thin shells at the boundaries between the core and the stiff matter shell as well as between that shell and the surrounding vacuum. The spacetime of a gravastar is free of curvature singularities and is regular in the sense that there is no horizon. Together with C\'eline Catto\"en and Matt Visser, I have generalized this layered gravastar to a model that exhibits a continuous pressure profile without distinguished layers. We showed that gravastars without any thin shells must have anisotropic pressure in their ``crust''. Therefore, the inner thin shell is unavoidable if one wants to model a gravastar with perfect fluid only.

Since the gravastar core is made out of matter similar to the cosmological constant, it is also said to be filled with ``dark energy'', a form of mass-energy that is widely accepted in cosmology where the dark energy density is very low. In a gravastar, this dark energy would have a density that is much higher than predicted by other standard astrophysical models.

A different, even more exotic alternative to black holes is the anti-gravastar. Matt Visser and I explored the properties of a compact object that has positive pressure everywhere but a core consisting of negative density matter. I argued that objects of this kind could also mimic the external gravitational field of a black hole. However, even though the anti-gravastar can consist entirely of a perfect fluid, it is more exotic in the sense that it consists of matter with a negative density. Unlike dark energy, there are no observational indications that matter of this kind might exist in the universe. 

While being agnostic about the existence of gravastars, anti-gravastars or black holes, I want to point out that there is as yet no \textit{direct} evidence for black holes with an event horizon and a central curvature singularity. The observations of the most compact objects are consistent with a black hole, but also with a gravastar or an anti-gravastar. However, before gravstars or anti-gravastars can be considered a serious alternative to black holes, their physical properties should be explored in more detail. Being a mathematical solution does not make the (anti-) gravastar a physically relevant object.

\paragraph{}
I hope the research presented in this thesis will make a valuable contribution to prevailing questions of astronomy, and I would like to thank again all people who helped me creating this work.

%% file: tc_bibliography.tex

\addcontentsline{toc}{chapter}{Bibliography}
\bibliographystyle{plain}
\bibliography{msc_references}

%% file: ta_refractive_index_arxiv.tex


\chapter[Weak-field refractive index tensor]{Effective refractive index tensor for weak-field gravity}
\label{sec:refindex}

\centerline{\bf
 Petarpa Boonserm,
C\'eline Catto\"en,
Tristan Faber, }
\centerline{\bf Matt Visser,
and
Silke Weinfurtner}

\vskip 0.5 cm
\centerline{{\footnotesize School of Mathematics, Statistics, and Computer Science,}} 
\centerline{{\footnotesize Victoria University of Wellington, }}
\centerline{{\footnotesize P.O.Box 600, Wellington, New Zealand}} 



\vskip 0.2 cm

\noindent
  {\small Gravitational lensing in a weak but otherwise arbitrary
  gravitational field can be described in terms of a $3\times3$
  tensor, the ``effective refractive index''. If the sources
  generating the gravitational field all have small internal fluxes,
  stresses, and pressures, then this tensor is automatically isotropic
  and the ``effective refractive index'' is simply a scalar that can
  be determined in terms of a classic result involving the Newtonian
  gravitational potential. In contrast if anisotropic stresses are
  ever important then the gravitational field acts similarly to an
  anisotropic crystal.  We derive simple formulae for the refractive
  index tensor, and indicate some situations in which this will be
  important.}

\vskip 0.1 cm
\noindent
{
\\
\vskip 0.3 cm
\centerline{{\small Published as {\bf Classical and Quantum Gravity 22 (2005) 1905-1915}}}}
\vskip 0.2 cm

\centerline{ {\footnotesize  \texttt{
petarpa.boonserm@mcs.vuw.ac.nz,
celine.cattoen@mcs.vuw.ac.nz,
tristan.faber@mcs.vuw.ac.nz,}}} 
\centerline{{\footnotesize  \texttt{matt.visser@mcs.vuw.ac.nz,
silke.weinfurtner@mcs.vuw.ac.nz}}}

\vskip 0.5cm
\centerline{
This article can be accessed online at \texttt{http://arxiv.org/abs/gr-qc/0411034}
}

\label{sec:refindex_static}
\label{sec:refindex_sphersymm}

%% file: ta_gravastar_anisotropy_arxiv.tex


\chapter[Gravastars \& anisotropic pressures]{Gravastars must have anisotropic pressures}
\label{sec:appgravastar}

\centerline{\bf
C\'eline Catto\"en,
Tristan Faber, and Matt Visser}
\centerline{ {\footnotesize School of Mathematics, Statistics, and Computer Science,} }
\centerline{{\footnotesize Victoria University of Wellington, }}
\centerline{{\footnotesize P.O.Box 600, Wellington, New Zealand}} 
\vskip 0.2 cm
{\small One of the very small number of serious alternatives to the usual concept of an astrophysical
black hole is the ``gravastar'' model developed by Mazur and Mottola; and a related phase-transition model due to Laughlin \emph{et al}. We consider a generalized class of similar models that exhibit continuous pressure --- without the presence of infinitesimally thin shells. By considering the usual TOV equation for static solutions with negative central pressure, we find that gravastars cannot be perfect fluids --- anisotropic pressures in the ``crust'' of a gravastar-like object are unavoidable. The anisotropic TOV equation can then be used to bound the pressure anisotropy.
The transverse stresses that support a gravastar permit a higher compactness than is given by the Buchdahl--Bondi bound for perfect fluid stars. Finally we comment on the qualitative features of the equation of state that gravastar material must have if it is to do the desired job of preventing horizon formation.}

\vskip 0.1 cm
\noindent
{
\\
\vskip 0.3 cm
\centerline{{\small Published as {\bf Classical and Quantum Gravity 22 (2005) 4189-4202}}}}
\vskip 0.2 cm

\centerline{ {\footnotesize  \texttt{
celine.cattoen@mcs.vuw.ac.nz,
tristan.faber@mcs.vuw.ac.nz, matt.visser@mcs.vuw.ac.nz}}}


\vskip 0.5cm
\centerline{
This article can be accessed online at \texttt{http://arxiv.org/abs/gr-qc/0505137}
}

\label{sec:gravastar_metric}
\label{sec:iso_fail}
\label{sec:GS_ss_interior}
\label{sec:gravastar_eos}

%% file: ta_galaxy_eos_arxiv.tex

\chapter[Measuring the equation of state of dark matter]{Combining rotation curves and gravitational lensing: How to measure the equation of state of dark matter in the galactic fluid}
\label{sec:galaxy_eos}

\centerline{\bf Tristan Faber and Matt Visser}

\vskip 0.5 cm
\centerline{{\footnotesize School of Mathematics, Statistics, and Computer Science,}} 
\centerline{{\footnotesize Victoria University of Wellington, }}
\centerline{{\footnotesize P.O.Box 600, Wellington, New Zealand}} 



\vskip 0.2 cm

\noindent
  {\small We argue that combined observations of galaxy rotation curves and gravitational lensing not only allow the deduction of a galaxy's mass profile, but also yield information about the pressure in the galactic fluid. We quantify this statement by enhancing the standard formalism for rotation curve and lensing measurements to a first post-Newtonian approximation. This enhanced formalism is compatible with currently employed and established data analysis techniques, and can in principle be used to reinterpret existing data in a more general context. The resulting density and pressure profiles from this new approach can be used to constrain the equation of state of the galactic fluid, and therefore might shed new light on the persistent question of the nature of dark matter.}

\vskip 0.1 cm
\noindent
{
\\
\vskip 0.3 cm
\centerline{{\small Submitted to {\bf Monthly Notices of the Royal Astronomical Society}}}}
\vskip 0.2 cm

\centerline{ {\footnotesize  \texttt{
tristan.faber@mcs.vuw.ac.nz,
matt.visser@mcs.vuw.ac.nz}}} 

\vskip 0.5cm
\centerline{
This article can be accessed online at \texttt{http://arxiv.org/abs/astro-ph/0512213}
}